\documentclass[10pt]{article}

\usepackage[left=20mm,right=20mm,top=20mm,bottom=25mm]{geometry}

\usepackage[english]{babel}
\usepackage[utf8]{inputenc}
\usepackage[autostyle]{csquotes}
\usepackage[T1]{fontenc}
\usepackage{hyperref}
\usepackage[affil-it]{authblk}

\usepackage{graphicx}%
\usepackage{multirow}%
\usepackage{amssymb,amsfonts}%
\usepackage{mathtools}
\usepackage{amsthm}%
\usepackage{mathrsfs}%
\usepackage[title]{appendix}%
\usepackage[dvipsnames,svgnames]{xcolor}%
\usepackage{textcomp}%
\usepackage{manyfoot}%
\usepackage{booktabs}%
\usepackage{algorithm}%
\usepackage{algorithmicx}%
\usepackage{algpseudocode}%
\usepackage{listings}%
\usepackage{verbatim}
\usepackage{tabularx}
\usepackage{comment}
\usepackage{siunitx}

\usepackage{tikz}
\usepackage{pgfplots}
\usepgfplotslibrary{groupplots}
\usepackage{pgfplotstable}
\usepackage{tikzscale}
\pgfplotsset{compat=newest}
\usepgfplotslibrary{fillbetween}
\usepgfplotslibrary{colormaps}
\usetikzlibrary{patterns}
\usetikzlibrary{shapes,arrows,matrix,positioning,arrows.meta,spy}

\usepackage{subcaption}
\usepackage[font=small]{caption}
\usepackage{wrapfig}

\usepackage[capitalise]{cleveref}
\usepackage[modulo,mathlines,switch]{lineno}

\usepackage[backend=biber, maxbibnames=20, maxcitenames=2, 
    natbib=true, style=numeric, date=year, 
    isbn=false, url=false,giveninits=true,sortcites]{biblatex} 
\theoremstyle{definition}   
\newtheorem{definition}{Definition}[section]
\newtheorem{remark}[definition]{Remark}

\newcommand{\R}{1.0}

\newcommand{\norm}[1]{\left\lVert#1\right\rVert}

\algnewcommand\alginput{\textbf{Input:}}
\algnewcommand\algoutput{\textbf{Output:}}

\newcommand{\SP}{\mathcal{SP}}
\newcommand{\lin}{\text{Lin}}

\newcommand{\SO}{\text{SO}}

\newcommand{\der}[2]{\frac{\partial #1}{\partial #2}} 

\newcommand{\initCS}{\boldsymbol{X}_{CS}}      
\newcommand{\currCS}{\hat{\boldsymbol{X}}_{CS}}

\newcommand{\matStr}{\boldsymbol{\epsilon}}  
\newcommand{\matCurv}{\boldsymbol{\kappa}}   

\newcommand{\spaForce}{\hat{\boldsymbol{n}}}         
\newcommand{\spaMoment}{\hat{\boldsymbol{m}}}        
\newcommand{\matForce}{\boldsymbol{n}}       
\newcommand{\matMoment}{\boldsymbol{m}}      

\newcommand{\vMatStrain}{\boldsymbol{p}}     
\newcommand{\vMatStress}{\boldsymbol{q}}     
\newcommand{\vMatStrainSym}{\boldsymbol{p}^{\text{sym}}}     
\newcommand{\vMatStrainTilde}{\tilde{\boldsymbol{p}}}
\newcommand{\vMatStressTilde}{\tilde{\boldsymbol{q}}}

\newcommand{\vMatStressSymTilde}{\tilde{\boldsymbol{q}}^{\text{sym}}}
\newcommand{\inputs}{\boldsymbol{\mathcal{I}}}
\newcommand{\inputsIso}{\inputs^{\text{ti}}}
\newcommand{\parameter}{\ensuremath{P}}
\newcommand{\pot}{\psi}                        
\newcommand{\potRod}{\psi}               
\newcommand{\potRodPANN}{\psi^{\text{PANN}}}               
\newcommand{\potNN}{\psi^{\text{NN}}}
\newcommand{\potEnergy}{\psi^{\text{energy}}}
\newcommand{\potStress}{\psi^{\text{stress}}}

\newcommand{\iden}{\boldsymbol{1}}     
\newcommand{\zero}{\boldsymbol{0}}      
\newcommand{\initDomain}{\Omega_0}  

\newcommand{\model}{\ensuremath{\cW}}
\newcommand{\modelsym}{\ensuremath{\cW^\text{sym}}}
\newcommand{\modeliso}{\ensuremath{\cW^\text{ti}}}

\newcommand{\symmodel}{\ensuremath{\model^{\text{sym}}}}

\newcommand{\ta}{\boldsymbol{a}}
\newcommand{\tb}{\boldsymbol{b}}

\newcommand{\td}{\boldsymbol{d}}

\newcommand{\tr}{\boldsymbol{r}}
\newcommand{\ts}{\boldsymbol{s}}
\newcommand{\tu}{\boldsymbol{u}}

\newcommand{\tx}{\boldsymbol{x}}

\newcommand{\tlambda}{\boldsymbol{\lambda}}
\newcommand{\tmu}{\boldsymbol{\mu}}

\newcommand{\tA}{\boldsymbol{A}}

\newcommand{\tD}{\boldsymbol{D}}
\newcommand{\tE}{\boldsymbol{E}}
\newcommand{\tF}{\boldsymbol{F}}

\newcommand{\tI}{\boldsymbol{I}}

\newcommand{\tN}{\boldsymbol{N}}

\newcommand{\tP}{\boldsymbol{P}}

\newcommand{\tR}{\boldsymbol{R}}

\newcommand{\tU}{\boldsymbol{U}}

\newcommand{\tW}{\boldsymbol{W}}
\newcommand{\tX}{\boldsymbol{X}}

\newcommand{\cD}{{\mathcal{D}}}

\newcommand{\cH}{{\mathcal{H}}}

\newcommand{\cL}{{\mathcal{L}}}

\newcommand{\cW}{{\mathcal{W}}}

\newcommand{\bbC}{{\mathbb{C}}}

\newcommand{\bbR}{{\mathbb{R}}}

\newcommand{\frakm}{\mathfrak{m}}

\usetikzlibrary{shapes}
\usetikzlibrary{arrows.meta} 

\tikzset{>=latex} 
\tikzstyle{node}        =[thick, circle, draw=black, text = black, minimum size=25, inner sep=0.5, outer sep=0.6]
\tikzstyle{node icnn}   =[node, color=2orange!10!black, fill=color22!25]
\tikzstyle{node in_out} =[node, color=2lightblue!10!black, fill=color21!25]
\tikzstyle{node inv_pot}=[node, color=2red!10!black, fill=color33!25]
\tikzstyle{node layer}  =[node, rectangle, color=2red!10!black, fill=color22!25]
\tikzstyle{node ffnn} = [node, rectangle, color=2red!10!black, fill=color22!25, minimum width=60]

\tikzstyle{connect}=[thick, black, shorten <=1, shorten >=1, very thick] 
\tikzstyle{connect arrow}=[connect, densely dashed, -{Latex[length=4,width=3.5]}, very thick]

\tikzstyle{connect2}=[thick, black, shorten <=1, shorten >=1, very thick] 
\tikzstyle{connect arrow2}=[connect2, -{Latex[length=4,width=3.5]}, very thick]

\tikzstyle{connect arrow constraint}=[connect arrow2, 2red, very thick]
\tikzstyle{connect constraint}=[connect2, 2red, very thick] 

\newcommand{\layernode}[6][1]{
    \node[node 2] (#4) at (#2, #3) {};
    \ifnum#6=1 
        \draw[domain=-4:4, shift={(#2, #3-0.2)}, thick, scale=0.05, smooth, variable=\x, black] plot ({\x}, {ln(1+exp(1.5*\x))});
    \else
        \ifnum#6=2 
            \draw[domain=-4:4, shift={(#2, #3-0.1)}, thick, scale=0.05, smooth, variable=\x, black] plot ({\x}, {\x});
        \else
            \ifnum#6=3
                \draw[domain=-4:4, shift={(#2, #3-0.2)}, thick, scale=0.05, smooth, variable=\x, black] plot ({\x}, {1.5*max(\x,0)});
            \else
                \draw[domain=-4:4, shift={(#2, #3)}, thick, scale=0.05, smooth, variable=\x, black] plot ({\x}, {3.5*tanh(\x)});
            \fi
        \fi
    \fi
    \
    \node[above=6pt, left=-2pt] at (#2, #3) {\small #5};
    \ifnum#1=1 
        \draw[connect arrow] (#2, #3+0.7) -- (#4);
    \fi
    \ifnum#1=2 
        \draw[connect arrow constraint] (#2, #3+0.7) -- (#4);
    \fi
}

\definecolor{CPSgreen}{RGB}{22,164,138}
\definecolor{CPSlightblue}{RGB}{104,143,198}
\definecolor{CPSdarkblue}{RGB}{67,83,132}
\definecolor{CPSgrey}{RGB}{204, 204, 204}
\definecolor{CPSorange}{RGB}{246,163,21}
\definecolor{CPSred}{RGB}{194,76,76}

\definecolor{1_blue}{RGB}{0,119,187}
\definecolor{1_cyan}{RGB}{51,187,238}
\definecolor{1_teal}{RGB}{0,153,136}
\definecolor{1_orange}{RGB}{238,119,51}
\definecolor{1_red}{RGB}{204,51,17}
\definecolor{1_magenta}{RGB}{238,51,119}
\definecolor{1_grey}{RGB}{187,187,187}

\definecolor{m_navy}{RGB}{0,51,102}
\definecolor{m_indigo}{RGB}{51,34,136}
\definecolor{m_cyan}{RGB}{136,204,238}
\definecolor{m_teal}{RGB}{68,170,153}
\definecolor{m_green}{RGB}{17,119,51}
\definecolor{m_olive}{RGB}{153,153,51}
\definecolor{m_sand}{RGB}{221,204,119}
\definecolor{m_rose}{RGB}{204,102,119}
\definecolor{m_wine}{RGB}{136,34,85}
\definecolor{m_purple}{RGB}{170,68,153}
\definecolor{m_grey}{RGB}{187,187,187}

\definecolor{color11a}{RGB}{236, 188, 0}
\definecolor{color12a}{RGB}{50, 67, 121}
\definecolor{color13}{RGB}{222, 222, 222}
\definecolor{color21a}{RGB}{92, 134, 196}
\definecolor{color22a}{RGB}{249, 156, 0}
\definecolor{color33a}{RGB}{191, 60, 60}
\definecolor{colorCPSa}{RGB}{0, 157, 129}

\colorlet{2yellow}{color11a!90!color13}
\colorlet{2darkblue}{color12a!90!color13}
\colorlet{2lightblue}{color21a!90!color13}
\colorlet{2orange}{color22a!90!color13}
\colorlet{2red}{color33a!90!color13}
\definecolor{2grey}{RGB}{222, 222, 222}
\colorlet{2green}{colorCPSa!90!color13}

\tikzstyle{node}=[thick,circle,draw=black,minimum size=22,inner sep=0.5,outer sep=0.6]
\tikzstyle{node icnn}=[color=2orange!10!black,node,draw=black,fill=2orange!25, text = black]
\tikzstyle{node in_out}=[node,2lightblue!10!black,draw=black,fill=2lightblue!25, text=black]
\tikzstyle{node inv_pot}=[rectangle, node,2red!10!black,draw=black,fill=2red!25, text=black]
\tikzstyle{connect}=[thick,black] 
\tikzstyle{connect arrow}=[-{Latex[length=4,width=3.5]},thick,black,shorten <=0.5,shorten >=1]

\pgfplotscreateplotcyclelist{colorlist_ROL_PE}{%
only marks,solid, mark options={solid,fill=1_blue,fill opacity=1,mark size=2.5pt},mark=halfsquare*,1_blue,mark repeat = 2, mark phase = 2\\
1_blue, very thick \\
only marks,solid, mark options={solid,fill=1_cyan,fill opacity=1,mark size=2.5pt},mark=halfsquare right*,1_cyan,mark repeat = 2, mark phase = 2\\
1_cyan, very thick\\
only marks,solid, mark options={solid,fill=1_teal,fill opacity=1,mark size=2.5pt},mark=halfsquare left*,1_teal,mark repeat = 2, mark phase = 2\\
1_teal, very thick \\
only marks,solid, mark options={solid,fill=1_red,fill opacity=1,mark size=2.5pt,rotate=-90},mark=halfcircle*,1_red,mark repeat = 2, mark phase = 2\\
1_red, very thick \\
only marks,solid, mark options={solid,fill=1_orange,fill opacity=1,mark size=2.5pt,rotate=90},mark=halfcircle*,1_orange,mark repeat = 2, mark phase = 2\\
1_orange, very thick \\
}

\pgfplotscreateplotcyclelist{colorlist_ROL_FE}{%
 thick, dashed,  mark options={solid,fill=1_red,fill opacity=1,mark size=2.5pt,rotate=-90},mark=halfcircle*,1_red,mark repeat = 2, mark phase = 2\\
 thick, dashed,  mark options={solid,fill=1_orange,fill opacity=1,mark size=2.5pt,rotate=90},mark=halfcircle*,1_orange,mark repeat = 2, mark phase = 2\\
 thick, dashed,  mark options={solid,fill=1_magenta,fill opacity=1,mark size=2.5pt,rotate=-180 },mark=halfcircle*,1_magenta,mark repeat = 2, mark phase = 2\\
 thick, dashed, mark options={solid,fill=1_blue,fill opacity=1,mark size=2.5pt},mark=halfsquare*,1_blue,mark repeat = 2, mark phase = 2\\
 thick, dashed,  mark options={solid,fill=1_cyan,fill opacity=1,mark size=2.5pt},mark=halfsquare right*,1_cyan,mark repeat = 2, mark phase = 2\\
 thick, dashed,  mark options={solid,fill=1_teal,fill opacity=1,mark size=2.5pt},mark=halfsquare left*,1_teal,mark repeat = 2, mark phase = 2\\
}

\pgfplotsset{
/pgfplots/colormap={temp}{rgb255=(36,0,217) rgb255=(25,29,247) rgb255=(41,87,255)
rgb255=(61,135,255) rgb255=(87,176,255) rgb255=(117,211,255) rgb255=(153,235,255)
rgb255=(189,249,255) rgb255=(235,255,255) rgb255=(255,255,235) rgb255=(255,242,189)
rgb255=(255,214,153) rgb255=(255,172,117) rgb255=(255,120,87) rgb255=(255,61,61)
rgb255=(247,40,54) rgb255=(217,22,48) rgb255=(166,0,33)}
}

\definecolor{color11a}{RGB}{236, 188, 0}
\definecolor{color12a}{RGB}{50, 67, 121}
\definecolor{color13}{RGB}{222, 222, 222}
\definecolor{color21a}{RGB}{92, 134, 196}
\definecolor{color22a}{RGB}{249, 156, 0}
\definecolor{color23}{RGB}{222, 222, 222}
\definecolor{color31}{RGB}{222, 222, 222}
\definecolor{color32}{RGB}{222, 222, 222}
\definecolor{color33a}{RGB}{191, 60, 60}

\colorlet{color11}{color11a!90!color13}
\colorlet{color12}{color12a!90!color13}
\colorlet{color21}{color21a!90!color13}
\colorlet{color22}{color22a!90!color13}
\colorlet{color33}{color33a!90!color13}
\definecolor{colorCPSa}{RGB}{0, 157, 129}

\colorlet{2yellow}{color11a!90!color13}
\colorlet{2darkblue}{color12a!90!color13}
\colorlet{2lightblue}{color21a!90!color13}
\colorlet{2orange}{color22a!90!color13}
\colorlet{2red}{color33a!90!color13}
\definecolor{2grey}{RGB}{222, 222, 222}
\colorlet{2green}{colorCPSa!90!color13}

\usetikzlibrary{external}
\newcommand{\revi}[1]{{\color{black}#1}}
\newcommand{\revii}[1]{{\color{black}#1}}

\title{Physics-augmented neural networks for constitutive modeling\\of hyperelastic geometrically exact beams}

\author[*]{Jasper~O.~Schommartz}
\author[ ]{Dominik~K.~Klein}
\author[ ]{\\Juan~C.~Alzate~Cobo}
\author[ ]{Oliver~Weeger}

\affil[ ]{\footnotesize Cyber-Physical Simulation, Department of Mechanical Engineering,\protect\\Technical University of Darmstadt, 64293 Darmstadt, Germany}
\affil[*]{\footnotesize Corresponding author, email: schommartz@cps.tu-darmstadt.de}

\date{November 20, 2024}

\begin{document}

\maketitle
\par\noindent\rule{\textwidth}{0.4pt}
\begin{abstract}
We present neural network-based constitutive models for hyperelastic geometrically exact beams. The proposed models are physics-augmented, i.e., formulated to fulfill important mechanical conditions by construction\revii{, which improves accuracy and generalization}. Strains and curvatures of the beam are used as input for feed-forward neural networks that represent the effective hyperelastic beam potential. Forces and moments are received as the gradients of the beam potential, ensuring thermodynamic consistency. 
\revii{Normalization conditions are considered via additional projection terms. Symmetry conditions are implemented by an invariant-based approach for transverse isotropy and a more flexible point symmetry constraint, which is included in transverse isotropy but poses fewer restrictions on the constitutive response. Furthermore, a data augmentation approach is proposed to improve the scaling behavior of the models for varying cross-section radii. Additionally, we introduce a parameterization with a scalar parameter to represent ring-shaped cross-sections with different ratios between the inner and outer radii.}
Formulating the beam potential as a neural network provides a highly flexible model. This enables efficient constitutive surrogate modeling for geometrically exact beams with nonlinear material behavior and cross-sectional deformation, which otherwise would require computationally much more expensive methods.
The models are calibrated \revii{and tested with} data generated for beams with circular \revii{and ring-shaped hyperelastic} deformable cross-sections \revii{at varying inner and outer radii}, showing excellent accuracy and generalization.
The applicability of the proposed \revii{point symmetric} model is further demonstrated by applying it in beam simulations. In all studied cases, the proposed model shows excellent performance.

\end{abstract}
\vspace*{2ex}

\par\noindent\rule{\textwidth}{0.4pt}
Accepted version of manuscript published in \emph{Computer Methods in Applied Mechanics and Engineering}.

\noindent Accepted: November 20, 2024. DOI: \href{https://doi.org/10.1016/j.cma.2024.117592}{10.1016/j.cma.2024.117592}. License: \href{http://creativecommons.org/licenses/by/4.0/}{CC BY 4.0}.\vspace*{-1.4mm}
\par\noindent\rule{\textwidth}{0.4pt}



\section{Introduction}


Recent advances in additive manufacturing have sparked the development of elastomeric beam-lattice structures. These metamaterials enable the design of architected mechanical properties unprecedented in classical materials, making them promising for many engineering applications~\cite{jiangHighlystretchable3DarchitectedMechanical2016, kochmannMultiscaleModelingOptimization2019, truszkiewiczMechanicalBehavior3D2022,surjadi19}. 
At the same time, numerical simulation of these metamaterials in both an \emph{accurate} and \emph{efficient} manner remains a significant challenge due to their multiscale nature and high nonlinearities on the material~\cite{zhang2024} as well as the structural scale~\cite{weeger2018c,fernandezAnisotropicHyperelasticConstitutive2021}. 
Accurate simulation of beam-lattice metamaterials starts at their smallest building block, i.e., the simulation of single beams. There, the highly nonlinear behavior of many 3D printing materials~\cite{zhang2024} and complex deformations of the beams~\cite{gaertner2024} pose significant challenges. Furthermore, the often parameterized geometry of beams within a structure, e.g., varying radii, must be considered~\cite{fernandezMaterialModelingParametric2022}.
Further applications and challenges for mechanical modeling of beam-like structures include flexible cables, hoses, and wires~\cite{dorlichFlexibleBeamLikeStructures2018}, soft robotic actuators~\cite{depayrebruneConstitutiveRelationsRodbased2016}, or textiles~\cite{orlik_optimization_2016,do2020}.

\medskip


The most accurate modeling approach for beams is their representation as volume continuum bodies. Thereby, no further assumptions on the behavior of the continuum body are made, rendering it a very non-intrusive approach. In theory, this allows the use of arbitrary nonlinear (and also inelastic) material models for the beam, as well as the representation of arbitrary complex deformations~\cite{jiangHighlystretchable3DarchitectedMechanical2016,truszkiewiczMechanicalBehavior3D2022}.
However, the numerical methods used to solve the associated boundary value problems quickly reach their limits, as slender beams and locking phenomena can lead to vast numbers of degrees of freedom required for accurate numerical solutions~\cite{echterNumericalEfficiencyLocking2010}. This can cause computation times to be far outside an acceptable range for parametric studies, optimization, and other development workflows in engineering design. 
Methods such as solid beam finite element (FE) formulations can alleviate these limitations to some degree~\cite{frischkornSolidbeamFiniteElement2013,shafqat2024robust}. However, the computational cost remains high if complex shapes and deformations of beam cross-sections are considered~\cite{frischkornSolidbeamFiniteElement2013}.

\medskip


The most efficient method for simulating slender structures is to model them as 1D bodies instead of 3D continua. These 1D bodies are referred to as \emph{beams}. The geometrically exact Simo-Reissner beam theory~\cite{simoThreedimensionalFinitestrainRod1986} has been recognized as one of the most versatile, as it enables modeling anisotropic rigid cross-sections and initial curvatures and incorporates deformation from shear, longitudinal, bending, and torsion strain at arbitrarily large deformations~\cite{meierGeometricallyExactFinite2019}. In the literature, this beam model is also referred to as special Cosserat rod~\cite{antmanNonlinearProblemsElasticity2005}, nonlinear Timoshenko beam~\cite{r.eugsterGeometricContinuumMechanics2015}, or simply as \emph{geometrically exact beam}.
The reduction from a 3D continuum to a 1D beam relies on several assumptions on the kinematics and constitutive behavior of the body. One assumption, which is particularly limiting, is the rigidity of the beam's cross-section. Consequently, the geometrically exact beam theory is generally limited to small strains and applied using linear elastic constitutive models. However, these assumptions fail if beams are thick, strains are large, or the base material is considered outside of its linear elastic range, quickly reducing the accuracy of simple beam modeling approaches. 
Therefore, some works have focused on including cross-sectional deformation in the geometrically exact beam theory by introducing additional degrees of freedom~\cite{simoGeometricallyexactRodModel1991,choi2021isogeometric}. However, these approaches are still restricted to relatively simple cross-sectional deformations.


\begin{figure}
    \centering
    \input{tikz_fe2_approach}
    \caption{The geometrically exact beam with deformable cross-section as a multiscale problem. A warping problem on the cross-section scale is solved at each quadrature point on the beam scale. Modified schematic adopted from~\cite{herrnbockTwoscaleOffandOnline2023}.}
    \label{fig:FE2-approach}
\end{figure}

To overcome this limitation, some works focus on treating the geometrically exact beam as a multiscale problem, where the cross-sectional deformation is computed separately from the beam deformation in an FE\textsuperscript{2} approach~\cite{aroraComputationalApproachObtain2019,kumarHelicalCauchyBornRule2016,leclezioNumericalTwoscaleApproach2023,herrnbockGeometricallyExactElastoplastic2021,herrnbockTwoscaleOffandOnline2023}, cf.~\cref{fig:FE2-approach}. This enables modeling of in- and out-of-plane deformation for arbitrary cross-sectional shapes. 
For given strains and curvatures of the beam, the resulting deformed cross-section is evaluated by solving a two-dimensional warping problem. From this, the resulting forces and moments can be calculated. Overall, this enables the simulation of very complex cross-sectional shapes and deformations and the use of nonlinear constitutive models for the beam.
In~\cite{kumarHelicalCauchyBornRule2016,aroraComputationalApproachObtain2019}, a warping problem is formulated that models the cross-sectional deformation for any combination of shear, longitudinal, bending, and twisting strain and is directly applicable in the framework of the geometrically exact Simo-Reissner beam theory.

While the accuracy of beam simulations is increased in a {concurrent} multiscale FE\textsuperscript{2} setting, these approaches require computing the solution of a warping problem on the cross-section scale for every quadrature point of the simulation on the beam scale, which has to be repeated for every iteration of the numerical solver~\cite{herrnbockHomogenizationFullyNonlinear2022}. Solving boundary value problems across scales leads to a drastic increase in computational cost, undermining the main benefit of using beam models, i.e., efficiency. In~\cite{leclezioNumericalTwoscaleApproach2023}, this issue was addressed with a {sequential} multiscale approach, where the warping problem is solved for some sampled values in the strain space before conducting simulations on the beam scale. Together with a multidimensional interpolation scheme, the resulting database is then used in the macroscale beam simulation. While pre-solving can speed up the macroscale simulation, a sufficiently dense sampling quickly leads to large datasets and expensive data generation. This holds particularly if all six strain measures of a shear-deformable beam -- and possibly additional geometric parameters describing the beam's cross-section -- are considered, in contrast to only four strain measures considered for the non-shearable beam in~\cite{leclezioNumericalTwoscaleApproach2023}.
To conclude, while these multiscale approaches for considering nonlinear material effects and warping offer excellent accuracy, their efficiency remains a drawback, which calls for efficient surrogate techniques for multiscale material modeling of beams.

\medskip


Shifting the focus to constitutive modeling, various constitutive models for 3D continuum bodies based on \emph{physics-augmented neural networks} (PANNs) have been proposed in recent years. These models combine the extraordinary flexibility of neural networks (NNs) with a sound mechanical basis, making them applicable to a wide range of materials~\cite{tac2023,kalina2024a,kalina2023}.
Furthermore, embedding mechanical knowledge in the models enables calibration with sparse datasets~\cite{lindenNeuralNetworksMeet2023}. 

In the finite elasticity theory of 3D continuum bodies, it is well understood how to formulate PANN constitutive models that fulfill all relevant mechanical conditions by construction. PANN models usually represent the corresponding energy potentials with an NN, which ensures thermodynamic consistency~\cite{lindenNeuralNetworksMeet2023}. To fulfill objectivity and material symmetry, the potentials are typically formulated in terms of strain invariants~\cite{lindenNeuralNetworksMeet2023}. However, some models only approximately fulfill these conditions by learning them through suitable calibration data~\cite{gaertner2021,kleinPolyconvexAnisotropicHyperelasticity2022}, \revii{which is referred to as data augmentation}. Then, the NN potential is complemented by additional growth and normalization terms~\cite{kleinPolyconvexAnisotropicHyperelasticity2022,lindenNeuralNetworksMeet2023}. For polyconvex potentials, input-convex neural network architectures can be applied~\cite{kleinPolyconvexAnisotropicHyperelasticity2022,amosInputConvexNeural2017}. Finally, different approaches for representing parameterized material behavior were proposed~\cite{fernandezMaterialModelingParametric2022,kleinParametrisedPolyconvexHyperelasticity2023,linka2020}. 

The extraordinary flexibility of these approaches enables their application to sequential multiscale methods of 3D solid bodies~\cite{kalina2023,Masi_Stefanou_2023,klein2024a}. Thereby, PANN constitutive models act as surrogates for the homogenized behavior of microstructured materials. Since this is often a highly nonlinear functional relationship, conventional constitutive models usually are not flexible enough for this task~\cite{kleinPolyconvexAnisotropicHyperelasticity2022}. Even for highly nonlinear material behavior, very shallow -- often single-layered -- NN architectures are sufficiently flexible and, thus, computationally efficient to evaluate.

\medskip

While PANN approaches are well established for 3D continuum material modeling, to the best of the authors' knowledge, they have not yet been applied to beam constitutive modeling. Consequently, the present work proposes a hyperelastic PANN constitutive model for geometrically exact, shear deformable beams. The model represents the relationship between strains/curvatures in the beam and the resulting forces/moments. Since the model is based on NNs, it can represent highly nonlinear behaviors, which are to be expected when considering hyperelastic base materials and warping in the beam's cross-section.
The model is also physics-augmented, i.e., it is formulated to fulfill important mechanical conditions relevant to beam theory. First, thermodynamic consistency is considered by introducing a beam potential represented by an NN. Furthermore, objectivity is guaranteed by the choice of input and output, while projection terms ensure stress and energy normalization.
For \revii{circular or ring-shaped} cross-sections \revii{without} \revi{handedness}, \revii{invariant-based transverse isotropy according to ~\cite{antmanNonlinearProblemsElasticity2005,healeyMaterialSymmetryChirality2002} is introduced in the model and compared to a more flexible point symmetry constraint. The latter can be derived from transverse isotropy but is found to be more suitable for deformable cross-sections}. Furthermore, a parameterization of the model \revii{for ring-shaped cross-sections with different ratios between the inner and outer radii is proposed}.

This model aims to bridge the gap between accuracy and efficiency in FE\textsuperscript{2} beam simulations. This is achieved by applying the PANN beam constitutive model as a surrogate for the warping problem proposed in~\cite{aroraComputationalApproachObtain2019}, ultimately providing a very \emph{accurate} model, which represents both material and geometrical nonlinearities occurring on the cross-section scale of the beam while also being computationally very \emph{efficient}.

\medskip

The outline of the work is as follows. In~\cref{sec:beam-model}, the fundamentals of the geometrically exact beam theory are introduced for both rigid and deformable cross-sections. In~\cref{sec:beam-model}, the formulation of the PANN constitutive model is proposed. In~\cref{sec:results}, the model's accuracy for highly nonlinear beam behavior, including nonlinear material behavior and cross-sectional deformations, is demonstrated. Furthermore, its applicability in macroscopic beam simulations is shown. This is followed by the conclusion in~\cref{sec:concl}.

\medskip

\paragraph{Notation}

Throughout this work, scalars are indicated by $a$, and first or second-order tensors are indicated by $\ta$ or $\tA$. The second-order identity is denoted as $\tI$. Transpose and determinant are denoted by $\tA^T$ and $\det \tA$, respectively. The special orthogonal group in $\bbR^3$ is $\SO(3):=\big\{\tX\in\allowbreak \mathbb{R}^{3\times 3}\;\rvert\allowbreak \;\tX^T\tX=\iden,\;\det \tX =1\big\}$. Latin indices $i$ denote the components $\{1,2,3\}$ while Greek indices $\alpha$ denote the components $\{1, 2\}$ in the Einstein summation notation. The axial vector of a skew-symmetric tensor is denoted by $\operatorname{ax}(\cdot)$.
The softplus and and linear activation functions are denoted by $\SP(x) = \ln{(1 + \exp{x})}$ and $\lin(x) = x$, respectively.
\section{Fundamentals of geometrically exact beam theory}
\label{sec:beam-model}

In this section, the geometrically exact, shear-deformable Cosserat rod, also called the Simo-Reissner beam theory, is introduced. After kinematics in~\cref{subsec:kinematics} and balance equations in~\cref{subsec:balance-equations}, constitutive relations for rigid and deformable cross-sections are described in~\ref{subsec:constitutive-relations}. Finally, in~\cref{subsec:warping-problem}, the cross-sectional warping problem is outlined.


\subsection{Kinematics of the Cosserat rod}
\label{subsec:kinematics}

\begin{figure}
    \centering
    \tikzsetnextfilename{tikz_beam}
\begin{tikzpicture}[xscale=0.8,yscale=-0.8]
\useasboundingbox(0,0) rectangle (91.2mm,78.5mm);
\begin{scope}[shift={(-37.6mm,-70.2mm)}]
\draw [fill=black,draw=black,line width=0.3mm] (75.8mm,99.8mm) circle (0.4mm);
\draw [-Stealth,fill=none,draw=black,line width=0.3mm] (86.3mm,89.4mm)
-- (86.3mm,71.6mm)
;
\draw [-Stealth,fill=none,draw=black,line width=0.3mm] (86.2mm,89.3mm)
-- (103.9mm,89.3mm)
;
\draw [-Stealth,fill=none,draw=black,line width=0.3mm] (86.3mm,89.2mm)
-- (75.1mm,100.7mm)
;
\draw [fill=none,draw=none,line width=0.2mm,dashed] (113.7mm,127.7mm) ellipse (13.6mm and 14.6mm) ;
\draw [fill=none,draw=black,line width=0.4mm] (46.3mm,129.3mm) ellipse (8.5mm and 8.6mm) ;

\definecolor{dc}{RGB}{22,164,138}

\draw [fill=none,draw=black,line width=0.4mm] (81.7mm,106.05mm) arc[start angle=45, end angle=-135, radius=8.55mm];
\draw [fill=none,draw=black,line width=0.4mm,dashed] (81.7mm,106.05mm) arc[start angle=45, end angle=225, radius=8.55mm];

\draw [fill=dc!20,draw=none] (61.2mm,114.5mm) ellipse (8.5mm and 8.6mm) ;
\draw [fill=none,draw=dc,line width=0.4mm] (67.2mm,120.5mm) arc[start angle=45, end angle=-135, radius=8.55mm];
\draw [fill=none,draw=dc,line width=0.4mm,dashed] (67.2mm,120.5mm) arc[start angle=45, end angle=225, radius=8.55mm];
\draw [fill=none,draw=black,line width=0.3mm] (40.5mm,122.9mm)
-- (69.9mm,93.6mm)
;
\draw [fill=none,draw=black,line width=0.3mm] (52.4mm,135.3mm)
-- (81.8mm,106.0mm)
;
\draw [fill=none,draw=black,line width=0.4mm] (88.9mm,139.9mm) ellipse (8.5mm and 8.6mm) ;
\draw [fill=none,draw=black,line width=0.4mm] (121.2mm,118.65mm) arc[start angle=69.5, end angle=-110.5, radius=8.55mm];
\draw [fill=none,draw=black,line width=0.4mm,dashed] (121.2mm,118.65mm) arc[start angle=69.5, end angle=249.5, radius=8.55mm];

\definecolor{dc}{RGB}{22,164,138}
\draw [fill=dc!20,draw=none] (102.2mm,122.45mm) ellipse (8.5mm and 8.6mm) ;
\draw [fill=none,draw=dc,line width=0.4mm] (109.2mm,127.3mm) arc[start angle=35, end angle=-140, radius=8.50mm];
\draw [fill=none,draw=dc,line width=0.4mm,dashed] (109.2mm,127.3mm) arc[start angle=35, end angle=220, radius=8.50mm];

\draw [-Stealth,fill=none,draw=black,line width=0.3mm] (61.1mm,114.4mm)
-- (61.1mm,98.9mm)
;
\definecolor{dc}{RGB}{194,76,76}
\draw [-Stealth,fill=none,draw=dc,line width=0.3mm] (61.2mm,114.4mm)
-- ++(-4.0mm,-6.0mm)
;
\definecolor{dc}{RGB}{194,76,76}
\draw [-Stealth,fill=none,draw=dc,line width=0.3mm] (103.1mm,122.7mm)
-- ++(0.0mm,-7.0mm)
;
\draw [-Stealth,fill=none,draw=black,line width=0.3mm] (61.1mm,114.4mm)
-- ++(15.5mm,0.0mm)
;
\draw [-Stealth,fill=none,draw=black,line width=0.3mm] (61.2mm,114.3mm)
-- ++(-10.9mm,10.9mm)
;
\begin{scope}[xshift=8.5mm]
\draw [fill=none,draw=black,line width=0.3mm] (85.0mm,147.2mm)
.. controls ++(5.0mm,-3.8mm) and ++(-3.1mm,5.4mm) .. ++(12.3mm,-14.1mm)
.. controls ++(0.9mm,-1.5mm) and ++(-1.0mm,1.5mm) .. ++(2.7mm,-4.6mm)
.. controls ++(3.0mm,-4.6mm) and ++(-5.2mm,1.7mm) .. ++(12.8mm,-9.8mm)
;
\end{scope}
\begin{scope}[shift={(3.7mm,-7.5mm)}]
\draw [fill=none,draw=black,line width=0.2mm,dashed] (85.2mm,147.4mm)
.. controls ++(4.8mm,-4.6mm) and ++(-3.8mm,5.5mm) .. ++(12.9mm,-15.3mm)
.. controls ++(0.9mm,-1.4mm) and ++(-1.0mm,1.3mm) .. ++(2.8mm,-4.0mm)
.. controls ++(3.5mm,-4.6mm) and ++(-5.4mm,2.0mm) .. ++(13.7mm,-10.1mm)
;
\end{scope}
\begin{scope}[xshift=8.5mm]
\draw [fill=none,draw=black,line width=0.3mm] (74.6mm,133.6mm)
.. controls ++(2.1mm,-1.9mm) and ++(-1.6mm,2.3mm) .. ++(5.6mm,-6.2mm)
.. controls ++(2.3mm,-3.2mm) and ++(-2.3mm,3.2mm) .. ++(6.4mm,-9.8mm)
.. controls ++(5.1mm,-7.0mm) and ++(-8.2mm,2.7mm) .. ++(20.6mm,-15.2mm)
;
\end{scope}
\begin{scope}[shift={(9.9mm,-0.3mm)}]
\definecolor{fc}{RGB}{128,128,128}
\draw [-Stealth,fill=fc,draw=black,line width=0.3mm] (93.2mm,123.0mm)
-- ++(5.6mm,12.7mm)
;
\end{scope}
\begin{scope}[rotate around={-156.00527:(98.1mm,121.8mm)}]
\definecolor{fc}{RGB}{128,128,128}
\draw [-Stealth,fill=fc,draw=black,line width=0.3mm] (93.2mm,123.0mm)
-- ++(15.3mm,0.2mm)
;
\end{scope}
\begin{scope}[rotate around={-65.896202:(97.9mm,115.3mm)}]
\definecolor{fc}{RGB}{128,128,128}
\draw [-Stealth,fill=fc,draw=black,line width=0.3mm] (93.2mm,123.0mm)
-- ++(15.3mm,0.2mm)
;
\end{scope}
\definecolor{dc}{RGB}{16,4,3}
\draw [fill=none,draw=dc,line width=0.2mm,dashed] (86.3mm,89.4mm)
-- (46.2mm,129.2mm)
;
\definecolor{fc}{RGB}{0,0,128}
\definecolor{dc}{RGB}{67,83,132}
\draw [-Stealth,fill=fc,draw=dc,line width=0.3mm] (86.3mm,89.4mm)
-- (61.2mm,114.4mm)
;
\draw [-Stealth,fill=black,draw=black,line width=0.3mm] (86.2mm,89.3mm)
-- (103.1mm,115.7mm)
;
\draw [-Stealth,fill=black,draw=black,line width=0.3mm] (86.3mm,89.3mm)
-- (57.2mm,108.4mm)
;
\definecolor{dc}{RGB}{64,83,138}
\draw [-Stealth,fill=black,draw=dc,line width=0.3mm] (86.2mm,89.3mm)
-- (103.3mm,123.0mm)
;
\definecolor{fc}{RGB}{0,0,128}\node [fc,font=\small] at (199.9mm,166.9mm) {  };
\draw [fill=black,draw=black,line width=0.3mm] (46.2mm,129.2mm) circle (0.4mm);
\draw [fill=black,draw=black,line width=0.3mm] (88.8mm,139.9mm) circle (0.4mm);
\draw [fill=black,draw=black,line width=0.3mm] (118.6mm,110.5mm) circle (0.4mm);
\end{scope}

\node at (65.0mm,22.5mm) {$\tE_1$};
\node at (52.5mm,3.0mm) {$\tE_2$};
\node at (42.5mm,30.0mm) {$\tE_3$};

\node at (40.0mm,47.5mm) {$\tD_1$};
\node at (24.0mm,26.5mm) {$\tD_2$};
\node at (19.0mm,53.0mm) {$\tD_3$};

\node at (67.5mm,40.0mm) {$\td_1$};
\node at (52.0mm,49.5mm) {$\td_2$};
\node at (74.0mm,63.0mm) {$\td_3$};

\node at (35.5mm,25.0mm) {$\tX$};
\node at (58.5mm,30.0mm) {$\tx$};

\node[text=CPSdarkblue] at (32.5mm,38.5mm) {$\ts$};
\node[text=CPSdarkblue] at (51.5mm,30.0mm) {$\tr$};
\end{tikzpicture}
    \caption{The geometrically exact beam in its initial (left) and current configuration (right).}
    \label{fig:cosserat-rod}
\end{figure}

In the Cosserat rod theory, a slender object is described via a centerline curve $\tr: s \mapsto \bbR^3$ and a rotation $\tR : s \mapsto \SO(3)$, where the arc length parameter $s \in [0, L]$ denotes the position on the centerline~\cite{simoThreedimensionalFinitestrainRod1986}. The orientation of the cross-section is represented by the 3D rotation matrix $\tR(s) = (\td_1, \td_2, \td_3)$, which consists of a set of orthonormal directors $\td_i$. Here, the out-of-plane component $\td_3(s) = \td_1(s) \times \td_2(s)$ is normal to the cross-sectional plane spanned by the in-plane directors $\td_{\alpha}(s),\, \alpha=1,2$. Without loss of generality, we consider an initially straight beam with a constant cross-section starting at the origin of the global coordinate system with $\tD_i = \tE_i$, i.e., the beam's directors in the undeformed configuration are aligned with the global coordinate system \revi{and the cross-section's principal axes of inertia}.

\subsubsection{Rigid cross-section}\label{sec:cross_rigid}

Assuming a rigid cross-section, a material point in the undeformed configuration is given by
\begin{equation}\label{eq:reference_configuration}
    \tX (s, \initCS) =  \ts (s) + \initCS\,,
\end{equation}
cf.~\cref{fig:cosserat-rod}. Here, the vector $\ts(s) = s \tE_3$ points to the centerline, and $\initCS = X_{\alpha} \tE_{\alpha}$ denotes the position vector in the cross-sectional plane. The directors in the initial and current configurations are related via
\begin{equation}
    \td_i(s) = \tR (s) \tD_i.
\end{equation}
Hence, a material point in the deformed configuration reads
\begin{equation}\label{eq:current_configuration}
    \tx(s,\initCS) = \tr(s) + \tR (s) \initCS.
\end{equation}
With $()' = \der{}{s}$ denoting the derivative w.r.t.\ the arc length parameter $s$, the deformation gradient follows as
\begin{equation}\label{eq:defgrad_rigid}
\begin{aligned}
    \tF(s,\matStr,\matCurv,\initCS) = \der{\tx}{\tX} &= \left(\left(\tr' - \td_3\right) + \tR' \initCS\right) \otimes \tE_3 + \td_{i} \otimes \tE_{i}
    \\
    &= \tR \left[ \left(\matStr + \matCurv \times \initCS\right) \otimes \tE_3 + \tI \right]\,.
\end{aligned}
\end{equation}
Here, $\matStr$ and $\matCurv$ denote the strain measures in the reference configuration
\begin{equation}\label{eq:strain_measures}
    \matStr = \tR^T \tr' - \tE_3\,,\qquad
    \matCurv = \text{ax}\,\left(\tR^T\tR'\right)\,,
\end{equation}
where $\epsilon_1$ and $\epsilon_2$ capture shear strain, while $\epsilon_3$ denotes axial stretch. Curvature and twist are represented by the components $\kappa_{1,2}$ and $\kappa_3$, respectively. Since the strain measures $\matStr$ and $\matCurv$ only depend on the arc-length parameter $s$, they prescribe the strain state for any position $\initCS$ on the cross-section. Hence, other authors have referred to them as \text{strain prescriptors}~\cite{herrnbockGeometricallyExactElastoplastic2021}. 
Furthermore, \cref{eq:defgrad_rigid} contains the decomposition of the deformation gradient $\tF = \tR \bar{\tU}$ into the rotation $\tR(s)$ and the \revi{non-symmetric strain} tensor $\bar{\tU}(\matStr, \matCurv, \initCS)$, where the latter depends exclusively on the position on the cross-section $\initCS$ and the vectors $\matStr(s)$ and $\matCurv(s)$. \revi{Note that the decomposition $\tF = \tR \bar{\tU}$ is different from the classical polar decomposition as $\bar{\tU}$ is not a pure stretch tensor~\cite{ortigosaComputationalFrameworkPolyconvex2016}.}

\subsubsection{Deformable cross-section}\label{sec:cross_deformable}

In general, the cross-section of a beam does not remain rigid but deforms depending on the beam's strain and curvature state.  Furthermore, the cross-sectional deformation at position $s$ does not only depend on the local strain state $(\matStr(s), \matCurv(s))$ but it is influenced by strains in the neighborhood of $s$ as well~\cite{kumarHelicalCauchyBornRule2016}. Assuming slowly varying strains along the length of the beam, the latter effect can be neglected. This assumption was initially proposed by~\cite{kumarHelicalCauchyBornRule2016} and is known as the \textit{helical Chauchy-Born rule}. 
Based on this assumption, the current configuration of a beam with a deformable cross-section is defined as
\begin{equation}\label{eq:current_configuration_warping}
    \tx(s, \initCS) = \tr(s) + \tR(s) \,\currCS(\initCS)\,.
\end{equation}
\revi{Here, $\currCS$ denotes the deformed configuration of the cross-section. It may be expressed as
\begin{equation}\label{eq:current_configuration_cs_warping}
    \currCS(\initCS) = \initCS + \tu(\initCS),
\end{equation}
where $u_{\alpha}(\initCS)$ denote the in-plane displacements and $u_3(\initCS)$ denotes the out-of-plane warping~\cite{aroraComputationalApproachObtain2019}. Since $\tu(\initCS)$ does not depend on $s$, which results from assuming the helical Cauchy-Born rule, the deformation gradient can be derived analogously to~\cref{eq:defgrad_rigid}. It reads
\begin{equation}\label{eq:defgrad_dhm}
    \tF \big(s, \matStr, \matCurv, \currCS(\initCS)\big) = \tR(s) \left[\big(\matStr + \matCurv \times \currCS  + \tE_3 \big) \otimes \tE_3 + \der{\currCS}{X_{\alpha}} \otimes \tE_{\alpha}\right].
\end{equation}
Here, we find the same strain measures as in~\cref{eq:strain_measures}. Note that this derivation does not include the implicit relation between $\currCS$ and the strain measures $\matStr$ and $\matCurv$. Finding it will be the topic of~\cref{subsec:warping-problem}.}


\subsection{Stress resultants and balance equations}
\label{subsec:balance-equations}

The quasi-static balance equations of a beam in spatial description are given as
\begin{equation}
\begin{split}
    \spaForce' + \bar{\spaForce} &= \zero ,\\
    \spaMoment' + r' \times \spaForce + \bar{\spaMoment} &= \zero.
\end{split}
\end{equation}
Here, $\spaForce(s) \in \bbR^3$ and $\spaMoment(s) \in \bbR^3$  denote forces and moments acting on the cross-section of the beam, with
\begin{equation}\label{eq:stress_resultants_spatial}
    \spaForce = n_i \td_i, \qquad \spaMoment = m_i \td_i.
\end{equation}
The components $n_1$ and $n_2$ denote shear forces, $n_3$ the normal force, $m_1$ and $m_2$ bending moments, and $m_3$ the twisting moment. 
Furthermore, $\bar{\spaForce}$ and $\bar{\spaMoment}$ represent external distributed forces and moments, respectively. The boundary value problem for determining the deformed configuration $(\tr, \tR)$ of the beam is complemented by Dirichlet and Neumann boundary conditions for $s=0$ and $s=L$:
\begin{equation}
\begin{split}
    &\tr(s) = \tr_s, \qquad \tR(s) = \tR_s ,\\
    &\spaForce(s) = \spaForce_s,\qquad \spaMoment(s) = \spaMoment_s.
\end{split}
\end{equation}
Here, $\tr_s$ and $\tR_s$ denote prescribed positions and rotations, respectively, while $\spaForce_s$ and $\spaMoment$ are externally applied forces and moments. 

The stress resultants in the material description read
\begin{equation}
    \matForce = n_i \tE_i \quad\text{and}\quad \matMoment = m_i \tE_i.
\end{equation}
They are related to the spatial description as
\begin{equation}
    \spaForce = \tR \matForce \quad\text{and}\quad \spaMoment = \tR \matMoment.
\end{equation}

\subsection{Constitutive relations}
\label{subsec:constitutive-relations}

A constitutive model is required to relate the beam's strains and curvatures to the resulting forces and moments. For this, assuming hyperelastic material behavior, the beam potential
\begin{equation}
\potRod:\bbR^3\times\bbR^3\rightarrow\bbR\,,\qquad (\matStr, \matCurv)\mapsto \potRod(\matStr, \matCurv)
\end{equation}
is introduced, from which stress resultants are derived as
\begin{equation}\label{eq:stress_from_potential}
    \matForce = \der{\potRod}{\matStr}\,, \qquad \matMoment = \der{\potRod}{\matCurv}.
\end{equation}
\revii{Furthermore, the stiffnesses can be derived from the beam potential as}
\begin{equation}\label{eq:stiffness_from_potential}
\revii{
    \bbC_{\matStr \matStr} = \frac{\partial^2\potRod}{\partial\matStr\partial\matStr}\,, \qquad 
    \bbC_{\matCurv \matCurv} = \frac{\partial^2\potRod}{\partial\matCurv\partial\matCurv}\,, \qquad 
    \bbC_{\matStr \matCurv} = \frac{\partial^2\potRod}{\partial\matStr\partial\matCurv} = \bbC_{\matCurv \matStr}^\top\,.
}
\end{equation}
The formulation of the beam potential $\potRod$ determines if rigid or deformable cross-sections are considered and defines the material behavior in the beam's cross-section. Note that $\potRod$ is associated with the beam scale and is not to be confused with the description of the beam's base material in terms of a hyperelastic continuum constitutive model, cf.~\cref{subsubsec:rhm}.

\subsubsection{Linear elastic model}
\label{subsubsec:lem}

In its simplest form, the beam potential is given by the quadratic expression
\begin{equation}\label{eq:potrod_lem}
    \potRod(\matStr, \matCurv) = \frac{1}{2} \matStr^T \bbC_{\matStr \matStr} \matStr + \frac{1}{2} \matCurv^T \bbC_{\matCurv \matCurv} \matCurv + \matStr^T \bbC_{\matStr \matCurv} \matCurv \,.
\end{equation}
Assuming \revii{circular or ring-shaped} homogeneous cross-sections, the constitutive matrices take the diagonal shapes
\begin{equation}\label{eq:constitutive_matrices}
    \bbC_{\matStr \matStr} = \text{diag}\left[k_1 G A, k_2 G A, E A\right]\,, \qquad
    \bbC_{\matCurv \matCurv} = \text{diag}\left[E I_1, E I_2, G J\right] \,, \qquad \bbC_{\matStr \matCurv} = \zero.
\end{equation}
Here, $E$ and  $G$ denote Young's and shear moduli, respectively, with the remaining parameters given as
\begin{equation} \label{eq:lem_geom}
    A=\pi \revii{\big(R^{2} - R_i^2\big)}\,,\qquad I_{1}=I_{2}=\frac{\pi}{4} \revii{\big(R^{4} - R_i^{4}\big)}\,,\qquad J=I_{1}+I_{2}\,,
\end{equation}
\revii{where $R$ is the outer and $R_i$ the inner radius of a circular, ring-shaped cross-section (with $R>R_i\geq0$),} and the shear correction factors are set to $k_1 = k_2 = \frac{5}{6}$. 
This choice of potential implies a linear relation between strains and stresses, i.e.,
\begin{equation}\label{eq:linear_el_stress_strain_realation}
    \matForce = \bbC_{\matStr \matStr} \matStr \,,\qquad
    \matMoment = \bbC_{\matCurv \matCurv} \matCurv\,.
\end{equation}
In the remainder of this work, this linear elastic constitutive model is referred to as ``LEM''.

\subsubsection{Hyperelastic model with rigid cross-section}
\label{subsubsec:rhm}

Let us consider a hyperelastic base material of the beam represented by the potential
\begin{equation}\label{eq:base_pot}
\Bar{\psi}:\operatorname{GL}^+(3)\rightarrow\bbR\,,\qquad\tF\mapsto\Bar{\psi}(\tF)\,.
\end{equation}
The potential $\Bar{\psi}$ describes the nonlinear material behavior {within} the beam and its cross-section. 
Furthermore assuming a rigid cross-section, as introduced in~\cref{sec:cross_rigid}, the beam potential is calculated by integrating the potential $\Bar{\psi}$ over the cross-section, i.e.,
\begin{equation}\label{eq:potrod_rhm}
    \potRod (s) = \int_{\initDomain} \Bar{\psi}(\tF(s, \matStr, \matCurv, \initCS)) \, d X_1 d X_2,
\end{equation}
where $\tF$ is the deformation gradient of the rigid cross-section as introduced in~\cref{eq:defgrad_rigid}.
The stress resultants follow as~\cite{simoFiniteStrainBeam1985}
\begin{equation}\label{eq:stress_resultants_rhm}
        \matForce = \int_{\initDomain} \tP \tE_3 \, d X_1 d X_2\,,\qquad
        \matMoment = \int_{\initDomain} \initCS \times \tP \tE_3 \, d X_1 d X_2\,,
\end{equation}
where $\tP(\tF)=\partial\bar{\psi}/\partial\tF$ is the first Piola-Kirchhoff stress tensor.
In the remainder of this work, this hyperelastic constitutive model with rigid cross-section is referred to as ``RHM''.

\begin{remark}
    While the RHM model seems reasonable for considering the nonlinearity of a beam's base material, it comes with a significant disadvantage over the LEM. This relates to the treatment of cross-sectional deformation. Strictly speaking, the LEM does not include any statement about the cross-section but instead depends exclusively on the values assigned to its parameters. However, in practice, the Young's modulus $E$ is determined for uniform cross-sectional deformation, and the shear correction factors $k_1$ and $k_2$ are chosen to prevent overly stiff behavior under shear. Hence, cross-sectional deformation is included implicitly by the LEM. The RHM, in contrast, strictly enforces cross-sectional rigidity through~\cref{eq:defgrad_rigid}, leading to undesired in-plane reaction forces~\cite{meierGeometricallyExactFinite2019}. The resulting overestimation of stiffness has been reported by several authors~\cite{aroraComputationalApproachObtain2019,kumarHelicalCauchyBornRule2016} and is demonstrated again in~\cref{subsec:model-evaluation}.
\end{remark}

\subsubsection{Hyperelastic model with deformable cross-section}
\label{subsubsec:dhm}

In this case, hyperelastic behavior of the beam's base material, as introduced in~\cref{eq:base_pot}, is considered in conjunction with a deformable cross-section as introduced in~\cref{sec:cross_deformable}. Based on these assumptions, the beam potential is derived as
\revi{\begin{equation}\label{eq:potrod_dhm}
    \potRod (s) = \int_{\initDomain} \Bar{\psi}(\tF(s, \matStr, \matCurv, \currCS(\initCS))) \, d X_1 d X_2\,,
\end{equation}}
where $\tF$ is the deformation gradient of the deformable cross-section as introduced in~\cref{eq:defgrad_dhm}. The stress resultants follow as
\begin{equation}
    \matForce = \int_{\initDomain} \tP \tE_3 \, d X_1 d X_2\,, \qquad
    \matMoment = \int_{\initDomain} \currCS \times \tP \tE_3 \, d X_1 d X_2\,.
\end{equation}
In the remainder of this work, this hyperelastic constitutive model with deformable cross-section is referred to as ``DHM''.


\subsection{Cross-sectional deformation with the warping problem}
\label{subsec:warping-problem}

Before evaluating the DHM, the deformed cross-section $\currCS$ for given strain measures $\matStr$, and $\matCurv$ must be calculated. In this work, the method proposed by~\cite{aroraComputationalApproachObtain2019} is applied and briefly summarized now. 
Note that in~\cite{aroraComputationalApproachObtain2019}, strains and stresses are assumed to vary slowly along the beam's arc length, cf.~\cref{subsubsec:dhm}.
The deformed configuration of the beam's cross-section, $\currCS$, is defined by the state of minimal strain energy
\revi{\begin{equation}\label{eq:minimization_problem}
    \underset{\currCS}{\mathrm{min}} \int_{\initDomain} \Bar{\psi} (\tF (\matStr, \matCurv, \currCS(\initCS))) \, d X_1 d X_2.
\end{equation}}
Here, the rotational part of the deformation gradient $\tF$ is set to $\tR(s) = \tI$, cf.~\cref{eq:defgrad_dhm}, as it does not influence the strain energy. Therefore, $\tF$ no longer depends explicitly on the arc-length parameter $s$.

Two sets of constraints are introduced to prevent rigid body motions in the solution of~\cref{eq:minimization_problem}. Firstly, the center of mass should remain in the origin of the local cross-sectional coordinate system. Therefore, three translational degrees of freedom are constrained by
\begin{equation}\label{eq:constraint_translation}
    \int_{\initDomain} \rho \currCS \, d X_1 d X_2 = \zero\,.
\end{equation}
Secondly, the directors of the local coordinate system should be aligned with the principal axes of the cross-section. Thus, the three rotational degrees of freedom are constrained by
\begin{equation}\label{eq:constraint_rotation}
    \int_{\initDomain} \boldsymbol{\frakm} \, d X_1 d X_2 = \zero,
\end{equation}
where
\begin{equation}\label{eq:constraint_rotation_m}
\boldsymbol{\frakm} = \begin{bmatrix}
    \rho \hat{X}_2 \hat{X}_3 \\
    \rho \hat{X}_1 \hat{X}_3 \\
    \rho \hat{X}_1 \hat{X}_2 \\
    \mathrm{arctan}\,\left(\cfrac{\hat{X}_2}{\hat{X}_1}\right) - \mathrm{arctan}\,\left(\cfrac{X_2}{X_1}\right)
\end{bmatrix}.
\end{equation}
Note that $\frakm$ has four components despite being used to constrain three degrees of freedom. \revi{According to~\cite{aroraComputationalApproachObtain2019}, including all four rotational constraints is relevant for circular cross-sections, while the first three constraints may be used for non-circular cross-sections. However, other authors have successfully forgone applying only constraints one, two, and four for circular and non-circular cross-sections~\cite{herrnbockGeometricallyExactElastoplastic2021,herrnbockTwoscaleOffandOnline2023}.}

\medskip

With the objective function in~\cref{eq:minimization_problem}, subject to the constraints in~\cref{eq:constraint_translation} and~\cref{eq:constraint_rotation}, the cross-sectional warping problem can be formulated as a constrained Lagrangian optimization problem. The Lagrange functional reads
\begin{equation}\label{eq:lagrange_functional}
\begin{aligned}
    \mathcal{L} = \int_{\initDomain} \left[ \bar{\pot}(\tF) + \tlambda \cdot \rho \currCS + \tmu \cdot \boldsymbol{\frakm} \right] d X_1 d X_2,
\end{aligned}
\end{equation}
where the Lagrange multipliers $\tlambda \in \bbR^3$ and $\tmu \in \bbR^4$ are defined globally on the domain. Since this work focuses on homogeneous cross-sections, the dummy density $\rho = 1\text{ kg}/\text{m}^3$ is set from now on. The first variation of~\cref{eq:lagrange_functional} yields the weak form~\cite{aroraComputationalApproachObtain2019}
\begin{equation}
    \begin{aligned}
    G &= \int_{\initDomain} \left[ \left[ \tP \tE_3 \times \matCurv + \tlambda + \boldsymbol{\mathfrak{M}} \tmu\right] \cdot \delta \currCS + \right. 
    \left. \tP \tE_{\alpha} \cdot \der{\delta \currCS}{X_{\alpha}} + \currCS \cdot \delta \tlambda + \boldsymbol{\frakm} \cdot \delta \tmu \right] d X_1 d X_2
    = 0\,,
    \end{aligned}
\end{equation}
with the abbreviation
\begin{equation}
    \boldsymbol{\mathfrak{M}}^T = \begin{bmatrix}
    0 & \hat{X}_3 & \hat{X}_2 \\
    \hat{X}_3 & 0 & \hat{X}_1 \\
    \hat{X}_2 & \hat{X}_1 & 0 \\
    -\cfrac{\hat{X}_2}{\hat{X}_1^2 + \hat{X}_2^2} & \cfrac{\hat{X}_1}{\hat{X}_1^2 + \hat{X}_2^2} & 0.
    \end{bmatrix}.
\end{equation}
For a detailed explanation of the cross-sectional warping problem and the derivation of the strong form, the stress resultants, and the constitutive matrix, the reader is referred to~\cite{aroraComputationalApproachObtain2019}. The process of computing the stress resultants $\matForce,\matMoment$, and the beam potential $\potRod$ for given strain measures $\matStr,\matCurv$ with the DHM and the warping problem is visualized in~\cref{fig:warping-problem}.

\begin{figure}[t!]
    \centering
        \resizebox{0.75\textwidth}{!}{
    \input{tikz_warping_problem}
    }
    \caption{Schematic visualization of the cross-sectional warping problem. The inputs are the strain measures $\matStr$ and $\matCurv$ and a discretization of the cross-section. The outputs are the beam potential $\potRod$ and the stress resultants $\matForce$ and $\matMoment$.}
    \label{fig:warping-problem}
\end{figure}

\section{Physics-augmented neural network constitutive model}
\label{sec:nn-model}

In this section, a physics-augmented neural network (PANN) constitutive model for hyperelastic beams is proposed. This model will be applied as a surrogate for the computationally expensive warping problem, and the DHM model discussed in~\cref{sec:beam-model}.
Ultimately, this PANN provides a very {accurate} and computationally {efficient} surrogate model that represents both material and geometrical nonlinearities occurring on the cross-section scale of the beam.
After discussing the mechanical conditions relevant to beam constitutive modeling in~\cref{subsec:mechanical-conditions}, the PANN model is introduced in~\cref{subsec:simple_pann}. This is followed by extensions of the PANN model to include \revii{transverse isotropy and a less restrictive point symmetry in~\cref{subsec:sym}} and \revii{ends with a} geometric parameterization \revii{for ring-shaped cross-sections} in~\cref{subsec:ring-parameterization}.


\subsection{Mechanical conditions}
\label{subsec:mechanical-conditions}

With the beam potential $\potRod$ introduced in~\cref{subsec:constitutive-relations} and the beam's forces and moments defined as the potential's gradients, \textit{thermodynamic consistency} is fulfilled by construction~\cite{smritiThermoelastoplasticTheorySpecial2019}. Moreover, \textit{objectivity} is guaranteed by the definition of the strain measures according to~\cref{eq:strain_measures}~\cite{antmanNonlinearProblemsElasticity2005}. 

Furthermore, in the strain-free state, the beam potential and the stress resultants should vanish, which is formalized as the \textit{normalization} conditions
\begin{equation}\label{eq:normalization_criteria}
    \potRod\left(\vMatStrain\right)\big\rvert_{\vMatStrain = \zero} = 0\,, \qquad
    \vMatStress\left(\vMatStrain\right)\big\rvert_{\vMatStrain = \zero}  = \zero\,.
\end{equation}
Here, the abbreviations $\vMatStrain = (\matStr, \matCurv)$ and $\vMatStress = (\matForce, \matMoment)$ for the strain measures and stress resultants are introduced. This condition is fulfilled trivially in the LEM model described in~\cref{subsubsec:lem}. For the RHM and DHM models introduced in~\cref{subsubsec:rhm,subsubsec:dhm}, the normalization conditions are fulfilled since the potential $\Bar{\psi}$ applied as a base material model for the beam fulfills equivalent normalization conditions, cf.~\cite{lindenNeuralNetworksMeet2023}.

\paragraph{Symmetry:} Symmetries in the beam's cross-section imply symmetries in the beam's constitutive response~\cite{antmanNonlinearProblemsElasticity2005,healeyMaterialSymmetryChirality2002}.
\revii{In this work, only beams with circular or ring-shaped cross-sections without chirality or handedness are considered. Under these circumstances, a transversely isotropic symmetry of the constitutive behavior is typically considered ~\cite{healeyMaterialSymmetryChirality2002,antmanNonlinearProblemsElasticity2005}. The corresponding invariant-based potential is then expressed as
\begin{equation}\label{eq:transverse-isotropy}
    \potRod(\inputsIso) = \Gamma\big(\epsilon_{\alpha}\epsilon_{\alpha}, \epsilon_3, (\epsilon_{\alpha} \kappa_{\alpha})^2, \kappa_{\alpha} \kappa_{\alpha}, \varepsilon_{\alpha \beta} \epsilon_{\alpha} \kappa_{\beta}, (\kappa_3)^2, \epsilon_{\alpha} \kappa_{\alpha} \kappa_3\big)\,,
\end{equation}
where the arguments of $\Gamma$ are invariants for transverse isotropy $\inputsIso \in \mathbb{R}^7$. Here, $\varepsilon_{\alpha \beta}$ denotes the Levi-Civita symbol, i.e., $\varepsilon_{\alpha \beta} \epsilon_{\alpha} \kappa_{\beta} = \epsilon_{1} \kappa_{2} - \epsilon_{2} \kappa_{1}$~\cite{healeyMaterialSymmetryChirality2002}. However, to the best of the authors' knowledge, the condition in~\cref{eq:transverse-isotropy} has only been shown for the RHM and not for the DHM undergoing large cross-sectional deformation.

It is is known for anisotropic hyperelasticity that symmetry properties can change between linear and finite deformations~\cite{kalina2024neuralnetworksmeetanisotropic}. Similarly, it seems reasonable that \cref{eq:transverse-isotropy}, which was derived for beams with rigid cross sections,  might not always hold for beams with deformable cross sections. This motivates the symmetry condition
\begin{equation}\label{eq:point-symmetry-condition-potential}
 \potRod(\epsilon_1, \epsilon_2, \epsilon_3, \kappa_1, \kappa_2, \kappa_3) = \potRod(-\epsilon_1, -\epsilon_2, \epsilon_3, -\kappa_1, -\kappa_2, \kappa_3)\,,
\end{equation}
which may be interpreted as point symmetry w.r.t.\ the origin of the $\mathbb{R}^4$ space of strain measures $(\epsilon_1, \epsilon_2, \kappa_1, \kappa_2)$. Clearly, \cref{eq:point-symmetry-condition-potential} fulfills the symmetry condition \cref{eq:transverse-isotropy} while being less restrictive. While the authors are unaware of any proof of~\cref{eq:point-symmetry-condition-potential} for circular or ring-shaped deformable cross-sections,~\cref{subsubsec:study-i} will investigate the utility of~\cref{eq:point-symmetry-condition-potential} in formulating neural network-based material models compared to~\cref{eq:transverse-isotropy}.}

\revii{
\paragraph{Scaling:}

Another important property of the DHM model is the scaling behavior of the beam potential when multiplying the cross-sectional dimensions by a factor $\lambda$. According to ~\cite{kumarHelicalCauchyBornRule2016}, the beam potential for a scaled deformable cross-section $\potRod_{\lambda}$ at given strain $\vMatStrain_{\lambda}$ can be computed by evaluating the beam potential of a reference cross-section $\potRod$ at the reference strain
\begin{equation}\label{eq:kumar-scaling-strain}
    \vMatStrain = \begin{bmatrix}
        \matStr_{\lambda} \\
        \lambda \matCurv_{\lambda}
    \end{bmatrix}.
\end{equation}
That is
\begin{equation}\label{eq:kumar-scaling-potential}
    \potRod_{\lambda}(\vMatStrain_{\lambda}) = \lambda^2 \potRod(\vMatStrain).
\end{equation}
}

\begin{remark}
In continuum hyperelasticity, more mechanical conditions than the ones introduced above are applied~\cite{holzapfelNonlinearSolidMechanics2000}. However, their application to beams with deformable cross-sections is not straightforward.
For example, a \textit{growth condition} is applied, ensuring that a material body can neither be compressed to zero volume nor dilated to an infinite volume. Similar growth conditions could be formulated for beam potentials. However,~\cref{subsec:kinematics} shows that for geometrically exact beams, this assumption translates not just to the material model but to a geometric problem. In~\cite{antmanNonlinearProblemsElasticity2005}, growth conditions for beams with rigid cross-sections that consider the cross-sectional shape are derived. However, to the best of the authors' knowledge, there is no equivalent condition for beams with deformable cross-sections.
Other essential conditions of continuum hyperelasticity are grounded in the concept of convexity, i.e., ellipticity and polyconvexity~\cite{neff2015}. Again, while similar convexity conditions can be formulated for beams with rigid cross-sections~\cite{ortigosaComputationalFrameworkPolyconvex2016}, to the best of the author's knowledge, there are no equivalent conditions for beams with deformable cross-sections.
\end{remark}


\subsection{NN-based beam potential}\label{subsec:simple_pann}

Based on the mechanical conditions for beam constitutive modeling introduced in~\cref{subsec:mechanical-conditions}, an NN-based beam potential similar to NN-based potentials of continuum hyperelasticity \cite{lindenNeuralNetworksMeet2023,kleinPolyconvexAnisotropicHyperelasticity2022} can be formulated. The overall PANN model consists of an NN part, which is complemented by projection terms that ensure the normalization conditions, i.e.,
\begin{equation}\label{eq:PANN-potential}
    \potRodPANN(\vMatStrain) = \potNN(\vMatStrain) + \potEnergy + \potStress(\vMatStrain)\,.
\end{equation}
Here, $\potNN$ denotes the NN part of the potential, which is given as a scalar-valued feed-forward neural network (FFNN) that receives the beam strain measures $\vMatStrain$ as input, i.e.,
\begin{equation}\label{eq:nn-potential}
    \potNN(\vMatStrain) = \text{FFNN}(\vMatStrain) \,.
\end{equation}
The normalization terms for energy $\potEnergy$ and stress $\potStress$ follow as the projection terms~\cite{huang2022,fernandezAnisotropicHyperelasticConstitutive2021}
\begin{equation}\label{eq:PANN_normalization}
    \potEnergy = - \left.\potNN\right\rvert_{\vMatStrain = \zero}\,,\qquad    \potStress(\vMatStrain) = - \left.\der{\potNN}{\vMatStrain}\right\rvert_{\vMatStrain = \zero} \cdot \vMatStrain\,.
\end{equation}
The model's flow and structure are visualized in~\cref{fig:PANN-beam-model}. In this work, this architecture is signified by $\model(\vMatStrain)=\potRodPANN(\vMatStrain)$.

\medskip

The NN part of the model $\potNN$ provides excellent flexibility to capture the nonlinear constitutive behavior of hyperelastic beams. In this work, fully connected FFNNs~\cite{kollmannsbergerDeepLearningComputational2021} are applied to represent $\potNN$. With input $\vMatStrain=\tx_0 \in \bbR^{6}$, output $\tx_{H+1} \in \bbR$ and $H$ hidden layers, the FFNNs are given as the mapping
\begin{equation}\label{eq:1st-hidden-layer}
    \tx_{h} = \sigma^{(h)}(\tW^{(h)} \tx_{h-1} + \tb^{(h)}) \in \mathbb{R}^{n_{h}}\,,\qquad h=1,...,H+1.
\end{equation}
Here, the weights $\tW^{(h)}$ and biases $\tb^{(h)}$ are the parameters of the NN potential, which are optimized to fit a given dataset. The activation functions of the FFNN are denoted by $\sigma^{(h)}$. In this work, softplus activation functions are used in the hidden layers, while linear activation is used in the output layer.

\revii{As the PANN-based beam potential represented by~\cref{eq:PANN-potential,eq:nn-potential,eq:PANN_normalization} is a twice continuously differentiable function, the stress resultants and stiffness matrix components can be easily calculated as $\vMatStress = \partial \potRodPANN / \partial \vMatStrain$ and $\mathbb{C} = \partial^2 \potRodPANN / \partial \vMatStrain^{2}$. This involves computing the first and second derivatives of the FFNN concerning its inputs, which can be done via automatic differentiation or analytically using the chain rule according to the derivations in~\cite{frankeAdvancedDiscretizationTechniques2023,rodiniAnalyticalDerivativesNeural2022}.
Finally, the PANN-based beam potential can be combined with the scaling law described by~\cref{eq:kumar-scaling-potential,eq:kumar-scaling-strain}~\cite{kumarHelicalCauchyBornRule2016}. In this work scaled models will be signified by $\model_{\lambda}(\vMatStrain_{\lambda}) = \lambda^2 \potRod^{\text{PANN}}(\vMatStrain)$, where $\lambda$ may be replaced with a cross-section specific scaling variable.}

\begin{figure}
    \centering
    \resizebox{0.75\textwidth}{!}{
    \tikzsetnextfilename{tikz_PANN_beam}
\begin{tikzpicture}[x=1.6cm,y=1.1cm]
  \large
  \def\NC{6} 
  \def\nstyle{int(\lay<\Nnodlen?(\lay<\NC?min(2,\lay):3):4)} 
  \tikzset{ 
    node 1/.style={node in_out},
    node 2/.style={node inv_pot},
    node 3/.style={node icnn},
    node 4/.style={node ffnn}
  }

  \draw[color33!40,fill=color33,fill opacity=0.02,rounded corners=4]
    (4.2,-0.1) rectangle++ (3.9,1.2);
 
  \node[node 1, outer sep=0.6] (p) at (0.5,0.5) {$\vMatStrain$};

  \node[node 2, outer sep=0.6] (I) at (1.6, 0.5) {$\inputs$};

  \draw[connect arrow2] (p) -- (I);

  \node (n) at (2.51, 0.5) {};
  \node[node 4, outer sep=0.6] (ffnn) at (3.1, 0.5) {FFNN};

  \draw[connect arrow2] (I) -- (n);
      
  \node[node 2, outer sep=0.6] (6-1) at (4.6,0.5) {$\psi^{\text{NN}}$};

  \draw[connect arrow2] (ffnn) -- (6-1);

  \node[align = center] (7-1) at (6.5,0.5) {$+\,\psi^{\text{energy}} + \psi^{\text{stress}}=:\psi^{\text{PANN}}$};

  \node (8-0) at (8.0,0.5) {};

  \node[node 2, outer sep=0.6] (8-1) at (9.0,0.5) {$\frac{\partial}{\partial \vMatStrain}$};
  \draw[connect arrow2]  (8-0) -- (8-1);    
      
  \node[node 1, outer sep=0.6] (9-1) at (10.1,0.5) {$\vMatStress$};
  \draw[connect arrow2]  (8-1) -- (9-1);
 
\end{tikzpicture}
    }
    \caption{Physics-augmented neural network-based beam constitutive model. The inputs are the strains and curvatures $\vMatStrain = (\matStr, \matCurv)$\revii{, which are converted to the FFNN inputs $\inputs$ -- in the simplest case $\inputs = \vMatStrain$}. The forces and moments $\vMatStress = (\matForce, \matMoment)$ are derived from the beam strain energy $\psi^\text{PANN}$, the model output.}
    \label{fig:PANN-beam-model}
\end{figure}

\revii{\subsection{Enforcing symmetry}}
\label{subsec:sym}

\revii{As discussed in~\cref{subsec:mechanical-conditions}, beams with circular deformable cross-sections may be expected to exhibit symmetric constitutive behavior, e.g., according to~\cref{eq:transverse-isotropy} or~\cref{eq:point-symmetry-condition-potential}. Here, we present how the PANN-based constitutive model presented in~\cref{subsec:simple_pann} can be extended to fulfill these conditions by construction.

\medskip

\paragraph{Transverse isotropy:}

To enforce transverse isotropy, the strain measure-based NN potential from \cref{eq:nn-potential} can be reformulated using the transversely isotropic invariants $\inputsIso$. This approach is in close analogy to the hyperelastic invariant-based PANN models that have been developed for 3D continuum materials~\cite{kleinFiniteElectroelasticityPhysicsaugmented2022,lindenNeuralNetworksMeet2023}. The transversely isotropic beam potential maintains the same basic structure as in~\cref{eq:PANN-potential}. However, the NN term of the PANN potential now receives the transversely isotropic invariants as inputs:
\begin{equation}\label{eq:trans-is-nn-potential}
    \potNN(\inputsIso) = \text{FFNN}\big(\epsilon_{\alpha}\epsilon_{\alpha}, \epsilon_3, (\epsilon_{\alpha} \kappa_{\alpha})^2, \kappa_{\alpha} \kappa_{\alpha}, \varepsilon_{\alpha \beta} \epsilon_{\alpha} \kappa_{\beta}, \kappa_3^2, \epsilon_{\alpha} \kappa_{\alpha} \kappa_3\big).
\end{equation}
Consequently, the normalization term simplifies to
\begin{equation}
    \potStress(\inputsIso) = -\left.\der{\potNN}{\epsilon_3}\right\rvert_{\vMatStrain = \zero} \epsilon_3,
\end{equation}
i.e., normalization must only be enforced for the stress resultant $n_3$, as the remaining stress resultants are automatically normalized through the quadratic and cubic nature of the invariants.
Finally, the corresponding energy normalization term is chosen as
\begin{equation}
    \potEnergy = -\left.\potNN(\inputsIso)\right\rvert_{\vMatStrain = \zero}.
\end{equation}
In the scope of this work, transversely isotropic models are signified by $\modeliso(\vMatStrain) = \potRodPANN(\inputsIso)$.}

\paragraph{Point symmetry:}

\revii{If no invariant-based formulation is available, symmetry can be enforced by modifying the strain measures before passing them to the PANN-based beam potential in~\cref{eq:PANN-potential}. For the symmetry described by~\cref{eq:point-symmetry-condition-potential} this can be done as follows:}
Consider the strain state \revii{$\vMatStrain = (\vMatStrainSym, \epsilon_3, \kappa_3)$} with the symmetric part of the strain vector \revii{$\vMatStrainSym = (\epsilon_1, \epsilon_2, \kappa_1, \kappa_2) \in\bbR^4$}. \revii{From~\cref{eq:point-symmetry-condition-potential}, we can deduce that there is a second strain state $\vMatStrainTilde = (-\vMatStrainSym, \epsilon_3, \kappa_3)$ whose material response is equivalent to $\vMatStrain$}.
To distinguish the two associated strain states, a hyperplane $\cH = \{\vMatStrainSym \in \bbR^\revii{4}: f(\vMatStrainSym)=0\}$ is introduced using the linear function $f(\vMatStrainSym)$. Here, $\cH$ could be any hyperplane that intersects the origin of the space \revii{$\bbR^4$}. For this work, the function
\revii{
\begin{equation}\label{eq:hyperplane_function}
    f(\vMatStrainSym) = \epsilon_1 + \epsilon_2 + \kappa_1 + \kappa_2
\end{equation}}
is used. With~\cref{eq:hyperplane_function}, the surrogate strain state $\vMatStrainTilde$ is defined as
\begin{equation}\label{eq:surrogate_strain_state}
\revii{
    \vMatStrainTilde = \left\{\begin{array}{lr}
    {\begin{bmatrix}
    -\vMatStrainSym \\
    \epsilon_3 \\
    \kappa_3
    \end{bmatrix}}\quad &\text{if}\quad f(\vMatStrainSym) < 0\,, \vspace*{2mm}\\
    {\begin{bmatrix}
    \vMatStrainSym \\
    \epsilon_3 \\
    \kappa_3
    \end{bmatrix}}\quad &\text{if}\quad f(\vMatStrainSym) \geq 0\,.
    \end{array}\right.
}
\end{equation}
After evaluating $\potRodPANN$ for $\vMatStrainTilde$ and taking the derivatives, the surrogate stress measures \revii{$\vMatStressTilde = (\vMatStressSymTilde, n_3, m_3)$} are obtained and have to be reflected analogously to get the final stress measures:
\begin{equation}\label{eq:surrogate_stress_state}
\revii{
    \vMatStress= \left\{\begin{array}{lr}
    {\begin{bmatrix}
    -\vMatStressSymTilde \\
    \epsilon_3 \\
    \kappa_3
    \end{bmatrix}}\quad &\text{if}\quad  f(\vMatStrainSym) < 0\,, \vspace*{2mm}\\
    {\begin{bmatrix}
    \vMatStressSymTilde \\
    \epsilon_3 \\
    \kappa_3
    \end{bmatrix}}\quad &\text{if}\quad f(\vMatStrainSym) \geq 0\,.
    \end{array}\right.
}
\end{equation}
\revii{Ultimately, the reflections conducted in ~\cref{eq:surrogate_strain_state,eq:surrogate_stress_state} for $f(\vMatStrainSym) < 0$ must also be applied to all components of the stiffness matrix $\mathbb{C}$ that contain mixed derivatives $\partial^2 \potRodPANN / (\partial \vMatStrainSym \partial \epsilon_3)$ or $\partial^2 \potRodPANN / (\partial \vMatStrainSym \partial \kappa_3)$ between symmetric and non-symmetric strain measures. In this work, point symmetric PANN models are signified by $\model^{\text{sym}}(\vMatStrain)=\potRodPANN(\vMatStrainTilde)$.}

\revii{
\subsection{Ring parameterization}
\label{subsec:ring-parameterization}

For practical application, a parameterization of the PANN-based beam potential with geometric parameters, e.g., the inner radius $R_i$ of a circular cross-section, is highly desirable. For the LEM, such a parameterization is trivial as the cross-sectional area and moments of inertia can be directly calculated for any $R_i$ according to~\cref{eq:lem_geom}. For the PANN model, building on ~\cref{eq:PANN-potential} or~\cref{eq:trans-is-nn-potential}, the parameterized potential takes the form\footnote{For the sake of a simple notation, the same symbols are applied for parameterized and non-parameterized PANN models in this work.}
\begin{equation}
\potRod:\bbR^d\times\bbR\rightarrow\bbR\,,\qquad (\inputs,\parameter)\mapsto \potRod(\inputs,\parameter),
\end{equation}
where $\inputs \in \mathbb{R}^d$ is the FFNN input vector containing strain measures or invariants, and the parameter $\parameter = R_i/R$ is the ratio between the ring's inner and outer radius with $0\leq P<1$. Note that the beam potential's functional dependency on $\parameter$ is not restricted by the mechanical conditions introduced in~\cref{subsec:mechanical-conditions}, cf.~\cite{kleinParametrisedPolyconvexHyperelasticity2023} for discussion.

The parameterization can be incorporated into the PANN-based beam potential from~\cref{eq:PANN-potential}, by passing $\parameter$ as an additional input to the FFNN, i.e,
\begin{equation}\label{eq:PANN-parameterized_1}
    \potRodPANN(\inputs,\parameter)= \potNN(\inputs,\parameter) + \potEnergy(\parameter) + \potStress(\inputs,\parameter).
\end{equation}
Consequently, the normalization terms now depend on the parameter $\parameter$, cf.~\cite{kleinParametrisedPolyconvexHyperelasticity2023} for a discussion. In this work, ring-parameterized architectures according to~\cref{eq:PANN-parameterized_1} are signified by $\model_{\parameter}(\inputs,\parameter) = \potRodPANN(\inputs,\parameter)$. The architecture is visualized in~\cref{fig:p-model}.
Note that the $\model_{\parameter}$ model can be extended with transverse isotropy or point symmetry as presented in~\cref{subsec:sym}. Furthermore, scaling according to~\cref{eq:kumar-scaling-potential} can be applied due to the dimensionless definition of $\parameter$.
}

\begin{figure*}
    \centering
    \resizebox{0.35\textwidth}{!}{
    \tikzsetnextfilename{tikz_PANN_beam_P}
\begin{tikzpicture}[x=1.6cm,y=1.1cm]
    \large
    
    \tikzset{ 
    node 1/.style={node in_out},
    node 2/.style={node layer},
    node 3/.style={node inv_pot},
    node 4/.style={node ffnn}
    }

    \node[node 1, outer sep=0.6] (X1) at (1.75, 1)  {$\vMatStrain$};
    \node[node 1, outer sep=0.6] (X2) at (1.75, 2.25) {$\parameter$};

    \node[node 3, outer sep=0.6] (I) at (2.85, 1) {$\inputs$};

    \draw[connect arrow] (X1) -- (I);

    \node (n) at (3.85, 1.) {};
    \node[node 4, outer sep=0.6] (ffnn) at (4.45, 1) {FFNN};
    
    \draw[connect arrow ] (I) -- (ffnn);
    \draw[connect, shorten >= 0, shorten <= 0] (X2) -- (2.85, 2.25);
    \draw[connect arrow2, shorten <= 0] (2.85, 2.25) -- (n);

    \node[node 3, outer sep=0.6] (Y) at (5.85, 1) {$\psi^{\text{NN}}$};

    \draw[connect arrow2] (ffnn) -- (Y);
         
  \end{tikzpicture}
    }
    \caption{\revii{Ring-parameterized PANN-based constitutive model $\model_{\parameter}$. Note that the FFNN parts of the models are complemented with normalization terms, cf.~\cref{fig:PANN-beam-model}.}}
    \label{fig:p-model}
\end{figure*}
\section{Numerical examples}
\label{sec:results}

In this section, the applicability of the PANN constitutive models for hyperelastic beams with highly nonlinear behavior is demonstrated. For this, data for a beam with nonlinear base material and a deformable cross-section is generated following the warping problem introduced in~\cite{aroraComputationalApproachObtain2019}, cf.~\cref{subsubsec:dhm,subsec:warping-problem}.
After the data generation in~\cref{subsec:data-generation}, the models are calibrated according to~\cref{subsec:model-calibration} and evaluated in~\cref{subsec:model-evaluation}. Finally, a non-parameterized PANN beam model is applied in beam simulations in~\cref{subsec:beam-simulation}.


\subsection{Data generation}
\label{subsec:data-generation}

Calibration of the PANN models requires datasets of the form
\begin{equation}
    \cD = \{(\prescript{1}{}{\vMatStrain}, \prescript{1}{}{\vMatStress}), (\prescript{2}{}{\vMatStrain}, \prescript{2}{}{\vMatStress}), ...\}
\end{equation}
for the non-parameterized models $\model$\revii{, $\modeliso$, and $\modelsym$ as well as}
\begin{equation}
    \revii{\cD_{\parameter}} = \{(\prescript{1}{}{\vMatStrain}, \prescript{1}{}{\vMatStress}, \prescript{1}{}{\revii{\parameter}}), (\prescript{2}{}{\vMatStrain}, \prescript{2}{}{\vMatStress}, \prescript{2}{}{\revii{\parameter}}), ...\}
\end{equation}
or
\begin{equation}
    \revii{\cD_{\lambda}} = \{(\prescript{1}{}{\vMatStrain}, \prescript{1}{}{\vMatStress}, \prescript{1}{}{\revii{\lambda}}), (\prescript{2}{}{\vMatStrain}, \prescript{2}{}{\vMatStress}, \prescript{2}{}{\revii{\lambda}}), ...\}
\end{equation}
for the parameterized \revii{and scaled} \revii{architecture $\model_{\parameter}$} \revii{and $\model_{\lambda}$, respectively}. Acquisition of these datasets involves a two-step process, which starts with sampling values for the input quantities $\vMatStrain$ and \revii{$\parameter$/$\lambda$} and ends with computing the corresponding outputs $\vMatStress$.

\subsubsection{Concentric sampling of strain measures}
\label{subsubsec:sampling}

Sampling of the inputs $\vMatStrain$ and \revii{$\parameter$} is conducted in a seven-dimensional space. We require the sampled strain states to be physically sensible in the sense that the volumetric compression ratios fulfill the impenetrability condition $J = \det \tF > 0$~\cite{holzapfelNonlinearSolidMechanics2000} anywhere on the cross-section. Furthermore, the sampled strain measures shall represent the space of admissible strain measures well while keeping the total number of samples as low as possible.

Concentric sampling is applied in this work to create a large variety of samples efficiently~\cite{kuncGenerationEnergyminimizingPoint2019}. The idea is to decompose a vector into a unit direction $\tN \in \bbR^6$ and an amplitude $t$, which are sampled separately and afterward joined to form a vector. Here, the decomposition of the strain measures reads
\begin{equation}\label{eq:sampling_directions}
    \vMatStrain = \norm{\vMatStrain} \cdot \frac{\vMatStrain}{\norm{\vMatStrain}} = t \tN.
\end{equation}
First, the directions are chosen to be uniformly distributed in the six-dimensional strain space $\bbR^6$. A method for generating uniformly distributed points on $n$-dimensional spheres using energy minimization is applied here~\cite{kuncGenerationEnergyminimizingPoint2019}. Next, the number of strain amplitudes between the minimum and maximum values $t_{\text{min}}$ and $t_{\text{max}}$ is defined.

After joining directions and amplitudes, random perturbations are applied to every data point. The perturbed strain measures read
\begin{equation}
    \vMatStrain_{\text{perturbed}} = \vMatStrain + \norm{\vMatStrain} \cdot \epsilon \cdot \tN_{\text{rand}},
\end{equation}
where $\tN_{\text{rand}} \in \bbR^6$ is a random unit direction, which is regenerated for every data point, and $\epsilon$ is a scalar that denotes the magnitude of the perturbation relative to the norm $\norm{\vMatStrain}$ of the initial strain state. In practice, $\epsilon$ should be chosen moderately to ensure that the jumps between two consecutive strain states are not too large during the numerical simulation. Lastly, the perturbed strain states are filtered by applying limits to every strain measure. 
For this, the limits $\varepsilon_1,\,\varepsilon_2\in[-0.2,\,0.2]$, $\varepsilon_3\in[-0.2,\,0.5]$, and $\kappa_1,\,\kappa_2,\,\kappa_3\in[-2\pi/L,\,2\pi/L]$ are applied. These values are motivated by the conception of a beam with length $L = 10$ subjected to bending or twist of 360\textdegree, 20\% compression, or 50\% elongation.
Lastly, the physical sensibility of the sampled strain measures is evaluated. For that, a reference rectangle is placed around the cross-section. Theoretical deformation gradients at the corner points $\tX_1$ to $\tX_4$ of the rectangle are computed with~\cref{eq:defgrad_rigid}. If all four points comply with the impenetrability of matter~\cite{holzapfelNonlinearSolidMechanics2000}, i.e.,
\begin{equation}
\begin{split}
    &\det\tF(\tX_i) > 0\quad i\in\{1,2,3,4\},
\end{split}
\end{equation}
the associated strain measures are added to the input dataset. An example of a concentrically sampled load path with perturbation is shown in~\cref{fig:train-prediction-concentric-perturbed}.

\subsubsection{Constitutive model evaluation}
\label{subsec:model-evaluation}
\begin{figure*}[t!]
    \centering
    \begin{subfigure}[t]{\linewidth}
        \centering
        \begin{subfigure}[b]{0.4\textwidth}
            \centering
            \resizebox{\linewidth}{!}{
                \tikzsetnextfilename{fig_rigid_vs_warping_epsilon_1}
\begin{tikzpicture}
\begin{axis}[
  name=plot1,
  xlabel={$\epsilon_1$},
  ylabel={$q_i$},
  transpose legend,
  legend style={at={(0.03,0.97)}, anchor=north west},
  legend columns=3,
  grid=major,
ytick={-10,0,10},
yticklabels={-10,0,10},
]
    
    \addplot [color=CPSred, dotted, very thick, no markers, forget plot] table [col sep=space, x index=0, y index=7, header=false]{fig_rigid_vs_warping_data_rigid_epsilon_1.txt};
    \addplot [color=CPSgrey, dotted, very thick, no markers, forget plot] table [col sep=space, x index=0, y index=8, header=false]{fig_rigid_vs_warping_data_rigid_epsilon_1.txt};
    \addplot [color=CPSdarkblue, dotted, very thick, no markers, forget plot] table [col sep=space, x index=0, y index=9, header=false]{fig_rigid_vs_warping_data_rigid_epsilon_1.txt};
    \addplot [color=CPSgrey, dotted, very thick, no markers, forget plot] table [col sep=space, x index=0, y index=10, header=false]{fig_rigid_vs_warping_data_rigid_epsilon_1.txt};
    \addplot [color=CPSgrey, dotted, very thick, no markers, forget plot] table [col sep=space, x index=0, y index=11, header=false]{fig_rigid_vs_warping_data_rigid_epsilon_1.txt};
    \addplot [color=CPSgrey, dotted, very thick, no markers, forget plot] table [col sep=space, x index=0, y index=12, header=false]{fig_rigid_vs_warping_data_rigid_epsilon_1.txt};

    \addplot [color=CPSred, solid, very thick, no markers] table [col sep=space, x index=0, y index=7, header=false]{fig_rigid_vs_warping_data_deformable_epsilon_1.txt};
    \addlegendentry{$n_1$}
    \addplot [color=CPSgrey, solid, very thick, no markers] table [col sep=space, x index=0, y index=8, header=false]{fig_rigid_vs_warping_data_deformable_epsilon_1.txt};
    \addlegendentry{$n_2$}
    \addplot [color=CPSdarkblue, solid, very thick, no markers] table [col sep=space, x index=0, y index=9, header=false]{fig_rigid_vs_warping_data_deformable_epsilon_1.txt};
    \addlegendentry{$n_3$}
    \addplot [color=CPSgrey, solid, very thick, no markers] table [col sep=space, x index=0, y index=10, header=false]{fig_rigid_vs_warping_data_deformable_epsilon_1.txt};
    \addlegendentry{$m_1$}
    \addplot [color=CPSgrey, solid, very thick, no markers] table [col sep=space, x index=0, y index=11, header=false]{fig_rigid_vs_warping_data_deformable_epsilon_1.txt};
    \addlegendentry{$m_2$}
    \addplot [color=CPSgrey, solid, very thick, no markers] table [col sep=space, x index=0, y index=12, header=false]{fig_rigid_vs_warping_data_deformable_epsilon_1.txt};
    \addlegendentry{$m_3$}

    \addplot[color=CPSred, dashed, very thick, domain=-0.20:0.20] {5/6*25*pi*\R^2*x};
    \addplot[color=CPSdarkblue, dashed, very thick, domain=-0.20:0.20] {0};

    \addplot [color=CPSred, dotted, very thick, no markers] table [col sep=space, x index=0, y index=7, header=false]{fig_rigid_vs_warping_data_rigid_epsilon_1.txt};
    \addplot [color=CPSdarkblue, dotted, very thick, no markers] table [col sep=space, x index=0, y index=9, header=false]{fig_rigid_vs_warping_data_rigid_epsilon_1.txt};

    \addplot [color=CPSred, solid, very thick, no markers] table [col sep=space, x index=0, y index=7, header=false]{fig_rigid_vs_warping_data_deformable_epsilon_1.txt};
    \addplot [color=CPSdarkblue, solid, very thick, no markers] table [col sep=space, x index=0, y index=9, header=false]{fig_rigid_vs_warping_data_deformable_epsilon_1.txt};
\end{axis}
\end{tikzpicture}
            }
        \end{subfigure}
        \qquad
        \begin{subfigure}[b]{0.35\textwidth}
            \centering
            \begin{subfigure}[t]{0.59\textwidth}
                \centering
                \includegraphics[width=1.0\textwidth]{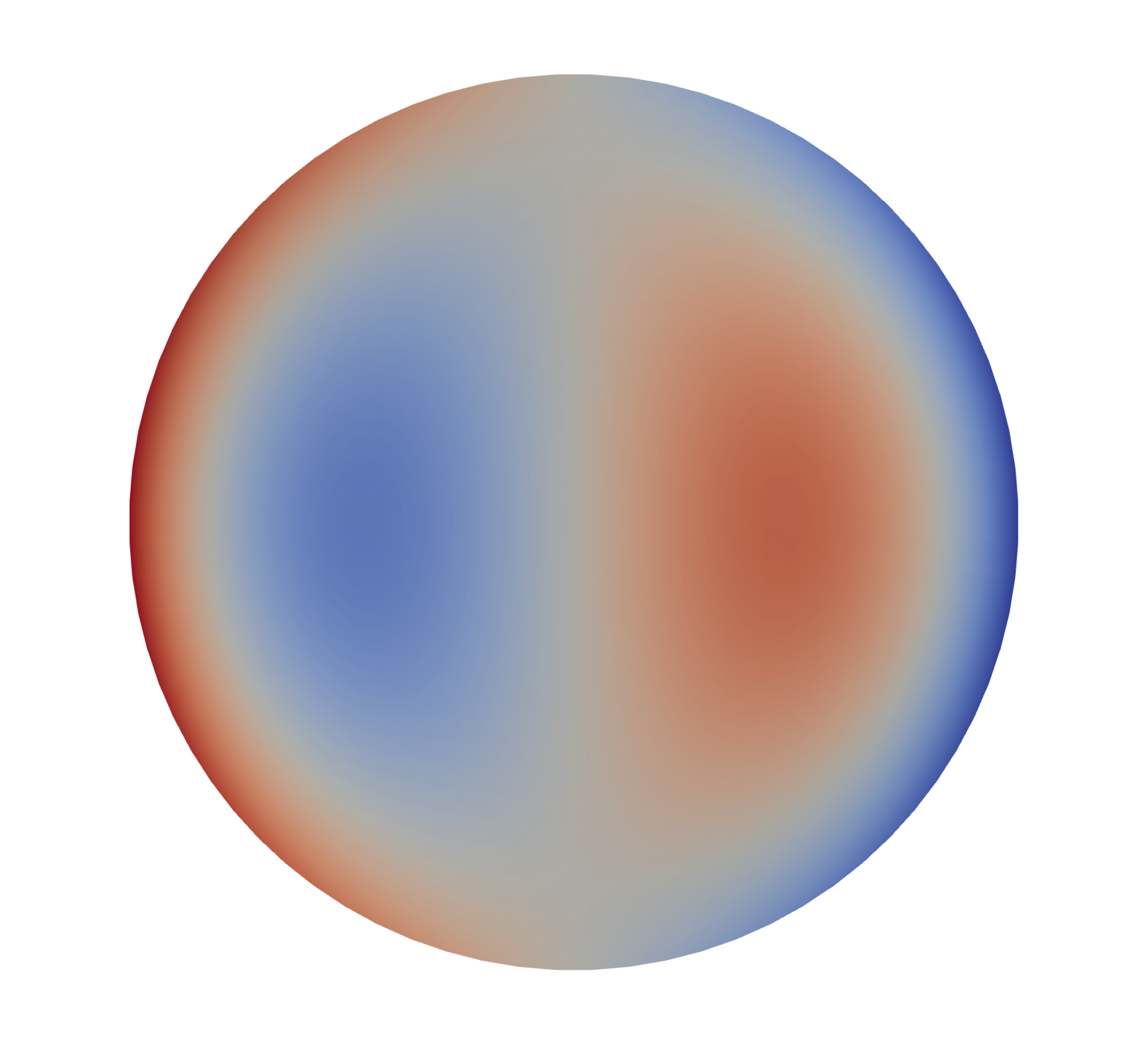}
            \end{subfigure}
            \begin{subfigure}[t]{0.39\textwidth}
                \centering
                \tikzsetnextfilename{fig_uz_colorbar}
\begin{tikzpicture}
\pgfplotscolorbardrawstandalone[ 
    point meta min=-0.27, 
    point meta max=0.27,
    colormap={cool-to-warm}{
        rgb(0)=(0.229999503869523,0.298998934049376,0.754000138575591)
        rgb(0.0158730158730159)=(0.248387005612446,0.326713084930128,0.778759751168141)
        rgb(0.0317460317460317)=(0.267060596781042,0.354213683475558,0.80247084431423)
        rgb(0.0476190476190476)=(0.286046233433907,0.381473413618758,0.825071993030171)
        rgb(0.0634920634920635)=(0.305359940427952,0.408457238050427,0.84650499453829)
        rgb(0.0793650793650794)=(0.325008919322423,0.435124782260566,0.866715008233051)
        rgb(0.0952380952380952)=(0.344992449467791,0.461431960153803,0.885650686307679)
        rgb(0.111111111111111)=(0.365302660727705,0.487332122586216,0.903264294298265)
        rgb(0.126984126984127)=(0.385925222694485,0.512776896138637,0.91951182090386)
        rgb(0.142857142857143)=(0.406839975594796,0.537716815138862,0.934353076530895)
        rgb(0.158730158730159)=(0.428021516636282,0.562101812270725,0.947751780091349)
        rgb(0.174603174603175)=(0.449439749029995,0.585881610291412,0.959675633658507)
        rgb(0.19047619047619)=(0.471060397373382,0.609006043154061,0.970096384653213)
        rgb(0.206349206349206)=(0.492845491291784,0.631425325735442,0.978989875298378)
        rgb(0.222222222222222)=(0.514753818467056,0.65309028541659,0.986336079140528)
        rgb(0.238095238095238)=(0.536741347968443,0.67395256479008,0.992119124494973)
        rgb(0.253968253968254)=(0.558761624861434,0.693964802064848,0.996327304725749)
        rgb(0.26984126984127)=(0.580766137238736,0.713080793870388,0.998953075322919)
        rgb(0.285714285714286)=(0.602704657000925,0.731255643849636,0.999993037787936)
        rgb(0.301587301587302)=(0.624525555865855,0.748445899495025,0.999447910382271)
        rgb(0.317460317460317)=(0.64617609818675,0.764609679007339,0.997322485835062)
        rgb(0.333333333333333)=(0.667602712206112,0.77970678946352,0.993625576141459)
        rgb(0.349206349206349)=(0.68875124137176,0.79369883721399,0.988369944613969)
        rgb(0.365079365079365)=(0.709567177301707,0.806549331155555,0.981572225373467)
        rgb(0.380952380952381)=(0.729995875916342,0.818223779316756,0.973252830483399)
        rgb(0.396825396825397)=(0.749982758168947,0.828689779030202,0.963435844938547)
        rgb(0.412698412698413)=(0.769473496707127,0.837917100837779,0.952148909716474)
        rgb(0.428571428571429)=(0.788414189694655,0.845877766169784,0.939423093082902)
        rgb(0.444444444444444)=(0.806751522920461,0.85254611875045,0.925292750308311)
        rgb(0.46031746031746)=(0.824432921222376,0.857898889604224,0.909795371897492)
        rgb(0.476190476190476)=(0.841406690160182,0.861915255464474,0.892971420350679)
        rgb(0.492063492063492)=(0.857622148786607,0.864576890314796,0.874864155355995)
        rgb(0.507936507936508)=(0.874239603676395,0.861703830626339,0.854248907429317)
        rgb(0.523809523809524)=(0.890678735446095,0.853320706019111,0.831640336391034)
        rgb(0.53968253968254)=(0.905552673939185,0.843617842774967,0.808523278638521)
        rgb(0.555555555555556)=(0.918869468148195,0.832615924366692,0.78495810425444)
        rgb(0.571428571428571)=(0.930635712635114,0.820337799431383,0.761004577523922)
        rgb(0.587301587301587)=(0.940857197090139,0.806808340933546,0.736721705094366)
        rgb(0.603174603174603)=(0.949539432266613,0.79205428793315,0.712167591757623)
        rgb(0.619047619047619)=(0.956688079767082,0.776104066879158,0.687399304196029)
        rgb(0.634920634920635)=(0.962309306085296,0.758987588429841,0.662472743015538)
        rgb(0.650793650793651)=(0.966410076233795,0.740736014556394,0.637442523375873)
        rgb(0.666666666666667)=(0.968998398592708,0.721381488984382,0.612361864520328)
        rgb(0.682539682539683)=(0.970083529892486,0.700956821674111,0.587282488508154)
        rgb(0.698412698412698)=(0.969676147212604,0.679495114746095,0.562254528462294)
        rgb(0.714285714285714)=(0.967788492347632,0.657029312579374,0.537326446666784)
        rgb(0.73015873015873)=(0.96443449272698,0.633591652063982,0.512544962884353)
        rgb(0.746031746031746)=(0.959629862179917,0.609212979090077,0.487954993319154)
        rgb(0.761904761904762)=(0.95339218414484,0.583921882550407,0.463599600726846)
        rgb(0.777777777777778)=(0.945740979381812,0.557743574505344,0.439519956280044)
        rgb(0.793650793650794)=(0.936697759823885,0.530698409733458,0.415755313939185)
        rgb(0.80952380952381)=(0.926286069868824,0.50279988089061,0.392342998266502)
        rgb(0.825396825396825)=(0.914531516148685,0.474051830893187,0.369318406866782)
        rgb(0.841269841269841)=(0.901461786605133,0.444444461402548,0.346715028958866)
        rgb(0.857142857142857)=(0.887106659532162,0.413948424491904,0.32456448199742)
        rgb(0.873015873015873)=(0.871498003116352,0.382505735188946,0.302896568802084)
        rgb(0.888888888888889)=(0.854669765901501,0.350015145998675,0.281739358344963)
        rgb(0.904761904761905)=(0.836657958524603,0.316307274081113,0.261119294239799)
        rgb(0.920634920634921)=(0.817500627010409,0.281099264139742,0.241061336118172)
        rgb(0.936507936507937)=(0.797237817870139,0.243904285414508,0.221589140526617)
        rgb(0.952380952380952)=(0.775911535225725,0.203826593912042,0.202725289788009)
        rgb(0.968253968253968)=(0.753565690175334,0.159000204558525,0.184491579471458)
        rgb(0.984126984126984)=(0.730246042632013,0.10443854422286,0.166909377664698)
        rgb(1)=(0.706000135911705,0.0159918240339807,0.1500000719222)
        },
    colorbar style={
        ylabel={$u_3$},
        ytick={-0.27,-0.2,-0.1,0,0.1,0.2,0.27}, 
        yticklabels={-0.027,,,0,,,0.027},
        height={25mm},
        width={2mm}
    }]
\end{tikzpicture}
                \caption*{}
            \end{subfigure}
            \vspace{0.1\linewidth}
        \end{subfigure}
        \caption{Shear}
        \label{fig:deformable_vs_rigid_shear}
    \end{subfigure}

    \vspace{0.03\linewidth}

    \begin{subfigure}[t]{\linewidth}
        \centering
        \begin{subfigure}[b]{0.4\textwidth}
            \centering
            \resizebox{\linewidth}{!}{
                \tikzsetnextfilename{fig_rigid_vs_warping_kappa_1}
\begin{tikzpicture}
\begin{axis}[
  name=plot1,
  xlabel={$\kappa_1$},
  ylabel={$q_i$},
  transpose legend,
  legend style={at={(0.03,0.97)}, anchor=north west},
  legend columns=3,
  grid=major,
ytick={-70,0,70},
yticklabels={-70,0,70},
xtick={-0.6,-0.3,0,0.3,0.6},
xticklabels={-0.6,-0.3,0,0.3,0.6},
]
    
    \addplot [color=CPSgrey, dotted, very thick, no markers, forget plot] table [col sep=space, x index=3, y index=7, header=false]{fig_rigid_vs_warping_data_rigid_kappa_1.txt};
    \addplot [color=CPSgrey, dotted, very thick, no markers, forget plot] table [col sep=space, x index=3, y index=8, header=false]{fig_rigid_vs_warping_data_rigid_kappa_1.txt};
    \addplot [color=CPSdarkblue, dotted, very thick, no markers, forget plot] table [col sep=space, x index=3, y index=9, header=false]{fig_rigid_vs_warping_data_rigid_kappa_1.txt};
    \addplot [color=CPSgreen, dotted, very thick, no markers, forget plot] table [col sep=space, x index=3, y index=10, header=false]{fig_rigid_vs_warping_data_rigid_kappa_1.txt};
    \addplot [color=CPSgrey, dotted, very thick, no markers, forget plot] table [col sep=space, x index=3, y index=11, header=false]{fig_rigid_vs_warping_data_rigid_kappa_1.txt};
    \addplot [color=CPSgrey, dotted, very thick, no markers, forget plot] table [col sep=space, x index=3, y index=12, header=false]{fig_rigid_vs_warping_data_rigid_kappa_1.txt};

    \addplot [color=CPSgrey, solid, very thick, no markers] table [col sep=space, x index=3, y index=7, header=false]{fig_rigid_vs_warping_data_deformable_kappa_1.txt};
    \addlegendentry{$n_1$}
    \addplot [color=CPSgrey, solid, very thick, no markers] table [col sep=space, x index=3, y index=8, header=false]{fig_rigid_vs_warping_data_deformable_kappa_1.txt};
    \addlegendentry{$n_2$}
    \addplot [color=CPSdarkblue, solid, very thick, no markers] table [col sep=space, x index=3, y index=9, header=false]{fig_rigid_vs_warping_data_deformable_kappa_1.txt};
    \addlegendentry{$n_3$}
    \addplot [color=CPSgreen, solid, very thick, no markers] table [col sep=space, x index=3, y index=10, header=false]{fig_rigid_vs_warping_data_deformable_kappa_1.txt};
    \addlegendentry{$m_1$}
    \addplot [color=CPSgrey, solid, very thick, no markers] table [col sep=space, x index=3, y index=11, header=false]{fig_rigid_vs_warping_data_deformable_kappa_1.txt};
    \addlegendentry{$m_2$}
    \addplot [color=CPSgrey, solid, very thick, no markers] table [col sep=space, x index=3, y index=12, header=false]{fig_rigid_vs_warping_data_deformable_kappa_1.txt};
    \addlegendentry{$m_3$}

    \addplot[color=CPSgreen, dashed, very thick, domain=-0.63:0.63] {70/4*pi*\R^4*x};
    \addplot[color=CPSdarkblue, dashed, very thick, domain=-0.63:0.63] {0};

    \addplot [color=CPSgreen, dotted, very thick, no markers] table [col sep=space, x index=3, y index=10, header=false]{fig_rigid_vs_warping_data_rigid_kappa_1.txt};
    \addplot [color=CPSdarkblue, dotted, very thick, no markers] table [col sep=space, x index=3, y index=9, header=false]{fig_rigid_vs_warping_data_rigid_kappa_1.txt};

    \addplot [color=CPSgreen, solid, very thick, no markers] table [col sep=space, x index=3, y index=10, header=false]{fig_rigid_vs_warping_data_deformable_kappa_1.txt};
    \addplot [color=CPSdarkblue, solid, very thick, no markers] table [col sep=space, x index=3, y index=9, header=false]{fig_rigid_vs_warping_data_deformable_kappa_1.txt};
    
\end{axis}
\end{tikzpicture}
            }
        \end{subfigure}
        \qquad
        \begin{subfigure}[b]{0.35\textwidth}
            \begin{subfigure}[t]{0.59\textwidth}
                \centering
                \includegraphics[width=1.0\textwidth]{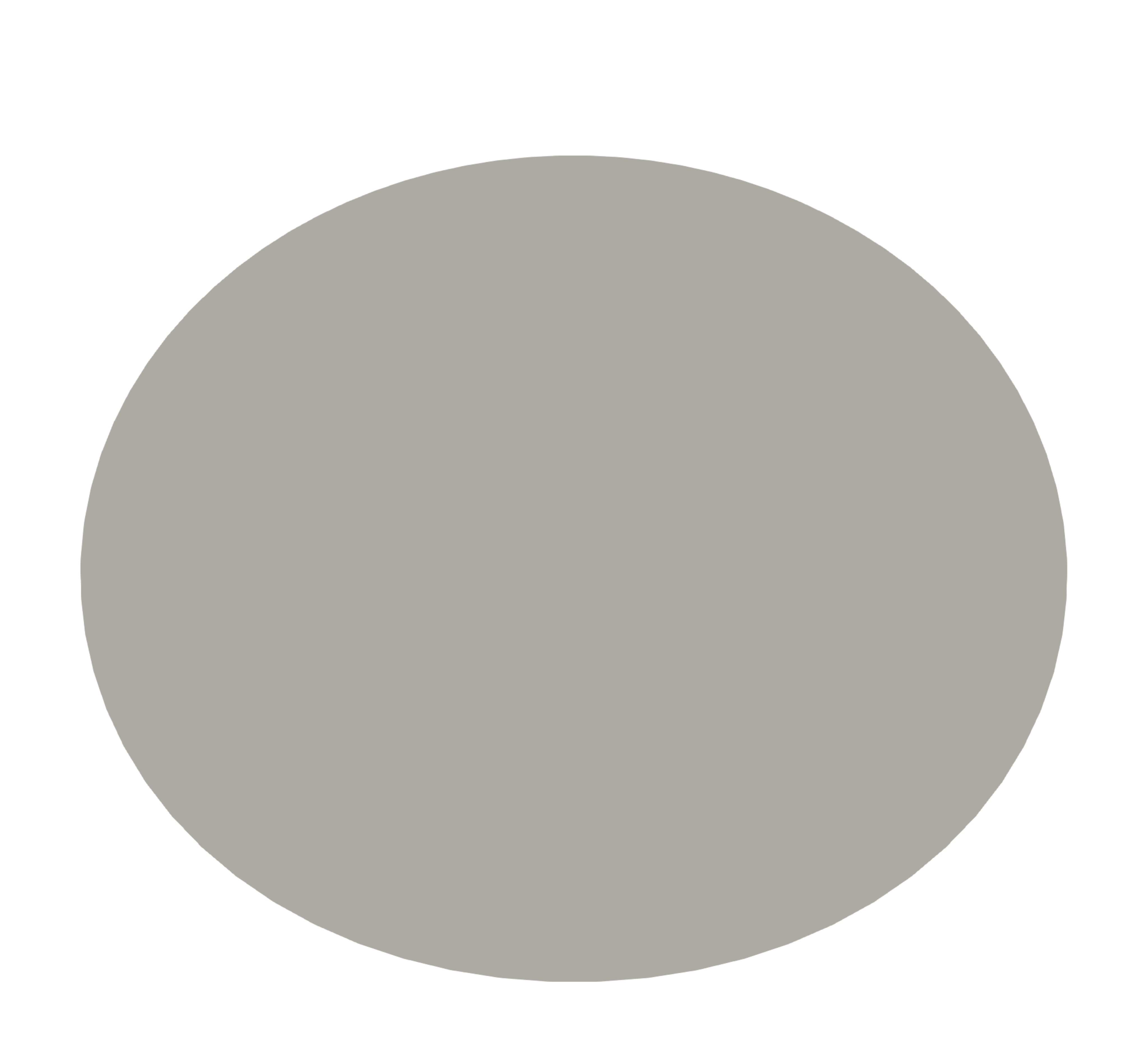}
            \end{subfigure}
            \begin{subfigure}[t]{0.39\textwidth}
                \centering
                \tikzsetnextfilename{fig_uz_colorbar}
\begin{tikzpicture}
\pgfplotscolorbardrawstandalone[ 
    point meta min=-0.27, 
    point meta max=0.27,
    colormap={cool-to-warm}{
        rgb(0)=(0.229999503869523,0.298998934049376,0.754000138575591)
        rgb(0.0158730158730159)=(0.248387005612446,0.326713084930128,0.778759751168141)
        rgb(0.0317460317460317)=(0.267060596781042,0.354213683475558,0.80247084431423)
        rgb(0.0476190476190476)=(0.286046233433907,0.381473413618758,0.825071993030171)
        rgb(0.0634920634920635)=(0.305359940427952,0.408457238050427,0.84650499453829)
        rgb(0.0793650793650794)=(0.325008919322423,0.435124782260566,0.866715008233051)
        rgb(0.0952380952380952)=(0.344992449467791,0.461431960153803,0.885650686307679)
        rgb(0.111111111111111)=(0.365302660727705,0.487332122586216,0.903264294298265)
        rgb(0.126984126984127)=(0.385925222694485,0.512776896138637,0.91951182090386)
        rgb(0.142857142857143)=(0.406839975594796,0.537716815138862,0.934353076530895)
        rgb(0.158730158730159)=(0.428021516636282,0.562101812270725,0.947751780091349)
        rgb(0.174603174603175)=(0.449439749029995,0.585881610291412,0.959675633658507)
        rgb(0.19047619047619)=(0.471060397373382,0.609006043154061,0.970096384653213)
        rgb(0.206349206349206)=(0.492845491291784,0.631425325735442,0.978989875298378)
        rgb(0.222222222222222)=(0.514753818467056,0.65309028541659,0.986336079140528)
        rgb(0.238095238095238)=(0.536741347968443,0.67395256479008,0.992119124494973)
        rgb(0.253968253968254)=(0.558761624861434,0.693964802064848,0.996327304725749)
        rgb(0.26984126984127)=(0.580766137238736,0.713080793870388,0.998953075322919)
        rgb(0.285714285714286)=(0.602704657000925,0.731255643849636,0.999993037787936)
        rgb(0.301587301587302)=(0.624525555865855,0.748445899495025,0.999447910382271)
        rgb(0.317460317460317)=(0.64617609818675,0.764609679007339,0.997322485835062)
        rgb(0.333333333333333)=(0.667602712206112,0.77970678946352,0.993625576141459)
        rgb(0.349206349206349)=(0.68875124137176,0.79369883721399,0.988369944613969)
        rgb(0.365079365079365)=(0.709567177301707,0.806549331155555,0.981572225373467)
        rgb(0.380952380952381)=(0.729995875916342,0.818223779316756,0.973252830483399)
        rgb(0.396825396825397)=(0.749982758168947,0.828689779030202,0.963435844938547)
        rgb(0.412698412698413)=(0.769473496707127,0.837917100837779,0.952148909716474)
        rgb(0.428571428571429)=(0.788414189694655,0.845877766169784,0.939423093082902)
        rgb(0.444444444444444)=(0.806751522920461,0.85254611875045,0.925292750308311)
        rgb(0.46031746031746)=(0.824432921222376,0.857898889604224,0.909795371897492)
        rgb(0.476190476190476)=(0.841406690160182,0.861915255464474,0.892971420350679)
        rgb(0.492063492063492)=(0.857622148786607,0.864576890314796,0.874864155355995)
        rgb(0.507936507936508)=(0.874239603676395,0.861703830626339,0.854248907429317)
        rgb(0.523809523809524)=(0.890678735446095,0.853320706019111,0.831640336391034)
        rgb(0.53968253968254)=(0.905552673939185,0.843617842774967,0.808523278638521)
        rgb(0.555555555555556)=(0.918869468148195,0.832615924366692,0.78495810425444)
        rgb(0.571428571428571)=(0.930635712635114,0.820337799431383,0.761004577523922)
        rgb(0.587301587301587)=(0.940857197090139,0.806808340933546,0.736721705094366)
        rgb(0.603174603174603)=(0.949539432266613,0.79205428793315,0.712167591757623)
        rgb(0.619047619047619)=(0.956688079767082,0.776104066879158,0.687399304196029)
        rgb(0.634920634920635)=(0.962309306085296,0.758987588429841,0.662472743015538)
        rgb(0.650793650793651)=(0.966410076233795,0.740736014556394,0.637442523375873)
        rgb(0.666666666666667)=(0.968998398592708,0.721381488984382,0.612361864520328)
        rgb(0.682539682539683)=(0.970083529892486,0.700956821674111,0.587282488508154)
        rgb(0.698412698412698)=(0.969676147212604,0.679495114746095,0.562254528462294)
        rgb(0.714285714285714)=(0.967788492347632,0.657029312579374,0.537326446666784)
        rgb(0.73015873015873)=(0.96443449272698,0.633591652063982,0.512544962884353)
        rgb(0.746031746031746)=(0.959629862179917,0.609212979090077,0.487954993319154)
        rgb(0.761904761904762)=(0.95339218414484,0.583921882550407,0.463599600726846)
        rgb(0.777777777777778)=(0.945740979381812,0.557743574505344,0.439519956280044)
        rgb(0.793650793650794)=(0.936697759823885,0.530698409733458,0.415755313939185)
        rgb(0.80952380952381)=(0.926286069868824,0.50279988089061,0.392342998266502)
        rgb(0.825396825396825)=(0.914531516148685,0.474051830893187,0.369318406866782)
        rgb(0.841269841269841)=(0.901461786605133,0.444444461402548,0.346715028958866)
        rgb(0.857142857142857)=(0.887106659532162,0.413948424491904,0.32456448199742)
        rgb(0.873015873015873)=(0.871498003116352,0.382505735188946,0.302896568802084)
        rgb(0.888888888888889)=(0.854669765901501,0.350015145998675,0.281739358344963)
        rgb(0.904761904761905)=(0.836657958524603,0.316307274081113,0.261119294239799)
        rgb(0.920634920634921)=(0.817500627010409,0.281099264139742,0.241061336118172)
        rgb(0.936507936507937)=(0.797237817870139,0.243904285414508,0.221589140526617)
        rgb(0.952380952380952)=(0.775911535225725,0.203826593912042,0.202725289788009)
        rgb(0.968253968253968)=(0.753565690175334,0.159000204558525,0.184491579471458)
        rgb(0.984126984126984)=(0.730246042632013,0.10443854422286,0.166909377664698)
        rgb(1)=(0.706000135911705,0.0159918240339807,0.1500000719222)
        },
    colorbar style={
        ylabel={$u_3$},
        ytick={-0.27,-0.2,-0.1,0,0.1,0.2,0.27}, 
        yticklabels={-0.027,,,0,,,0.027},
        height={25mm},
        width={2mm}
    }]
\end{tikzpicture}
                \caption*{}
            \end{subfigure}
            \vspace{0.1\linewidth}
        \end{subfigure}
        \caption{Bending}
        \label{fig:deformable_vs_rigid_bending}
    \end{subfigure}

    \vspace{3mm}
    
    \begin{subfigure}[b]{\linewidth}
        \centering
        \tikzsetnextfilename{fig_rigid_vs_warping_legend}
\begin{tikzpicture}
\begin{axis}[
  hide axis,
  xmin=0.0,
  xmax=0.001,
  ymin=0.0,
  ymax=0.001,
  legend columns=3,
legend style={/tikz/every even column/.append style={column sep=0.25cm,
row sep=0.1cm}}
]
    \addlegendimage{dashed, very thick, black}
    \addlegendentry{LEM}
    \addlegendimage{dotted, very thick, black}
    \addlegendentry{RHM}
    \addlegendimage{solid, very thick, black}
    \addlegendentry{DHM}
    
\end{axis}
\end{tikzpicture}
    \end{subfigure}
    \caption{Stress resultants and deformed cross-sections obtained from solving the hyperelastic model with deformable cross-section (DHM). Solid lines denote the DHM. Dotted lines denote the hyperelastic model with rigid cross-section (RHM). Dashed lines denote the linear elastic model (LEM). Note that no out-of-plane deformation is present under bending. All results are computed for $R=1.0$}
    \label{fig:deformable_vs_rigid}
\end{figure*}

Following the sampling of strain measures, training and testing datasets were generated using the DHM and the warping problem presented in~\cref{subsubsec:dhm} and~\cref{subsec:warping-problem}. For this purpose, they were implemented using the open-source software FEniCS~\cite{alnaesFEniCSProjectVersion2015}, where the circular cross-section was discretized using 2089 linear triangle elements. The number of elements was obtained from a convergence study. \revi{In the FEniCS implementation of the warping problem, the rotational degrees of freedom were constrained following the practice presented in~\cite{aroraComputationalApproachObtain2019}, i.e., all four constraints from~\cref{eq:constraint_rotation_m} were used, which yielded reliable convergence.} On the cross-section scale, a Neo-Hookean hyperelastic material model with the following strain energy function was employed:
\begin{equation}\label{eq:hyperelastic_potential}
    \bar\pot = \frac{\bar\mu}{2} \left(I_1 - \log\left(J^2\right) - 3\right) + \frac{\bar\lambda}{8} \log\left(J^2\right)^2
\end{equation}
with
\begin{equation}
    \bar\mu = \frac{E}{2(1+\nu)}, \qquad \bar\lambda = \frac{2 \bar\mu \nu}{1 - 2 \nu},
\end{equation}
\revii{where $\bar\lambda$ refers to a Lamé parameter and is not to be confused with the scaling parameter presented in Eq. (\ref{eq:kumar-scaling-potential}). Additionally,} $I_1=\text{tr}(\tF^T\tF)$ and $J=\det(\tF)$ denote the first invariant \revii{of the right Cauchy-Green tensor $\tF^T\tF$} and the Jacobi determinant of the deformation gradient, respectively. The material parameters $E = \qty{70}{\MPa}$ and $\nu = 0.4$ correspond to a thermoplastic polyurethane used for 3D printing with fused deposition modeling. The units will not be considered in the remainder of this work.

\medskip

To understand the effect of cross-sectional deformation on the constitutive behavior, the stress resultants for the DHM, RHM, and LEM are plotted for pure shear and pure bending in~\cref{fig:deformable_vs_rigid}. These two simple deformations were chosen because they already show the most relevant effects present in more complicated deformation states. 

The comparison demonstrates that the LEM approximates the DHM well in the small strain regime with amplitude $t < 0.2$. In this range, the LEM accurately captures the dominating transversal force $n_1$ and the twisting moments $m_1$ for shear and bending, respectively. It shows that the LEM implicitly accounts for some cross-sectional deformation through its constitutive parameters. In contrast, the RHM drastically overestimates the dominating stress resultants $n_1$ and $m_1$ in both cases.

Looking at larger strains with $t > 0.2$, the differences in the stress resultants between the RHM and the DHM keep increasing. Furthermore, significant differences between the DHM and the LEM become observable. While the dominating stress resultants $n_1$ and $m_1$ are still captured decently by the LEM, it does not capture any of the distinct couplings with the normal force $n_3$. Simultaneously, the RHM only models the coupling for the pure bending case but not for pure shear.

The implications of allowing for cross-sectional deformation also become apparent in the right column of~\cref{fig:deformable_vs_rigid}, where the deformed configurations computed by the warping problem are shown for the last load steps of the two load paths. Under pure shear strain, the cross-section exhibits out-of-plane warping and almost no in-plane deformation. In contrast, bending leads to large in-plane but no out-of-plane deformation.

Although the deformations in~\cref{fig:deformable_vs_rigid} are specific to homogeneous circular cross-sections, it is evident that accounting for cross-sectional deformation is necessary when applying hyperelastic material models and large strains in the framework of geometrically exact beams. It leads to a more accurate representation of the beam stiffness and enables the modeling of coupling effects that the LEM neglects entirely. Nonetheless, coupling and nonlinearities are only relevant for large strains in the studied examples.

\revii{
\paragraph{Data generation for ring-shaped cross-sections:}
In addition to fully circular cross-sections, the DHM with the above implementation was used to generate data for ring-shaped cross-sections with different ratios $\parameter = R_i / R$ between the inner and outer beam radii $R_i$ and $R$, respectively. Thereby, the same material model, i.e.,~\cref{eq:hyperelastic_potential}, was employed on the cross-sectional scale and the outer radius was fixed to $R=1.0$.
~\cref{fig:ring-data} shows the obtained stress resultants for an exemplary mixed strain path at $\parameter=0.0$ and $\parameter=0.5$. As expected, an increase of the inner radius results in stronger nonlinearity~\cite{leclezioNumericalTwoscaleApproach2023} and qualitative changes in the stress response. This motivates using a parameterization for modeling beams with different ratios $\parameter$, which will be presented in~\cref{subsec:ring-parameterization}.
}

\begin{figure}[t]
    \centering
        \begin{subfigure}[t]{0.4\textwidth}
        \centering
        \resizebox{\linewidth}{!}{
                \tikzsetnextfilename{fig_ring_00_19}
\begin{tikzpicture}
\begin{axis}[
  name=plot1,
  xlabel={$t$},
  ylabel={$q_i$},
  transpose legend,
  legend style={at={(0.03,0.03)}, anchor=south west},
  legend columns=3,
  grid=major,
]

    \addplot [color=CPSred, solid, very thick, mark=*, mark options={solid}, mark repeat=10] table [col sep=comma, x=t, y=n_1]{fig_ring_data_00_19.txt};
    \addlegendentry{$n_1$}
    \addplot [color=CPSred, solid, very thick, mark=triangle*, mark options={solid}, mark repeat=10] table [col sep=comma, x=t, y=n_2]{fig_ring_data_00_19.txt};
    \addlegendentry{$n_2$}
    \addplot [color=CPSdarkblue, solid, very thick, mark=square*, mark options={solid}, mark repeat=10] table [col sep=comma, x=t, y=n_3]{fig_ring_data_00_19.txt};
    \addlegendentry{$n_3$}
    \addplot [color=CPSgreen, solid, very thick, mark=*, mark options={solid}, mark repeat=10] table [col sep=comma, x=t, y=m_1]{fig_ring_data_00_19.txt};
    \addlegendentry{$m_1$}
    \addplot [color=CPSgreen, solid, very thick, mark=triangle*, mark options={solid}, mark repeat=10] table [col sep=comma, x=t, y=m_2]{fig_ring_data_00_19.txt};
    \addlegendentry{$m_2$}
    \addplot [color=CPSorange, solid, very thick, mark=diamond*, mark options={solid}, mark repeat=10] table [col sep=comma, x=t, y=m_3]{fig_ring_data_00_19.txt};
    \addlegendentry{$m_3$}

\end{axis}
\end{tikzpicture}
        }
       \caption{$\parameter = 0.0$}
        \label{fig:ring_00}
    \end{subfigure}
    \hspace{0.03\linewidth}
    \begin{subfigure}[t]{0.4\textwidth}
        \centering
        \resizebox{\linewidth}{!}{
                \tikzsetnextfilename{fig_ring_05_19}
\begin{tikzpicture}
\begin{axis}[
  name=plot1,
  xlabel={$t$},
  ylabel={$q_i$},
  transpose legend,
  legend style={at={(0.03,0.03)}, anchor=south west},
  legend columns=3,
  grid=major,
]

    \addplot [color=CPSred, solid, very thick, mark=*, mark options={solid}, mark repeat=10] table [col sep=comma, x=t, y=n_1]{fig_ring_data_05_19.txt};
    \addlegendentry{$n_1$}
    \addplot [color=CPSred, solid, very thick, mark=triangle*, mark options={solid}, mark repeat=10] table [col sep=comma, x=t, y=n_2]{fig_ring_data_05_19.txt};
    \addlegendentry{$n_2$}
    \addplot [color=CPSdarkblue, solid, very thick, mark=square*, mark options={solid}, mark repeat=10] table [col sep=comma, x=t, y=n_3]{fig_ring_data_05_19.txt};
    \addlegendentry{$n_3$}
    \addplot [color=CPSgreen, solid, very thick, mark=*, mark options={solid}, mark repeat=10] table [col sep=comma, x=t, y=m_1]{fig_ring_data_05_19.txt};
    \addlegendentry{$m_1$}
    \addplot [color=CPSgreen, solid, very thick, mark=triangle*, mark options={solid}, mark repeat=10] table [col sep=comma, x=t, y=m_2]{fig_ring_data_05_19.txt};
    \addlegendentry{$m_2$}
    \addplot [color=CPSorange, solid, very thick, mark=diamond*, mark options={solid}, mark repeat=10] table [col sep=comma, x=t, y=m_3]{fig_ring_data_05_19.txt};
    \addlegendentry{$m_3$}

\end{axis}
\end{tikzpicture}
        }        
        \caption{$\parameter = 0.5$}
        \label{fig:ring_05}
    \end{subfigure}
    \caption{\revii{Stress resultants $q_i$ plotted over the strain amplitude $t$ for an exemplary mixed strain path. The data is obtained from the DHM constitutive model for ring-shaped cross-sections with different ratios $\parameter = R_i/R$ between the inner radius $R_i$ and the outer radius $R$. All results were computed for $R=1.0$ using a Neo-Hookean hyperelastic base material model.}}
    \label{fig:ring-data}
\end{figure}


\subsection{PANN model calibration}
\label{subsec:model-calibration}

Next, the PANN models are fitted to the datasets $\cD$\revii{, $\cD_{\parameter}$, or $\cD_{\lambda}$} obtained with the DHM and the data generation procedure presented in~\cref{subsec:data-generation}. In the context of NNs, this process is known as training or calibration, where a loss function $\cL$ is minimized w.r.t.\ the model parameters.

In the supervised learning scenario relevant to this work, the loss represents the error between the predicted and the actual stress resultants for a given set of strain measures. However, the absolute error can be very unevenly distributed between the different stress resultants depending on \revii{ the parameter $\parameter$ and the scaling factor $\lambda$, which in this work correspond to the outer beam radius, i.e., $\lambda = R$. For example, we} know that $n_i \sim R^2$ and $m_i \sim R^4$ hold for the LEM. Consequently, $n_i$ and $m_i$ can have different orders of magnitude depending on the beam radius. If the radius is large, as in~\cref{fig:deformable_vs_rigid}, forces and moment have the same magnitude. Meanwhile, if the radius is reduced to $R/10$, the forces and moments are reduced by $10^{-2}$ and $10^{-4}$, respectively, for a given strain.

While the radius-dependency may differ slightly for the DHM, different orders of magnitude pose a significant challenge to the training process. To enable equally accurate predictions for all stress resultants, the loss function is formulated as
\begin{equation}
    \cL = \frac{1}{6 n} \sum_{i=1}^{6} \sum_{j=1}^{n} \prescript{ij}{}{w} \sum_{k=1}^{\prescript{j}{}{m}} \norm{\prescript{jk}{}{q_i} - q_i(\prescript{jk}{}{\vMatStrain)}}^2.
\end{equation}
Here, $n$ denotes the number of unique \revii{tuples $\prescript{j}{}{(R, \parameter)}$} in the dataset, and $\prescript{j}{}{m}$ is the number of data points \revii{belonging to a tuple $\prescript{j}{}{(R, \parameter)}$}. The weights $\prescript{ij}{}{w}$ are defined as
\begin{equation}
    \prescript{ij}{}{w} = \frac{1}{\prescript{j}{}{m}} \sum_{k=1}^{\prescript{j}{}{m}} \norm{\prescript{jk}{}{q_i}}^2,
\end{equation}
resulting in one weight per combination of \revii{$\prescript{j}{}{(R, \parameter)}$} and stress resultant $q_i$. As the loss term is formulated in terms of \revi{the first derivative} of the FFNN outputs w.r.t.\ its inputs, the training strategy used in this work represents a form of Sobolev training~\cite{czarneckiSobolevTrainingNeural2017}. Note that \revi{neither the} the value of the strain energy $\psi$, which is the immediate model output\revi{, nor the stiffnesses/elasticity tensors as second derivatives were} considered during training in this work, as this was not deemed necessary or beneficial in previous studies\revi{~\cite{kleinPolyconvexAnisotropicHyperelasticity2022,lindenNeuralNetworksMeet2023,frankeAdvancedDiscretizationTechniques2023}}.

The PANN models presented in this work were implemented in Python 3.11 with Tensorflow 2.15 and calibrated using the ADAM optimizer, with a mini-batch size of 32 and a learning rate of 0.002. All calibrations were performed on an Intel i7-13700K CPU or comparable desktop PC hardware.


\subsection{PANN model evaluation}

\revii{A total of four} studies were conducted to assess the performance of the PANN models. \revi{All data was generated using the implementation of the DHM presented in~\cref{subsec:model-evaluation}. Network architecture and data studies for the different neural network-based models are provided in~\cref{sec:node-study} and~\cref{sec:data study}, respectively.}

\revii{
\subsubsection{Study I -- Symmetry analysis}
\label{subsubsec:study-i}
}

\revii{To test the implementation and evaluate the practical applicability of the transverse isotropy and point symmetry constraints introduced in~\cref{subsec:sym}, the non-parameterized and non-scaled architectures $\model$, $\modelsym$, and $\modeliso$ for $(R = 1.0, \parameter = 0.0)$ were calibrated to a mixed load path with $\vMatStrain = t / 3.5 \cdot (0.5, 1, 0.5, 1.5, 2.5, 1.5)^T$, and $t \in [0, 0.59]$ and afterward evaluated on the same load path for $t \in [-0.59, 0]$. The calibration and test datasets consist of 186 data points each. The FFNN has one hidden layer with 32 neurons calibrated for 10000 epochs.

\Cref{fig:symmetry} shows the results of Study I. Both the non-symmetric model $\model$ and the point symmetric model $\modelsym$ perfectly fit the calibration data for $t \geq 0$ in~\cref{fig:symmetry-none,fig:symmetry-point-symmetric}, respectively. This statement is supported by~\cref{fig:symmetry-loss}, which shows that the calibration loss for both models lies in the order of $10^{-6}$. When extrapolating to the symmetric counterpart for $t < 0$, the point symmetric model $\modelsym$ generalizes perfectly and yields exactly symmetric results. Consequently, the test loss is identical to the calibration loss. In contrast, the non-symmetric model $\model$ completely fails to generalize to the unseen data.

Shifting the focus to~\cref{fig:symmetry-isotropic}, the transversely isotropic model shows some visible deviations even for the calibration data. In particular, the model prediction of the moments deviates for increasing strain amplitudes. Accordingly, the test and calibration loss of $\modeliso$ is approximately two orders of magnitude larger than $\modelsym$ despite the fully converged model calibration, cf.~\cref{fig:symmetry-loss}. These findings suggest that despite being transversely isotropic for small strains, the symmetry group for circular cross-sections changes for finite deformations, similar to the change of symmetry in anisotropic 3D continuum hyperelasticity~\cite{kalina2024neuralnetworksmeetanisotropic}. 
%
As we are particularly interested in large strains and finite cross-sectional deformations in this work, the point symmetric model $\modelsym$ will be the basis for all subsequent studies.
}

\begin{figure}[t]
    \centering
    \begin{subfigure}[t]{0.45\textwidth}
        \centering
        \resizebox{0.89\linewidth}{!}{
            \tikzsetnextfilename{fig_sym_PANN}
\begin{tikzpicture}
\begin{axis}[
  name=plot1,
  xlabel={$t$},
  ylabel={$q_i$},
  transpose legend,
  legend style={at={(0.03,0.97)}, anchor=north west},
  legend columns=3,
  grid=major,
ymin=-14,ymax=14
]
    
    \addplot [color=CPSred, dashed, very thick, mark=*, mark options={solid}, mark repeat=50] table [col sep=comma, x=t, y=n_1]{fig_sym_data_PANN_1.txt};
    \addlegendentry{$n_1$}
    \addplot [color=CPSred, dashed, very thick, mark=triangle*, mark options={solid}, mark repeat=50] table [col sep=comma, x=t, y=n_2]{fig_sym_data_PANN_1.txt};
    \addlegendentry{$n_2$}
    \addplot [color=CPSdarkblue, dashed, very thick, mark=square*, mark options={solid}, mark repeat=50] table [col sep=comma, x=t, y=n_3]{fig_sym_data_PANN_1.txt};
    \addlegendentry{$n_3$}
    \addplot [color=CPSgreen, dashed, very thick, mark=*, mark options={solid}, mark repeat=50] table [col sep=comma, x=t, y=m_1]{fig_sym_data_PANN_1.txt};
    \addlegendentry{$m_1$}
    \addplot [color=CPSgreen, dashed, very thick, mark=triangle*, mark options={solid}, mark repeat=50] table [col sep=comma, x=t, y=m_2]{fig_sym_data_PANN_1.txt};
    \addlegendentry{$m_2$}
    \addplot [color=CPSorange, dashed, very thick, mark=diamond*, mark options={solid}, mark repeat=50] table [col sep=comma, x=t, y=m_3]{fig_sym_data_PANN_1.txt};
    \addlegendentry{$m_3$}

    \addplot [color=CPSred, solid, very thick, no markers] table [col sep=comma, x=t, y=n_1_pred]{fig_sym_data_PANN_1.txt};
    \addplot [color=CPSred, solid, very thick, no markers] table [col sep=comma, x=t, y=n_2_pred]{fig_sym_data_PANN_1.txt};
    \addplot [color=CPSdarkblue, solid, very thick, no markers] table [col sep=comma, x=t, y=n_3_pred]{fig_sym_data_PANN_1.txt};
    \addplot [color=CPSgreen, solid, very thick, no markers] table [col sep=comma, x=t, y=m_1_pred]{fig_sym_data_PANN_1.txt};
    \addplot [color=CPSgreen, solid, very thick, no markers] table [col sep=comma, x=t, y=m_2_pred]{fig_sym_data_PANN_1.txt};
    \addplot [color=CPSorange, solid, very thick, no markers] table [col sep=comma, x=t, y=m_3_pred]{fig_sym_data_PANN_1.txt};

    \addplot [color=CPSred, dashed, very thick, mark=*, mark options={solid}, mark repeat=50] table [col sep=comma, x=t, y=n_1]{fig_sym_data_PANN_2.txt};
    \addplot [color=CPSred, dashed, very thick, mark=triangle*, mark options={solid}, mark repeat=50] table [col sep=comma, x=t, y=n_2]{fig_sym_data_PANN_2.txt};
    \addplot [color=CPSdarkblue, dashed, very thick, mark=square*, mark options={solid}, mark repeat=50] table [col sep=comma, x=t, y=n_3]{fig_sym_data_PANN_2.txt};
    \addplot [color=CPSgreen, dashed, very thick, mark=*, mark options={solid}, mark repeat=50] table [col sep=comma, x=t, y=m_1]{fig_sym_data_PANN_2.txt};
    \addplot [color=CPSgreen, dashed, very thick, mark=triangle*, mark options={solid}, mark repeat=50] table [col sep=comma, x=t, y=m_2]{fig_sym_data_PANN_2.txt};
    \addplot [color=CPSorange, dashed, very thick, mark=diamond*, mark options={solid}, mark repeat=50] table [col sep=comma, x=t, y=m_3]{fig_sym_data_PANN_2.txt};

    \addplot [color=CPSred, solid, very thick, no markers] table [col sep=comma, x=t, y=n_1_pred]{fig_sym_data_PANN_2.txt};
    \addplot [color=CPSred, solid, very thick, no markers] table [col sep=comma, x=t, y=n_2_pred]{fig_sym_data_PANN_2.txt};
    \addplot [color=CPSdarkblue, solid, very thick, no markers] table [col sep=comma, x=t, y=n_3_pred]{fig_sym_data_PANN_2.txt};
    \addplot [color=CPSgreen, solid, very thick, no markers] table [col sep=comma, x=t, y=m_1_pred]{fig_sym_data_PANN_2.txt};
    \addplot [color=CPSgreen, solid, very thick, no markers] table [col sep=comma, x=t, y=m_2_pred]{fig_sym_data_PANN_2.txt};
    \addplot [color=CPSorange, solid, very thick, no markers] table [col sep=comma, x=t, y=m_3_pred]{fig_sym_data_PANN_2.txt};

    \draw[color=black, ultra thick,loosely dashed] (0,-14) -- (0,14);

\end{axis}
\end{tikzpicture}
        }
       \caption{Non-symmetric model $\model$}
        \label{fig:symmetry-none}
    \end{subfigure}
    \hspace{0.03\linewidth}
    \begin{subfigure}[t]{0.45\textwidth}
        \centering
        \resizebox{0.89\linewidth}{!}{
            \tikzsetnextfilename{fig_sym_PANN_sym}
\begin{tikzpicture}
\begin{axis}[
  name=plot1,
  xlabel={$t$},
  ylabel={$q_i$},
  transpose legend,
  legend style={at={(0.03,0.97)}, anchor=north west},
  legend columns=3,
  grid=major,
ymin=-14,ymax=14
]
    
    \addplot [color=CPSred, dashed, very thick, mark=*, mark options={solid}, mark repeat=50] table [col sep=comma, x=t, y=n_1]{fig_sym_data_PANN_sym_1.txt};
    \addlegendentry{$n_1$}
    \addplot [color=CPSred, dashed, very thick, mark=triangle*, mark options={solid}, mark repeat=50] table [col sep=comma, x=t, y=n_2]{fig_sym_data_PANN_sym_1.txt};
    \addlegendentry{$n_2$}
    \addplot [color=CPSdarkblue, dashed, very thick, mark=square*, mark options={solid}, mark repeat=50] table [col sep=comma, x=t, y=n_3]{fig_sym_data_PANN_sym_1.txt};
    \addlegendentry{$n_3$}
    \addplot [color=CPSgreen, dashed, very thick, mark=*, mark options={solid}, mark repeat=50] table [col sep=comma, x=t, y=m_1]{fig_sym_data_PANN_sym_1.txt};
    \addlegendentry{$m_1$}
    \addplot [color=CPSgreen, dashed, very thick, mark=triangle*, mark options={solid}, mark repeat=50] table [col sep=comma, x=t, y=m_2]{fig_sym_data_PANN_sym_1.txt};
    \addlegendentry{$m_2$}
    \addplot [color=CPSorange, dashed, very thick, mark=diamond*, mark options={solid}, mark repeat=50] table [col sep=comma, x=t, y=m_3]{fig_sym_data_PANN_sym_1.txt};
    \addlegendentry{$m_3$}

    \addplot [color=CPSred, solid, very thick, no markers] table [col sep=comma, x=t, y=n_1_pred]{fig_sym_data_PANN_sym_1.txt};
    \addplot [color=CPSred, solid, very thick, no markers] table [col sep=comma, x=t, y=n_2_pred]{fig_sym_data_PANN_sym_1.txt};
    \addplot [color=CPSdarkblue, solid, very thick, no markers] table [col sep=comma, x=t, y=n_3_pred]{fig_sym_data_PANN_sym_1.txt};
    \addplot [color=CPSgreen, solid, very thick, no markers] table [col sep=comma, x=t, y=m_1_pred]{fig_sym_data_PANN_sym_1.txt};
    \addplot [color=CPSgreen, solid, very thick, no markers] table [col sep=comma, x=t, y=m_2_pred]{fig_sym_data_PANN_sym_1.txt};
    \addplot [color=CPSorange, solid, very thick, no markers] table [col sep=comma, x=t, y=m_3_pred]{fig_sym_data_PANN_sym_1.txt};

    \addplot [color=CPSred, dashed, very thick, mark=*, mark options={solid}, mark repeat=50] table [col sep=comma, x=t, y=n_1]{fig_sym_data_PANN_sym_2.txt};
    \addplot [color=CPSred, dashed, very thick, mark=triangle*, mark options={solid}, mark repeat=50] table [col sep=comma, x=t, y=n_2]{fig_sym_data_PANN_sym_2.txt};
    \addplot [color=CPSdarkblue, dashed, very thick, mark=square*, mark options={solid}, mark repeat=50] table [col sep=comma, x=t, y=n_3]{fig_sym_data_PANN_sym_2.txt};
    \addplot [color=CPSgreen, dashed, very thick, mark=*, mark options={solid}, mark repeat=50] table [col sep=comma, x=t, y=m_1]{fig_sym_data_PANN_sym_2.txt};
    \addplot [color=CPSgreen, dashed, very thick, mark=triangle*, mark options={solid}, mark repeat=50] table [col sep=comma, x=t, y=m_2]{fig_sym_data_PANN_sym_2.txt};
    \addplot [color=CPSorange, dashed, very thick, mark=diamond*, mark options={solid}, mark repeat=50] table [col sep=comma, x=t, y=m_3]{fig_sym_data_PANN_sym_2.txt};

    \addplot [color=CPSred, solid, very thick, no markers] table [col sep=comma, x=t, y=n_1_pred]{fig_sym_data_PANN_sym_2.txt};
    \addplot [color=CPSred, solid, very thick, no markers] table [col sep=comma, x=t, y=n_2_pred]{fig_sym_data_PANN_sym_2.txt};
    \addplot [color=CPSdarkblue, solid, very thick, no markers] table [col sep=comma, x=t, y=n_3_pred]{fig_sym_data_PANN_sym_2.txt};
    \addplot [color=CPSgreen, solid, very thick, no markers] table [col sep=comma, x=t, y=m_1_pred]{fig_sym_data_PANN_sym_2.txt};
    \addplot [color=CPSgreen, solid, very thick, no markers] table [col sep=comma, x=t, y=m_2_pred]{fig_sym_data_PANN_sym_2.txt};
    \addplot [color=CPSorange, solid, very thick, no markers] table [col sep=comma, x=t, y=m_3_pred]{fig_sym_data_PANN_sym_2.txt};

    \draw[color=black, ultra thick,loosely dashed] (0,-14) -- (0,14);

\end{axis}
\end{tikzpicture}
        }        
        \caption{Point symmetric model $\modelsym$}
        \label{fig:symmetry-point-symmetric}
    \end{subfigure}

    \vspace{0.01\linewidth}

    \begin{subfigure}[t]{0.45\textwidth}
        \centering
        \resizebox{0.89\linewidth}{!}{
                \tikzsetnextfilename{fig_sym_PANN_iso}
\begin{tikzpicture}
\begin{axis}[
  name=plot1,
  xlabel={$t$},
  ylabel={$q_i$},
  transpose legend,
  legend style={at={(0.03,0.97)}, anchor=north west},
  legend columns=3,
  grid=major,
ymin=-14,ymax=14
]
    
    \addplot [color=CPSred, dashed, very thick, mark=*, mark options={solid}, mark repeat=50] table [col sep=comma, x=t, y=n_1]{fig_sym_data_PANN_iso_1.txt};
    \addlegendentry{$n_1$}
    \addplot [color=CPSred, dashed, very thick, mark=triangle*, mark options={solid}, mark repeat=50] table [col sep=comma, x=t, y=n_2]{fig_sym_data_PANN_iso_1.txt};
    \addlegendentry{$n_2$}
    \addplot [color=CPSdarkblue, dashed, very thick, mark=square*, mark options={solid}, mark repeat=50] table [col sep=comma, x=t, y=n_3]{fig_sym_data_PANN_iso_1.txt};
    \addlegendentry{$n_3$}
    \addplot [color=CPSgreen, dashed, very thick, mark=*, mark options={solid}, mark repeat=50] table [col sep=comma, x=t, y=m_1]{fig_sym_data_PANN_iso_1.txt};
    \addlegendentry{$m_1$}
    \addplot [color=CPSgreen, dashed, very thick, mark=triangle*, mark options={solid}, mark repeat=50] table [col sep=comma, x=t, y=m_2]{fig_sym_data_PANN_iso_1.txt};
    \addlegendentry{$m_2$}
    \addplot [color=CPSorange, dashed, very thick, mark=diamond*, mark options={solid}, mark repeat=50] table [col sep=comma, x=t, y=m_3]{fig_sym_data_PANN_iso_1.txt};
    \addlegendentry{$m_3$}

    \addplot [color=CPSred, solid, very thick, no markers] table [col sep=comma, x=t, y=n_1_pred]{fig_sym_data_PANN_iso_1.txt};
    \addplot [color=CPSred, solid, very thick, no markers] table [col sep=comma, x=t, y=n_2_pred]{fig_sym_data_PANN_iso_1.txt};
    \addplot [color=CPSdarkblue, solid, very thick, no markers] table [col sep=comma, x=t, y=n_3_pred]{fig_sym_data_PANN_iso_1.txt};
    \addplot [color=CPSgreen, solid, very thick, no markers] table [col sep=comma, x=t, y=m_1_pred]{fig_sym_data_PANN_iso_1.txt};
    \addplot [color=CPSgreen, solid, very thick, no markers] table [col sep=comma, x=t, y=m_2_pred]{fig_sym_data_PANN_iso_1.txt};
    \addplot [color=CPSorange, solid, very thick, no markers] table [col sep=comma, x=t, y=m_3_pred]{fig_sym_data_PANN_iso_1.txt};

    \addplot [color=CPSred, dashed, very thick, mark=*, mark options={solid}, mark repeat=50] table [col sep=comma, x=t, y=n_1]{fig_sym_data_PANN_iso_2.txt};
    \addplot [color=CPSred, dashed, very thick, mark=triangle*, mark options={solid}, mark repeat=50] table [col sep=comma, x=t, y=n_2]{fig_sym_data_PANN_iso_2.txt};
    \addplot [color=CPSdarkblue, dashed, very thick, mark=square*, mark options={solid}, mark repeat=50] table [col sep=comma, x=t, y=n_3]{fig_sym_data_PANN_iso_2.txt};
    \addplot [color=CPSgreen, dashed, very thick, mark=*, mark options={solid}, mark repeat=50] table [col sep=comma, x=t, y=m_1]{fig_sym_data_PANN_iso_2.txt};
    \addplot [color=CPSgreen, dashed, very thick, mark=triangle*, mark options={solid}, mark repeat=50] table [col sep=comma, x=t, y=m_2]{fig_sym_data_PANN_iso_2.txt};
    \addplot [color=CPSorange, dashed, very thick, mark=diamond*, mark options={solid}, mark repeat=50] table [col sep=comma, x=t, y=m_3]{fig_sym_data_PANN_iso_2.txt};

    \addplot [color=CPSred, solid, very thick, no markers] table [col sep=comma, x=t, y=n_1_pred]{fig_sym_data_PANN_iso_2.txt};
    \addplot [color=CPSred, solid, very thick, no markers] table [col sep=comma, x=t, y=n_2_pred]{fig_sym_data_PANN_iso_2.txt};
    \addplot [color=CPSdarkblue, solid, very thick, no markers] table [col sep=comma, x=t, y=n_3_pred]{fig_sym_data_PANN_iso_2.txt};
    \addplot [color=CPSgreen, solid, very thick, no markers] table [col sep=comma, x=t, y=m_1_pred]{fig_sym_data_PANN_iso_2.txt};
    \addplot [color=CPSgreen, solid, very thick, no markers] table [col sep=comma, x=t, y=m_2_pred]{fig_sym_data_PANN_iso_2.txt};
    \addplot [color=CPSorange, solid, very thick, no markers] table [col sep=comma, x=t, y=m_3_pred]{fig_sym_data_PANN_iso_2.txt};

    \draw[color=black, ultra thick,loosely dashed] (0,-14) -- (0,14);

\end{axis}
\end{tikzpicture}
        }
       \caption{Transversely isotropic model $\modeliso$}
        \label{fig:symmetry-isotropic}
    \end{subfigure}
    \hspace{0.03\linewidth}
    \begin{subfigure}[t]{0.45\textwidth}
        \centering
        \resizebox{0.89\linewidth}{!}{
                \tikzsetnextfilename{fig_sym_loss_bar}
\begin{tikzpicture}
\begin{axis}[
    ylabel={$\mathcal{L}$},
    ymode=log,
    symbolic x coords={a,b,c},
    xtick=data,
    xticklabels={$\model$, $\modelsym$, $\modeliso$},
    ybar=5pt,
    bar width=20pt,
    legend style={at={(0.5,0.03)},anchor=south, legend columns = -1},
    ymajorgrids=true,
    ymin=4E-9,
    grid style=dashed,
    enlarge x limits=0.2,
]

\addplot[fill=blue!30,error bars/.cd,y dir=both,y explicit]
    coordinates {
    (a,1.4770612395409444e-6) +- (7.393006846712069e-6,1.2544799901093029e-6)
    (b,5.643815072176039e-7) +- (1.105472934170848e-6,4.3583362554500125e-7)
    (c,0.000171019011759199) +- (9.368523024024477e-7,5.990441422907601e-7)
    };

\addplot[fill=red!30,error bars/.cd,y dir=both,y explicit]
    coordinates {
    (a,2.0592748403549193) +- (0.17272002696990985,0.10563533306121808)
    (b,5.643815072176039e-7) +- (1.105472934170848e-6,4.3583362554500125e-7)
    (c,0.000171019011759199) +- (9.368523024024477e-7,5.990441422907601e-7)
    };

\legend{Calibration, Testing}

\end{axis}
\end{tikzpicture}
        }        
        \caption{Loss of ten consecutive training runs and evaluations}
        \label{fig:symmetry-loss}
    \end{subfigure}
    \caption{\revii{Analysis of the symmetry constraints} by comparing (a) a non-symmetric with (b) a point symmetric\revii{, and (c) a transversely isotropic} PANN beam model for $R = 1.0$. The models (solid lines) are calibrated for $t\geq 0$. For $t<0$, the models are extrapolated along a symmetric counterpart. \revi{The ground-truth data was generated with the DHM and a Neo-Hookean strain energy density, cf. \cref{subsec:model-evaluation}}.}
    \label{fig:symmetry}
\end{figure}

\subsubsection{Study II -- Point symmetric PANN beam model}
\label{subsubsec:study-ii}

\begin{figure*}[t]
    \centering
    \begin{subfigure}[t]{0.4\textwidth}
        \centering
        \resizebox{\linewidth}{!}{
            \tikzsetnextfilename{fig_min_max_prediction_247_perturbed}
\begin{tikzpicture}
\begin{axis}[
  name=plot1,
  xlabel={$t$},
  ylabel={$q_i$},
  transpose legend,
  legend style={at={(0.03,0.03)}, anchor=south west},
  legend columns=3,
  grid=major,
]
    
    \addplot[color=CPSred, name path=min, solid, thick, no markers, forget plot] table [col sep=tab, x=t, y=n_1_pred_min]{fig_min_max_data_247_perturbed.txt};
    \addplot[color=CPSred, name path=max, solid, thick, no markers, forget plot] table [col sep=tab, x=t, y=n_1_pred_max]{fig_min_max_data_247_perturbed.txt};
    \addplot[CPSred, opacity=0.4, forget plot] fill between [of=min and max];

    \addplot[color=CPSred, name path=min, solid, thick, no markers, forget plot] table [col sep=tab, x=t, y=n_2_pred_min]{fig_min_max_data_247_perturbed.txt};
    \addplot[color=CPSred, name path=max, solid, thick, no markers, forget plot] table [col sep=tab, x=t, y=n_2_pred_max]{fig_min_max_data_247_perturbed.txt};
    \addplot[CPSred, opacity=0.4, forget plot] fill between [of=min and max];

    \addplot[color=CPSdarkblue, name path=min, solid, thick, no markers, forget plot] table [col sep=tab, x=t, y=n_3_pred_min]{fig_min_max_data_247_perturbed.txt};
    \addplot[color=CPSdarkblue, name path=max, solid, thick, no markers, forget plot] table [col sep=tab, x=t, y=n_3_pred_max]{fig_min_max_data_247_perturbed.txt};
    \addplot[CPSdarkblue, opacity=0.4, forget plot] fill between [of=min and max];
    
    \addplot[color=CPSgreen, name path=min, solid, thick, no markers, forget plot] table [col sep=tab, x=t, y=m_1_pred_min]{fig_min_max_data_247_perturbed.txt};
    \addplot[color=CPSgreen, name path=max, solid, thick, no markers, forget plot] table [col sep=tab, x=t, y=m_1_pred_max]{fig_min_max_data_247_perturbed.txt};
    \addplot[CPSgreen, opacity=0.4, forget plot] fill between [of=min and max];

    \addplot[color=CPSgreen, name path=min, solid, thick, no markers, forget plot] table [col sep=tab, x=t, y=m_2_pred_min]{fig_min_max_data_247_perturbed.txt};
    \addplot[color=CPSgreen, name path=max, solid, thick, no markers, forget plot] table [col sep=tab, x=t, y=m_2_pred_max]{fig_min_max_data_247_perturbed.txt};
    \addplot[CPSgreen, opacity=0.4, forget plot] fill between [of=min and max];

    \addplot[color=CPSorange, name path=min, solid, thick, no markers, forget plot] table [col sep=tab, x=t, y=m_3_pred_min]{fig_min_max_data_247_perturbed.txt};
    \addplot[color=CPSorange, name path=max, solid, thick, no markers, forget plot] table [col sep=tab, x=t, y=m_3_pred_max]{fig_min_max_data_247_perturbed.txt};
    \addplot[CPSorange, opacity=0.4, forget plot] fill between [of=min and max];

    
    \addplot [color=CPSred, dashed, very thick, mark=*, mark options={solid}, mark repeat=10] table [col sep=tab, x=t, y=n_1]{fig_min_max_data_247_perturbed.txt};
    \addlegendentry{$n_1$}
    \addplot [color=CPSred, dashed, very thick, mark=triangle*, mark options={solid}, mark repeat=10] table [col sep=tab, x=t, y=n_2]{fig_min_max_data_247_perturbed.txt};
    \addlegendentry{$n_2$}
    \addplot [color=CPSdarkblue, dashed, very thick, mark=square*, mark options={solid}, mark repeat=10] table [col sep=tab, x=t, y=n_3]{fig_min_max_data_247_perturbed.txt};
    \addlegendentry{$n_3$}
    \addplot [color=CPSgreen, dashed, very thick, mark=*, mark options={solid}, mark repeat=10] table [col sep=tab, x=t, y=m_1]{fig_min_max_data_247_perturbed.txt};
    \addlegendentry{$m_1$}
    \addplot [color=CPSgreen, dashed, very thick, mark=triangle*, mark options={solid}, mark repeat=10] table [col sep=tab, x=t, y=m_2]{fig_min_max_data_247_perturbed.txt};
    \addlegendentry{$m_2$}
    \addplot [color=CPSorange, dashed, very thick, mark=diamond*, mark options={solid}, mark repeat=10] table [col sep=tab, x=t, y=m_3]{fig_min_max_data_247_perturbed.txt};
    \addlegendentry{$m_3$}

\end{axis}
\end{tikzpicture}
        }
        \caption{247\textsuperscript{th} perturbed, concentric (calibration)}
        \label{fig:train-prediction-concentric-perturbed}
    \end{subfigure}
    \hspace{0.03\linewidth}
    \begin{subfigure}[t]{0.4\textwidth}
        \centering
        \resizebox{\linewidth}{!}{
            \tikzsetnextfilename{fig_min_max_19}
\begin{tikzpicture}
\begin{axis}[
  name=plot1,
  xlabel={$t$},
  ylabel={$q_i$},
  transpose legend,
  legend style={at={(0.03,0.03)}, anchor=south west},
  legend columns=3,
  grid=major,
]
    
    \addplot[color=CPSred, name path=min, solid, thick, no markers, forget plot] table [col sep=tab, x=t, y=n_1_pred_min]{fig_min_max_data_19.txt};
    \addplot[color=CPSred, name path=max, solid, thick, no markers, forget plot] table [col sep=tab, x=t, y=n_1_pred_max]{fig_min_max_data_19.txt};
    \addplot[CPSred, opacity=0.4, forget plot] fill between [of=min and max];

    \addplot[color=CPSred, name path=min, solid, thick, no markers, forget plot] table [col sep=tab, x=t, y=n_2_pred_min]{fig_min_max_data_19.txt};
    \addplot[color=CPSred, name path=max, solid, thick, no markers, forget plot] table [col sep=tab, x=t, y=n_2_pred_max]{fig_min_max_data_19.txt};
    \addplot[CPSred, opacity=0.4, forget plot] fill between [of=min and max];

    \addplot[color=CPSdarkblue, name path=min, solid, thick, no markers, forget plot] table [col sep=tab, x=t, y=n_3_pred_min]{fig_min_max_data_19.txt};
    \addplot[color=CPSdarkblue, name path=max, solid, thick, no markers, forget plot] table [col sep=tab, x=t, y=n_3_pred_max]{fig_min_max_data_19.txt};
    \addplot[CPSdarkblue, opacity=0.4, forget plot] fill between [of=min and max];
    
    \addplot[color=CPSgreen, name path=min, solid, thick, no markers, forget plot] table [col sep=tab, x=t, y=m_1_pred_min]{fig_min_max_data_19.txt};
    \addplot[color=CPSgreen, name path=max, solid, thick, no markers, forget plot] table [col sep=tab, x=t, y=m_1_pred_max]{fig_min_max_data_19.txt};
    \addplot[CPSgreen, opacity=0.4, forget plot] fill between [of=min and max];

    \addplot[color=CPSgreen, name path=min, solid, thick, no markers, forget plot] table [col sep=tab, x=t, y=m_2_pred_min]{fig_min_max_data_19.txt};
    \addplot[color=CPSgreen, name path=max, solid, thick, no markers, forget plot] table [col sep=tab, x=t, y=m_2_pred_max]{fig_min_max_data_19.txt};
    \addplot[CPSgreen, opacity=0.4, forget plot] fill between [of=min and max];

    \addplot[color=CPSorange, name path=min, solid, thick, no markers, forget plot] table [col sep=tab, x=t, y=m_3_pred_min]{fig_min_max_data_19.txt};
    \addplot[color=CPSorange, name path=max, solid, thick, no markers, forget plot] table [col sep=tab, x=t, y=m_3_pred_max]{fig_min_max_data_19.txt};
    \addplot[CPSorange, opacity=0.4, forget plot] fill between [of=min and max];

    
    \addplot [color=CPSred, dashed, very thick, mark=*, mark options={solid}, mark repeat=10] table [col sep=tab, x=t, y=n_1]{fig_min_max_data_19.txt};
    \addlegendentry{$n_1$}
    \addplot [color=CPSred, dashed, very thick, mark=triangle*, mark options={solid}, mark repeat=10] table [col sep=tab, x=t, y=n_2]{fig_min_max_data_19.txt};
    \addlegendentry{$n_2$}
    \addplot [color=CPSdarkblue, dashed, very thick, mark=square*, mark options={solid}, mark repeat=10] table [col sep=tab, x=t, y=n_3]{fig_min_max_data_19.txt};
    \addlegendentry{$n_3$}
    \addplot [color=CPSgreen, dashed, very thick, mark=*, mark options={solid}, mark repeat=10] table [col sep=tab, x=t, y=m_1]{fig_min_max_data_19.txt};
    \addlegendentry{$m_1$}
    \addplot [color=CPSgreen, dashed, very thick, mark=triangle*, mark options={solid}, mark repeat=10] table [col sep=tab, x=t, y=m_2]{fig_min_max_data_19.txt};
    \addlegendentry{$m_2$}
    \addplot [color=CPSorange, dashed, very thick, mark=diamond*, mark options={solid}, mark repeat=10] table [col sep=tab, x=t, y=m_3]{fig_min_max_data_19.txt};
    \addlegendentry{$m_3$}

\end{axis}
\end{tikzpicture}
        }
        \caption{\revii{19}\textsuperscript{th} concentric (testing)}
        \label{fig:test-prediction-concentric}
    \end{subfigure}

    \vspace{0.03\linewidth}
    
    \begin{subfigure}[t]{0.4\textwidth}
        \centering
        \resizebox{\linewidth}{!}{
                \tikzsetnextfilename{fig_min_max_prediction_epsilon_1}
\begin{tikzpicture}
\begin{axis}[
  name=plot1,
  xlabel={$\epsilon_1$},
  ylabel={$q_i$},
  transpose legend,
  legend style={at={(0.03,0.97)}, anchor=north west},
  legend columns=3,
  grid=major,
]
    

    \addplot[color=CPSgrey, name path=min, solid, thick, no markers, forget plot] table [col sep=tab, x=t, y=n_2_pred_min]{fig_min_max_data_shear.txt};
    \addplot[color=CPSgrey, name path=max, solid, thick, no markers, forget plot] table [col sep=tab, x=t, y=n_2_pred_max]{fig_min_max_data_shear.txt};
    \addplot[CPSgrey, opacity=0.4, forget plot] fill between [of=min and max];
    
    \addplot[color=CPSgrey, name path=min, solid, thick, no markers, forget plot] table [col sep=tab, x=t, y=m_1_pred_min]{fig_min_max_data_shear.txt};
    \addplot[color=CPSgrey, name path=max, solid, thick, no markers, forget plot] table [col sep=tab, x=t, y=m_1_pred_max]{fig_min_max_data_shear.txt};
    \addplot[CPSgrey, opacity=0.4, forget plot] fill between [of=min and max];

    \addplot[color=CPSgrey, name path=min, solid, thick, no markers, forget plot] table [col sep=tab, x=t, y=m_2_pred_min]{fig_min_max_data_shear.txt};
    \addplot[color=CPSgrey, name path=max, solid, thick, no markers, forget plot] table [col sep=tab, x=t, y=m_2_pred_max]{fig_min_max_data_shear.txt};
    \addplot[CPSgrey, opacity=0.4, forget plot] fill between [of=min and max];

    \addplot[color=CPSgrey, name path=min, solid, thick, no markers, forget plot] table [col sep=tab, x=t, y=m_3_pred_min]{fig_min_max_data_shear.txt};
    \addplot[color=CPSgrey, name path=max, solid, thick, no markers, forget plot] table [col sep=tab, x=t, y=m_3_pred_max]{fig_min_max_data_shear.txt};
    \addplot[CPSgrey, opacity=0.4, forget plot] fill between [of=min and max];
    
    \addplot [color=CPSred, dashed, very thick, mark=*, mark options={solid}, mark repeat=10] table [col sep=tab, x=t, y=n_1]{fig_min_max_data_shear.txt};
    \addlegendentry{$n_1$}
    \addplot [color=CPSgrey, dashed, very thick, mark=triangle*, mark options={solid}, mark repeat=10] table [col sep=tab, x=t, y=n_2]{fig_min_max_data_shear.txt};
    \addlegendentry{$n_2$}
    \addplot [color=CPSdarkblue, dashed, very thick, mark=square*, mark options={solid}, mark repeat=10] table [col sep=tab, x=t, y=n_3]{fig_min_max_data_shear.txt};
    \addlegendentry{$n_3$}
    \addplot [color=CPSgrey, dashed, very thick, mark=*, mark options={solid}, mark repeat=10] table [col sep=tab, x=t, y=m_1]{fig_min_max_data_shear.txt};
    \addlegendentry{$m_1$}
    \addplot [color=CPSgrey, dashed, very thick, mark=triangle*, mark options={solid}, mark repeat=10] table [col sep=tab, x=t, y=m_2]{fig_min_max_data_shear.txt};
    \addlegendentry{$m_2$}
    \addplot [color=CPSgrey, dashed, very thick, mark=diamond*, mark options={solid}, mark repeat=10] table [col sep=tab, x=t, y=m_3]{fig_min_max_data_shear.txt};
    \addlegendentry{$m_3$}


    \addplot [color=CPSred, dashed, very thick, mark=*, mark options={solid}, mark repeat=10] table [col sep=tab, x=t, y=n_1]{fig_min_max_data_shear.txt};
    \addplot [color=CPSdarkblue, dashed, very thick, mark=*, mark options={solid}, mark repeat=10] table [col sep=tab, x=t, y=n_3]{fig_min_max_data_shear.txt};

    \addplot[color=CPSred, name path=min, solid, thick, no markers, forget plot] table [col sep=tab, x=t, y=n_1_pred_min]{fig_min_max_data_shear.txt};
    \addplot[color=CPSred, name path=max, solid, thick, no markers, forget plot] table [col sep=tab, x=t, y=n_1_pred_max]{fig_min_max_data_shear.txt};
    \addplot[CPSred, opacity=0.4, forget plot] fill between [of=min and max];

    \addplot[color=CPSdarkblue, name path=min, solid, thick, no markers, forget plot] table [col sep=tab, x=t, y=n_3_pred_min]{fig_min_max_data_shear.txt};
    \addplot[color=CPSdarkblue, name path=max, solid, thick, no markers, forget plot] table [col sep=tab, x=t, y=n_3_pred_max]{fig_min_max_data_shear.txt};
    \addplot[CPSdarkblue, opacity=0.4, forget plot] fill between [of=min and max];

\end{axis}
\end{tikzpicture}
        }
        \caption{Shear (testing)}
        \label{fig:test-prediction-shear}
    \end{subfigure}
    \hspace{0.03\linewidth}
    \begin{subfigure}[t]{0.4\textwidth}
        \centering
        \resizebox{\linewidth}{!}{
                \tikzsetnextfilename{fig_min_max_bending}
\begin{tikzpicture}
\begin{axis}[
  name=plot1,
  xlabel={$\kappa_1$},
  ylabel={$q_i$},
  transpose legend,
  legend style={at={(0.03,0.97)}, anchor=north west},
  legend columns=3,
  grid=major,
ytick={-30,0,30},
yticklabels={-30,0,30},
xtick={-0.6,-0.3,0,0.3,0.6},
xticklabels={-0.6,-0.3,0,0.3,0.6},
]
    
    \addplot[color=CPSgrey, name path=min, solid, thick, no markers, forget plot] table [col sep=comma, x=t, y=n_1_pred_min]{fig_min_max_data_bending.txt};
    \addplot[color=CPSgrey, name path=max, solid, thick, no markers, forget plot] table [col sep=comma, x=t, y=n_1_pred_max]{fig_min_max_data_bending.txt};
    \addplot[CPSgrey, opacity=0.4, forget plot] fill between [of=min and max];

    \addplot[color=CPSgrey, name path=min, solid, thick, no markers, forget plot] table [col sep=comma, x=t, y=n_2_pred_min]{fig_min_max_data_bending.txt};
    \addplot[color=CPSgrey, name path=max, solid, thick, no markers, forget plot] table [col sep=comma, x=t, y=n_2_pred_max]{fig_min_max_data_bending.txt};
    \addplot[CPSgrey, opacity=0.4, forget plot] fill between [of=min and max];

    \addplot[color=CPSgrey, name path=min, solid, thick, no markers, forget plot] table [col sep=comma, x=t, y=m_2_pred_min]{fig_min_max_data_bending.txt};
    \addplot[color=CPSgrey, name path=max, solid, thick, no markers, forget plot] table [col sep=comma, x=t, y=m_2_pred_max]{fig_min_max_data_bending.txt};
    \addplot[CPSgrey, opacity=0.4, forget plot] fill between [of=min and max];

    \addplot[color=CPSgrey, name path=min, solid, thick, no markers, forget plot] table [col sep=comma, x=t, y=m_3_pred_min]{fig_min_max_data_bending.txt};
    \addplot[color=CPSgrey, name path=max, solid, thick, no markers, forget plot] table [col sep=comma, x=t, y=m_3_pred_max]{fig_min_max_data_bending.txt};
    \addplot[CPSgrey, opacity=0.4, forget plot] fill between [of=min and max];
    
    \addplot [color=CPSgrey, dashed, very thick, mark=*, mark options={solid}, mark repeat=10] table [col sep=comma, x=t, y=n_1]{fig_min_max_data_bending.txt};
    \addlegendentry{$n_1$}
    \addplot [color=CPSgrey, dashed, very thick, mark=triangle*, mark options={solid}, mark repeat=10] table [col sep=comma, x=t, y=n_2]{fig_min_max_data_bending.txt};
    \addlegendentry{$n_2$}
    \addplot [color=CPSdarkblue, dashed, very thick, mark=square*, mark options={solid}, mark repeat=10] table [col sep=comma, x=t, y=n_3]{fig_min_max_data_bending.txt};
    \addlegendentry{$n_3$}
    \addplot [color=CPSgreen, dashed, very thick, mark=*, mark options={solid}, mark repeat=10] table [col sep=comma, x=t, y=m_1]{fig_min_max_data_bending.txt};
    \addlegendentry{$m_1$}
    \addplot [color=CPSgrey, dashed, very thick, mark=triangle*, mark options={solid}, mark repeat=10] table [col sep=comma, x=t, y=m_2]{fig_min_max_data_bending.txt};
    \addlegendentry{$m_2$}
    \addplot [color=CPSgrey, dashed, very thick, mark=diamond*, mark options={solid}, mark repeat=10] table [col sep=comma, x=t, y=m_3]{fig_min_max_data_bending.txt};
    \addlegendentry{$m_3$}


    \addplot [color=CPSdarkblue, dashed, very thick, mark=*, mark options={solid}, mark repeat=10] table [col sep=comma, x=t, y=n_3]{fig_min_max_data_bending.txt};
    \addplot [color=CPSgreen, dashed, very thick, mark=*, mark options={solid}, mark repeat=10] table [col sep=comma, x=t, y=m_1]{fig_min_max_data_bending.txt};
    

    \addplot[color=CPSdarkblue, name path=min, solid, thick, no markers, forget plot] table [col sep=comma, x=t, y=n_3_pred_min]{fig_min_max_data_bending.txt};
    \addplot[color=CPSdarkblue, name path=max, solid, thick, no markers, forget plot] table [col sep=comma, x=t, y=n_3_pred_max]{fig_min_max_data_bending.txt};
    \addplot[CPSdarkblue, opacity=0.4, forget plot] fill between [of=min and max];

    \addplot[color=CPSgreen, name path=min, solid, thick, no markers, forget plot] table [col sep=comma, x=t, y=m_1_pred_min]{fig_min_max_data_bending.txt};
    \addplot[color=CPSgreen, name path=max, solid, thick, no markers, forget plot] table [col sep=comma, x=t, y=m_1_pred_max]{fig_min_max_data_bending.txt};
    \addplot[CPSgreen, opacity=0.4, forget plot] fill between [of=min and max];

\end{axis}
\end{tikzpicture}
        }
        \caption{Bending (testing)}
        \label{fig:test-prediction-bending}
    \end{subfigure}
    
    \caption{Stress measure prediction of the $\modelsym$ model (solid lines) on one training and three selected testing load paths. Dotted lines and markers denote data. Colored surfaces show the minimum-maximum prediction range of ten training runs bounded by two solid lines. The mixed load path \revi{(b)} represents the test case with the highest evaluation loss. \revi{The ground-truth data was generated with the DHM \revii{for a full circle cross-section with $R = 1.0$} and a Neo-Hookean strain energy density, cf. \cref{subsec:model-evaluation}}.}
    \label{fig:test-prediction}
\end{figure*}

To assess the prediction quality of the point symmetric architecture $\modelsym$, it is calibrated separately ten times for \revii{$(\parameter=0.0, R=1.0)$}. The training dataset comprises 256 concentric load paths with $\epsilon = 0.1$ perturbation and 8297 data points. The test dataset consists of six univariate and 25 concentric load paths with $\epsilon = 0.0$ and 1415 data points. The validation dataset consists of 13 concentric load paths with $\epsilon = 0.0$ and 445 data points. The FFNN has one hidden layer with 32 neurons. It is calibrated for a maximum of 10000 epochs with an additional early stopping call back that terminates training if the validation loss does not improve for more than 500 epochs.

The 247\textsuperscript{th} perturbed, concentric calibration case, as well as the \revii{19}\textsuperscript{th} concentric, pure shear, and pure bending test cases are selected for the evaluation of Study II. The corresponding predicted stress resultants are shown in~\cref{fig:test-prediction}. Together, the three test cases capture the most important results, where the \revii{19}\textsuperscript{th} concentric case has overall the highest loss in the test dataset. Excellent accuracy is achieved on all three testing load paths for $t \leq 0.5$ with minimal variance between the individual training runs. In particular, the model correctly captures the pronounced nonlinearities of the bending moments $m_{1,2}$ and the normal force $n_3$ in~\cref{fig:test-prediction-concentric} and~\cref{fig:test-prediction-bending}, as well as the coupling with the normal force $n_3$. The latter is most observable for shear and bending in~\cref{fig:test-prediction-shear,fig:test-prediction-bending}, respectively.

Some deviations from the data are observable for $t > 0.5$ under bending and in the concentric load path in~\cref{fig:test-prediction-bending} and~\cref{fig:test-prediction-concentric}, respectively. On the one hand, this could indicate some issues with the generalization of the PANN beam model. On the other hand, the validity of the data itself may be questioned, as the volumetric compression ratios in the cross-section reach values of $J < 0.8$ in these cases. In light of this, the PANN beam model's accuracy is very promising, especially since only one hidden layer with 32 nodes is required in the FFNN.

\revii{
\subsubsection{Study III -- Scaling behavior}
\label{subsubsec:study-iii}

In this study, we investigate the scaling behavior of the point symmetric architecture $\modelsym$ regarding the scaling law described by~\cite{kumarHelicalCauchyBornRule2016}, cf.~\cref{subsec:mechanical-conditions}. Therefore, models were calibrated using two different calibration approaches. The first set of models is trained on data for unity radius $R_{\text{cal}} = 1.0$ with the same load paths as in Study II, cf.~\cref{subsubsec:study-ii}. According to the theory, this should be sufficient to predict the stress resultants for arbitrary beam radii~\cite{kumarHelicalCauchyBornRule2016}. The second set of models is calibrated using a data augmentation approach, where the data of four radii, i.e., $R_{\text{cal}} \in \{0.1, 0.4, 0.7, 1.0\}$, is included in the training dataset with 33294 data points. The models are calibrated and evaluated in analogy to Study II with one validation data set per calibration radius $R_{\text{cal}}$ and one test dataset per $R_{\text{test}} \in \{0.1, 0.2, 0.3, 0.4., 0.5, 0.6, 0.7, 0.8, 0.9, 1.0\}$. The validation and test datasets have 13940 and 1780 data points, respectively. As before, the study is performed 10 times to account for the randomness of weight initialization and optimization.  

\Cref{fig:kumar_scaling_loss} shows the minimum, maximum, and average test losses plotted over the radius for the two calibration approaches. Results for non-augmented models are displayed in blue, while data-augmented models are shown in red. As expected, all models yield similar and excellent accuracy for $R_{\text{test}} > 0.5$ regardless of the calibration approach. Hence, the displayed test losses for $R_{\text{test}} = 1.0$ are in close agreement with Study II, cf.~\cref{subsubsec:study-ii}. Meanwhile, for $R_{\text{test}} < 0.5$, the data-augmented models outperform the non-augmented models, where the test loss for $R_{\text{test}} = 1.0$ is reduced by nearly one order of magnitude as a result of the data augmentation. 
In~\cref{fig:kumar_scaling_bending,fig:kumar_scaling_bending_augmented_2}, the predictions of the best model with and without data augmentation, respectively, are plotted for the pure bending case. This test case is challenging since the bending and extension-bending stiffnesses depend massively on the beam radius. The data-augmented model outperforms the non-augmented model in this scenario.

We conclude that simply applying the scaling law is not necessarily sufficient for obtaining accurate results for arbitrary beam radii using the PANN beam potential $\modelsym$. In the present case, the non-augmented model was trained on the largest beam radius $R = 1.0$. Hence, all test cases with $R < 1.0$ are essentially interpolations of the model at smaller strains. As the bending stiffness for $R = 0.1$ is approximately four orders of magnitude smaller than for $R=1.0$, errors that appear small for $R=1.0$ can become large for $R=0.1$ in comparison. This effect can be mitigated by including data for different radii, which changes the weighting of data points during training. Instead of generating data for different radii, an alternative approach could be to apply different weights to data points in a single load path depending on their strain state. 
}

\begin{figure*}[t!]
    \centering
    \begin{subfigure}[t]{0.6\textwidth}
        \centering
        \resizebox{0.7\linewidth}{!}{
            \tikzsetnextfilename{fig_lor_kumar_scaling}
\begin{tikzpicture}
\begin{semilogyaxis}[
  name=plot1,
  xlabel={$R$},
  ylabel={$\mathcal{L}$},
  legend style={at={(0.5,-0.23)}, anchor=north},
  legend columns=2,
  grid=major,
]
    \addplot[color=CPSdarkblue, name path=min, solid, thick, no markers, forget plot] table [col sep=comma, x=R, y=min]{fig_scaling_data_lor.txt};
    \addplot[color=CPSdarkblue, name path=max, solid, thick, no markers, forget plot] table [col sep=comma, x=R, y=max]{fig_scaling_data_lor.txt};
    \addplot[CPSdarkblue, opacity=0.4, forget plot] fill between [of=min and max];
    \addplot [color=CPSdarkblue, dashed, very thick] table [col sep=comma, x=R, y=mean]{fig_scaling_data_lor.txt};
    \addlegendentry{$R_{\text{cal}}=1.0$}
    
    \addplot[color=CPSred, name path=min, solid, thick, no markers, forget plot] table [col sep=comma, x=R, y=min]{fig_scaling_data_lor_augmented_2.txt};
    \addplot[color=CPSred, name path=max, solid, thick, no markers, forget plot] table [col sep=comma, x=R, y=max]{fig_scaling_data_lor_augmented_2.txt};
    \addplot[CPSred, opacity=0.4, forget plot] fill between [of=min and max];
    \addplot [color=CPSred, dashed, very thick] table [col sep=comma, x=R, y=mean]{fig_scaling_data_lor_augmented_2.txt};
    \addlegendentry{$R_{\text{cal}}\in\{0.1, 0.4, 0.7, 1.0\}$}
    
\end{semilogyaxis}
\end{tikzpicture}
        }
        \caption{\revii{Test loss for models with and without and without data augmentation. Dotted lines denote the average of ten training runs. Colored surfaces show the minimum-maximum range.}}
        \label{fig:kumar_scaling_loss}
    \end{subfigure}

    \vspace{0.03\linewidth}

    \begin{subfigure}[t]{0.4\textwidth}
        \centering
        \resizebox{\linewidth}{!}{
                \tikzsetnextfilename{fig_lor_kumar_scaling_bending}
\begin{tikzpicture}
\begin{axis}[
  name=plot1,
  xlabel={$\kappa_1$},
  ylabel={$q_i$},
  transpose legend,
  legend style={at={(0.03,0.97)}, anchor=north west},
  legend columns=3,
  grid=major,
  xtick={0,0.3,0.6},
  xticklabels={0,0.3,0.6},
]
    
    \addplot [color=CPSgrey, dashed, very thick, mark=*, mark options={solid}, mark repeat=10] table [col sep=comma, x=t, y=n_1]{fig_scaling_data_bending.txt};
    \addlegendentry{$n_1$}
    \addplot [color=CPSgrey, dashed, very thick, mark=triangle*, mark options={solid}, mark repeat=10] table [col sep=comma, x=t, y=n_2]{fig_scaling_data_bending.txt};
    \addlegendentry{$n_2$}
    \addplot [color=CPSdarkblue, dashed, very thick, mark=square*, mark options={solid}, mark repeat=10] table [col sep=comma, x=t, y=n_3]{fig_scaling_data_bending.txt};
    \addlegendentry{$n_3$}
    \addplot [color=CPSgreen, dashed, very thick, mark=*, mark options={solid}, mark repeat=10] table [col sep=comma, x=t, y=m_1]{fig_scaling_data_bending.txt};
    \addlegendentry{$m_1$}
    \addplot [color=CPSgrey, dashed, very thick, mark=triangle*, mark options={solid}, mark repeat=10] table [col sep=comma, x=t, y=m_2]{fig_scaling_data_bending.txt};
    \addlegendentry{$m_2$}
    \addplot [color=CPSgrey, dashed, very thick, mark=diamond*, mark options={solid}, mark repeat=10] table [col sep=comma, x=t, y=m_3]{fig_scaling_data_bending.txt};
    \addlegendentry{$m_3$}

    \addplot [color=CPSdarkblue, dashed, very thick, mark=*, mark options={solid}, mark repeat=10] table [col sep=comma, x=t, y=n_3]{fig_scaling_data_bending.txt};
    \addplot [color=CPSgreen, dashed, very thick, mark=*, mark options={solid}, mark repeat=10] table [col sep=comma, x=t, y=m_1]{fig_scaling_data_bending.txt};
    
    \addplot [color=CPSgrey, solid, very thick, no markers] table [col sep=comma, x=t, y=n_1_pred]{fig_scaling_data_bending.txt};
    \addplot [color=CPSgrey, solid, very thick, no markers] table [col sep=comma, x=t, y=n_2_pred]{fig_scaling_data_bending.txt};
    \addplot [color=CPSgrey, solid, very thick, no markers] table [col sep=comma, x=t, y=m_2_pred]{fig_scaling_data_bending.txt};
    \addplot [color=CPSgrey, solid, very thick, no markers] table [col sep=comma, x=t, y=m_3_pred]{fig_scaling_data_bending.txt};

    \addplot [color=CPSdarkblue, solid, very thick, no markers] table [col sep=comma, x=t, y=n_3_pred]{fig_scaling_data_bending.txt};
    \addplot [color=CPSgreen, solid, very thick, no markers] table [col sep=comma, x=t, y=m_1_pred]{fig_scaling_data_bending.txt};

\end{axis}
\end{tikzpicture}
        }
        \caption{\revii{Best non-augmented model tested for $R = 0.1$}}
        \label{fig:kumar_scaling_bending}
    \end{subfigure}
    \hspace{0.03\linewidth}
    \begin{subfigure}[t]{0.4\textwidth}
        \centering
        \resizebox{\linewidth}{!}{
                \tikzsetnextfilename{fig_lor_kumar_scaling_bending}
\begin{tikzpicture}
\begin{axis}[
  name=plot1,
  xlabel={$\kappa_1$},
  ylabel={$q_i$},
  transpose legend,
  legend style={at={(0.03,0.97)}, anchor=north west},
  legend columns=3,
  grid=major,
  xtick={-0.6,-0.3,0,0.3,0.6},
  xticklabels={-0.6,-0.3,0,0.3,0.6},
]
    
    \addplot [color=CPSgrey, dashed, very thick, mark=*, mark options={solid}, mark repeat=10] table [col sep=comma, x=t, y=n_1]{fig_scaling_data_bending_augmented_2.txt};
    \addlegendentry{$n_1$}
    \addplot [color=CPSgrey, dashed, very thick, mark=triangle*, mark options={solid}, mark repeat=10] table [col sep=comma, x=t, y=n_2]{fig_scaling_data_bending_augmented_2.txt};
    \addlegendentry{$n_2$}
    \addplot [color=CPSdarkblue, dashed, very thick, mark=square*, mark options={solid}, mark repeat=10] table [col sep=comma, x=t, y=n_3]{fig_scaling_data_bending_augmented_2.txt};
    \addlegendentry{$n_3$}
    \addplot [color=CPSgreen, dashed, very thick, mark=*, mark options={solid}, mark repeat=10] table [col sep=comma, x=t, y=m_1]{fig_scaling_data_bending_augmented_2.txt};
    \addlegendentry{$m_1$}
    \addplot [color=CPSgrey, dashed, very thick, mark=triangle*, mark options={solid}, mark repeat=10] table [col sep=comma, x=t, y=m_2]{fig_scaling_data_bending_augmented_2.txt};
    \addlegendentry{$m_2$}
    \addplot [color=CPSgrey, dashed, very thick, mark=diamond*, mark options={solid}, mark repeat=10] table [col sep=comma, x=t, y=m_3]{fig_scaling_data_bending_augmented_2.txt};
    \addlegendentry{$m_3$}

    \addplot [color=CPSdarkblue, dashed, very thick, mark=*, mark options={solid}, mark repeat=10] table [col sep=comma, x=t, y=n_3]{fig_scaling_data_bending_augmented_2.txt};
    \addplot [color=CPSgreen, dashed, very thick, mark=*, mark options={solid}, mark repeat=10] table [col sep=comma, x=t, y=m_1]{fig_scaling_data_bending_augmented_2.txt};
    
    \addplot [color=CPSgrey, solid, very thick, no markers] table [col sep=comma, x=t, y=n_1_pred]{fig_scaling_data_bending_augmented_2.txt};
    \addplot [color=CPSgrey, solid, very thick, no markers] table [col sep=comma, x=t, y=n_2_pred]{fig_scaling_data_bending_augmented_2.txt};
    \addplot [color=CPSgrey, solid, very thick, no markers] table [col sep=comma, x=t, y=m_2_pred]{fig_scaling_data_bending_augmented_2.txt};
    \addplot [color=CPSgrey, solid, very thick, no markers] table [col sep=comma, x=t, y=m_3_pred]{fig_scaling_data_bending_augmented_2.txt};

    \addplot [color=CPSdarkblue, solid, very thick, no markers] table [col sep=comma, x=t, y=n_3_pred]{fig_scaling_data_bending_augmented_2.txt};
    \addplot [color=CPSgreen, solid, very thick, no markers] table [col sep=comma, x=t, y=m_1_pred]{fig_scaling_data_bending_augmented_2.txt};

\end{axis}
\end{tikzpicture}
        }
        \caption{\revii{Best data augmented model tested for $R = 0.1$}}
        \label{fig:kumar_scaling_bending_augmented_2}
    \end{subfigure}
    \caption{\revii{Investigation of the effect of data augmentation on the scaling accuracy of models with the symmetric and non-parameterized architecture $\modelsym$ for full circle cross-sections. Non-augmented models were calibrated with data for radius $R_{\text{cal}} = 1.0$. Augmented models were calibrated with data for radii $R_{\text{cal}} \in \{0.1, 0.4, 0.7, 1.0\}$. Ground-truth data was generated from the DHM with a Neo-Hookean strain energy density for every $R_{\text{cal}}$, cf. \cref{subsec:model-evaluation}.}}
    \label{fig:kumar_scaling}
\end{figure*}

\revii{
\subsubsection{Study IV -- Ring parameterization}
\label{subsubsec:study-iv}

In this study, the performance of the ring-parameterized and point symmetric architecture $\modelsym_{\parameter}$ is to be evaluated on ring-shaped cross-sections with varying ratios between the inner and outer radius, i.e., $ \parameter = R_i / R$. The outer radius is fixed to $R = 1.0$. 
Calibration and evaluation are conducted in analogy to Study II, cf.~\cref{subsubsec:study-ii}, with one training and one validation dataset per ratio $\parameter_{\text{cal}} \in \{0.1, 0.2, 0.3, 0.4, 0.5\}$ and one test dataset per $\parameter_{\text{test}} \in \{0.0, 0.05, 0.1, 0.15, 0.2, 0.25, 0.3, 0.35, 0.4, 0.45, 0.5\}$. The calibration, validation, and test data sets consist of 48396, 2676, and 49199 data points, respectively. Since the data generation for the ring-shaped cross-sections required smaller load increments than for the full circle data used in Study I, every fourth load step was included in the calibration and validation datasets, resulting in the numbers above. Two hidden layers with 32 nodes each are used for the FFNN. For details, cf.~\cref{sec:node-study}.

The minimum, maximum, and average test losses from ten consecutive training runs are plotted over the parameter $\parameter$ in~\cref{fig:lor_ring}. For $\parameter = 0.0$, i.e., the full circle, the test losses are even lower than the values reported for the non-parameterized models Study II, cf.~\cref{tab:node-study-pann}. 
}

\begin{figure*}[t!]
    \centering
    \begin{subfigure}[t]{0.4\textwidth}
        \centering
        \resizebox{\linewidth}{!}{
            \tikzsetnextfilename{fig_lor_ring}
\begin{tikzpicture}
\begin{semilogyaxis}[
  name=plot1,
  xlabel={$P = R/R_i$},
  ylabel={$\mathcal{L}$},
  legend style={at={(0.5,-0.23)}, anchor=north},
  legend columns=2,
  grid=major,
]
    \addplot[color=CPSdarkblue, name path=min, solid, thick, no markers, forget plot] table [col sep=comma, x=P, y=min]{fig_lor_data.txt};
    \addplot[color=CPSdarkblue, name path=max, solid, thick, no markers, forget plot] table [col sep=comma, x=P, y=max]{fig_lor_data.txt};
    \addplot[CPSdarkblue, opacity=0.4, forget plot] fill between [of=min and max];
    \addplot [color=CPSdarkblue, dashed, very thick] table [col sep=comma, x=P, y=mean]{fig_lor_data.txt};
    
\end{semilogyaxis}

\draw[fill=m_grey] 
  (0.58,5.7) circle (0.5);

\draw[fill=m_olive] [even odd rule]
  (3.43,5.7) circle (0.5)
  (3.43,5.7) circle (0.125);

\draw[fill=m_navy] [even odd rule]
  (6.28,5.7) circle (0.5)
  (6.28,5.7) circle (0.25);

\end{tikzpicture}
        }
        \caption{\revii{Test loss plotted over the parameter $\parameter$. Dotted lines denote the average of ten training runs. Colored surfaces show the minimum-maximum range.}}
        \label{fig:lor_ring}
    \end{subfigure}
    \hspace{0.03\linewidth}
    \begin{subfigure}[t]{0.4\textwidth}
        \centering
        \resizebox{\linewidth}{!}{
            \tikzsetnextfilename{fig_ring_min_max_05_23}
\begin{tikzpicture}
\begin{axis}[
  name=plot1,
  xlabel={$t$},
  ylabel={$q_i$},
  transpose legend,
  legend style={at={(0.03,0.03)}, anchor=south west},
  legend columns=3,
  grid=major,
]
    
    \addplot[color=CPSred, name path=min, solid, thick, no markers, forget plot] table [col sep=comma, x expr=abs(\thisrow{t}), y=n_1_pred_min]{fig_ring_data_min_max_0.5_23.txt};
    \addplot[color=CPSred, name path=max, solid, thick, no markers, forget plot] table [col sep=comma, x expr=abs(\thisrow{t}), y=n_1_pred_max]{fig_ring_data_min_max_0.5_23.txt};
    \addplot[CPSred, opacity=0.4, forget plot] fill between [of=min and max];

    \addplot[color=CPSred, name path=min, solid, thick, no markers, forget plot] table [col sep=comma, x expr=abs(\thisrow{t}), y=n_2_pred_min]{fig_ring_data_min_max_0.5_23.txt};
    \addplot[color=CPSred, name path=max, solid, thick, no markers, forget plot] table [col sep=comma, x expr=abs(\thisrow{t}), y=n_2_pred_max]{fig_ring_data_min_max_0.5_23.txt};
    \addplot[CPSred, opacity=0.4, forget plot] fill between [of=min and max];

    \addplot[color=CPSdarkblue, name path=min, solid, thick, no markers, forget plot] table [col sep=comma, x expr=abs(\thisrow{t}), y=n_3_pred_min]{fig_ring_data_min_max_0.5_23.txt};
    \addplot[color=CPSdarkblue, name path=max, solid, thick, no markers, forget plot] table [col sep=comma, x expr=abs(\thisrow{t}), y=n_3_pred_max]{fig_ring_data_min_max_0.5_23.txt};
    \addplot[CPSdarkblue, opacity=0.4, forget plot] fill between [of=min and max];
    
    \addplot[color=CPSgreen, name path=min, solid, thick, no markers, forget plot] table [col sep=comma, x expr=abs(\thisrow{t}), y=m_1_pred_min]{fig_ring_data_min_max_0.5_23.txt};
    \addplot[color=CPSgreen, name path=max, solid, thick, no markers, forget plot] table [col sep=comma, x expr=abs(\thisrow{t}), y=m_1_pred_max]{fig_ring_data_min_max_0.5_23.txt};
    \addplot[CPSgreen, opacity=0.4, forget plot] fill between [of=min and max];

    \addplot[color=CPSgreen, name path=min, solid, thick, no markers, forget plot] table [col sep=comma, x expr=abs(\thisrow{t}), y=m_2_pred_min]{fig_ring_data_min_max_0.5_23.txt};
    \addplot[color=CPSgreen, name path=max, solid, thick, no markers, forget plot] table [col sep=comma, x expr=abs(\thisrow{t}), y=m_2_pred_max]{fig_ring_data_min_max_0.5_23.txt};
    \addplot[CPSgreen, opacity=0.4, forget plot] fill between [of=min and max];

    \addplot[color=CPSorange, name path=min, solid, thick, no markers, forget plot] table [col sep=comma, x expr=abs(\thisrow{t}), y=m_3_pred_min]{fig_ring_data_min_max_0.5_23.txt};
    \addplot[color=CPSorange, name path=max, solid, thick, no markers, forget plot] table [col sep=comma, x expr=abs(\thisrow{t}), y=m_3_pred_max]{fig_ring_data_min_max_0.5_23.txt};
    \addplot[CPSorange, opacity=0.4, forget plot] fill between [of=min and max];

    \addplot [color=CPSred, dashed, very thick, mark=*, mark options={solid}, mark repeat=50] table [col sep=comma, x expr=abs(\thisrow{t}), y=n_1]{fig_ring_data_min_max_0.5_23.txt};
    \addlegendentry{$n_1$}
    \addplot [color=CPSred, dashed, very thick, mark=triangle*, mark options={solid}, mark repeat=50] table [col sep=comma, x expr=abs(\thisrow{t}), y=n_2]{fig_ring_data_min_max_0.5_23.txt};
    \addlegendentry{$n_2$}
    \addplot [color=CPSdarkblue, dashed, very thick, mark=square*, mark options={solid}, mark repeat=50] table [col sep=comma, x expr=abs(\thisrow{t}), y=n_3]{fig_ring_data_min_max_0.5_23.txt};
    \addlegendentry{$n_3$}
    \addplot [color=CPSgreen, dashed, very thick, mark=*, mark options={solid}, mark repeat=50] table [col sep=comma, x expr=abs(\thisrow{t}), y=m_1]{fig_ring_data_min_max_0.5_23.txt};
    \addlegendentry{$m_1$}
    \addplot [color=CPSgreen, dashed, very thick, mark=triangle*, mark options={solid}, mark repeat=50] table [col sep=comma, x expr=abs(\thisrow{t}), y=m_2]{fig_ring_data_min_max_0.5_23.txt};
    \addlegendentry{$m_2$}
    \addplot [color=CPSorange, dashed, very thick, mark=diamond*, mark options={solid}, mark repeat=50] table [col sep=comma, x expr=abs(\thisrow{t}), y=m_3]{fig_ring_data_min_max_0.5_23.txt};
    \addlegendentry{$m_3$}

\end{axis}
\end{tikzpicture}
        }
        \caption{\revii{23\textsuperscript{rd} concentric test case for $P_{\text{test}} = 0.5$ corresponding to the largest loss within the test dataset. Dotted lines denote data. Colored surfaces show the minimum-maximum prediction range.}}
        \label{fig:ring_05_23}
    \end{subfigure}
    \caption{\revii{Evaluation of the ring-parameterized and point symmetric architecture $\modelsym_{\parameter}$. The models were calibrated on data for $P_{\text{cal}} \in \{0.0, 0.1, 0.2, 0.3, 0.4, 0.5\}$. Ground-truth data was generated with the DHM hyperelastic model for ring-shaped cross-sections with outer radius $R = 1.0$ according to~\cref{subsec:model-evaluation}.}}
    \label{fig:ring}
\end{figure*}

\begin{figure*}[t!]
    \centering
    \begin{subfigure}[t]{0.4\textwidth}
        \centering
        \resizebox{\linewidth}{!}{
            \tikzsetnextfilename{fig_deviation_bending_m_1}
\begin{tikzpicture}[
  spy using outlines={rectangle, magnification=2.5,
                       size=2.5cm, connect spies}
]
\begin{axis}[
  name=plot1,
  xlabel={Data $m_1$},
  ylabel={Prediction $m_1$},
  legend style={at={(0.5,-0.25)}, anchor=north},
  legend columns=5,
  grid=major,
]
    
    \addplot [m_grey, only marks, semithick, mark=x] table [col sep=comma, x=m_1_0.0, y=m_1_pred_0.0]{fig_deviation_data_bending_m_1.txt};
    \addplot [m_purple, only marks, semithick, mark=x] table [col sep=comma, x=m_1_0.05, y=m_1_pred_0.05]{fig_deviation_data_bending_m_1.txt};
    \addplot [m_wine, only marks, semithick, mark=x] table [col sep=comma, x=m_1_0.1, y=m_1_pred_0.1]{fig_deviation_data_bending_m_1.txt};
    \addplot [m_rose, only marks, semithick, mark=x] table [col sep=comma, x=m_1_0.15, y=m_1_pred_0.15]{fig_deviation_data_bending_m_1.txt};
    \addplot [m_sand, only marks, semithick, mark=x] table [col sep=comma, x=m_1_0.2, y=m_1_pred_0.2]{fig_deviation_data_bending_m_1.txt};
    \addplot [m_olive, only marks, semithick, mark=x] table [col sep=comma, x=m_1_0.25, y=m_1_pred_0.25]{fig_deviation_data_bending_m_1.txt};
    \addplot [m_green, only marks, semithick, mark=x] table [col sep=comma, x=m_1_0.3, y=m_1_pred_0.3]{fig_deviation_data_bending_m_1.txt};
    \addplot [m_teal, only marks, semithick, mark=x] table [col sep=comma, x=m_1_0.35, y=m_1_pred_0.35]{fig_deviation_data_bending_m_1.txt};
    \addplot [m_cyan, only marks, semithick, mark=x] table [col sep=comma, x=m_1_0.4, y=m_1_pred_0.4]{fig_deviation_data_bending_m_1.txt};
    \addplot [m_indigo, only marks, semithick, mark=x] table [col sep=comma, x=m_1_0.45, y=m_1_pred_0.45]{fig_deviation_data_bending_m_1.txt};
    \addplot [m_navy, only marks, semithick, mark=x] table [col sep=comma, x=m_1_0.5, y=m_1_pred_0.5]{fig_deviation_data_bending_m_1.txt};

    \draw[black]
            (axis cs:\pgfkeysvalueof{/pgfplots/xmin},\pgfkeysvalueof{/pgfplots/xmin}) -- 
            (axis cs:\pgfkeysvalueof{/pgfplots/xmax},\pgfkeysvalueof{/pgfplots/xmax});

\end{axis}

    \spy on (3.5,2.9) in node at (1.5,4.1);
    \spy on (5.9,4.9) in node at (5.4,1.5);

\end{tikzpicture}
        }
        \caption{Bending moment $m_1$}
    \end{subfigure}
    \hspace{0.03\linewidth}
    \begin{subfigure}[t]{0.4\textwidth}
        \centering
        \resizebox{\linewidth}{!}{
            \tikzsetnextfilename{fig_deviation_bending_n_3}
\begin{tikzpicture}[
  spy using outlines={rectangle, magnification=2.5,
                       size=2.5cm, connect spies}
]
\begin{axis}[
  name=plot1,
  xlabel={Data $n_3$},
  ylabel={Prediction $n_3$},
  legend style={at={(0.5,-0.25)}, anchor=north},
  legend columns=5,
  grid=major,
]
    
    \addplot [m_grey, only marks, semithick, mark=x] table [col sep=comma, x=n_3_0.0, y=n_3_pred_0.0]{fig_deviation_data_bending_n_3.txt};
    \addplot [m_purple, only marks, semithick, mark=x] table [col sep=comma, x=n_3_0.05, y=n_3_pred_0.05]{fig_deviation_data_bending_n_3.txt};
    \addplot [m_wine, only marks, semithick, mark=x] table [col sep=comma, x=n_3_0.1, y=n_3_pred_0.1]{fig_deviation_data_bending_n_3.txt};
    \addplot [m_rose, only marks, semithick, mark=x] table [col sep=comma, x=n_3_0.15, y=n_3_pred_0.15]{fig_deviation_data_bending_n_3.txt};
    \addplot [m_sand, only marks, semithick, mark=x] table [col sep=comma, x=n_3_0.2, y=n_3_pred_0.2]{fig_deviation_data_bending_n_3.txt};
    \addplot [m_olive, only marks, semithick, mark=x] table [col sep=comma, x=n_3_0.25, y=n_3_pred_0.25]{fig_deviation_data_bending_n_3.txt};
    \addplot [m_green, only marks, semithick, mark=x] table [col sep=comma, x=n_3_0.3, y=n_3_pred_0.3]{fig_deviation_data_bending_n_3.txt};
    \addplot [m_teal, only marks, semithick, mark=x] table [col sep=comma, x=n_3_0.35, y=n_3_pred_0.35]{fig_deviation_data_bending_n_3.txt};
    \addplot [m_cyan, only marks, semithick, mark=x] table [col sep=comma, x=n_3_0.4, y=n_3_pred_0.4]{fig_deviation_data_bending_n_3.txt};
    \addplot [m_indigo, only marks, semithick, mark=x] table [col sep=comma, x=n_3_0.45, y=n_3_pred_0.45]{fig_deviation_data_bending_n_3.txt};
    \addplot [m_navy, only marks, semithick, mark=x] table [col sep=comma, x=n_3_0.5, y=n_3_pred_0.5]{fig_deviation_data_bending_n_3.txt};

    \draw[black]
            (axis cs:\pgfkeysvalueof{/pgfplots/xmin},\pgfkeysvalueof{/pgfplots/xmin}) -- 
            (axis cs:\pgfkeysvalueof{/pgfplots/xmax},\pgfkeysvalueof{/pgfplots/xmax});
\end{axis}

    \spy on (0.95,0.85) in node at (1.5,4.1);
    \spy on (4.2,3.5) in node at (5.4,1.5);

\end{tikzpicture}
        }
        \caption{Normal force $n_3$}
        \label{fig:deviation-bending-ring}
    \end{subfigure}

    \vspace{0.02\linewidth}
    
    \begin{subfigure}[b]{0.5\linewidth}
        \centering
        \resizebox{\linewidth}{!}{
            \tikzsetnextfilename{fig_deviation_legend_ring}
\begin{tikzpicture}
\begin{axis}[
  hide axis,
  xmin=0.0,
  xmax=0.001,
  ymin=0.0,
  ymax=0.001,
  transpose legend,
  legend columns=2,
legend style={/tikz/every even column/.append style={column sep=0.25cm,
row sep=0.1cm}}
]
    \addlegendimage{only marks, ultra thick, mark=square*, m_grey}
    \addlegendentry{$P = 0.0$}
    \addlegendimage{only marks, ultra thick, mark=square*, m_purple}
    \addlegendentry{$P = 0.05$}
    \addlegendimage{only marks, ultra thick, mark=square*, m_wine}
    \addlegendentry{$P = 0.1$}
    \addlegendimage{only marks, ultra thick, mark=square*, m_rose}
    \addlegendentry{$P = 0.15$}
    \addlegendimage{only marks, ultra thick, mark=square*, m_sand}
    \addlegendentry{$P = 0.2$}
    \addlegendimage{only marks, ultra thick, mark=square*, m_olive}
    \addlegendentry{$P = 0.25$}
    \addlegendimage{only marks, ultra thick, mark=square*, m_green}
    \addlegendentry{$P = 0.3$}
    \addlegendimage{only marks, ultra thick, mark=square*, m_teal}
    \addlegendentry{$P = 0.35$}
    \addlegendimage{only marks, ultra thick, mark=square*, m_cyan}
    \addlegendentry{$P = 0.4$}
    \addlegendimage{only marks, ultra thick, mark=square*, m_indigo}
    \addlegendentry{$P = 0.45$}
    \addlegendimage{only marks, ultra thick, mark=square*, m_navy}
    \addlegendentry{$P = 0.5$}
    
\end{axis}
\end{tikzpicture}
        }
    \end{subfigure}
    \caption{\revii{Correlation of stress resultant data and prediction for a ring-parameterized model, $\modelsym_{\parameter}$, on the pure bending test case. The model was calibrated according to~\cref{subsubsec:study-iv}. The results stem from the best-performing model out of ten training runs. To make the figures more readable and due to the model's symmetry, every fourth interpolated data point and only predictions for $\kappa_1 \geq 0$ are shown.}}
    \label{fig:deviation-ring}
\end{figure*}

\revii{
With increasing $\parameter$, the prediction error grows significantly. This is to be expected since in~\cref{subsec:model-evaluation}, we reported higher nonlinearity for increasing inner radii. 
Looking at~\cref{fig:ring_05_23}, which shows the minimum-maximum prediction range for the test case with the largest overall loss, we observe that the error for $\parameter=0.5$ primarily results from data points with $t > 0.4$. It indicates that these points lie in the extrapolatory range, whereas the points with $t \leq 0.4$ are represented well by the training data. To back up these observations, the correlation of predicted and actual strain measures is plotted for the pure bending test case in~\cref{fig:deviation-bending-ring}. Here, the best out of the ten trained models is plotted. It shows that the dominating stress resultants $m_1$ and $n_3$ are predicted very well for all tested parameters. The deviations are slightly larger for larger values of $\parameter$ and start to increase for larger values of the stress resultants.

These results are expected as rings with larger $\parameter$ are more prone to buckling of the beam's cross-section. Including such instabilities on the cross-sectional scale could easily lead to numerical instabilities of simulations on the beam scale and would require further treatment. As this is not within the scope of our work, we applied deformations only within the stable regime of the cross-section. It seems that in the test dataset, more load paths were on the verge of cross-sectional instabilities, especially in the case of pure bending, than in the calibration dataset. This is a possible explanation for the reduced accuracy for $t > 0.4$.
}


\subsection{PANN model application}
\label{subsec:beam-simulation}

In this section, we illustrate the direct applicability of the PANN beam material model in beam simulations. \revi{The aim is to showcase that considering nonlinear behavior and cross-section deformation in beams can lead to significantly different results than the standard assumption of linear elastic material behavior.} \revi{The DHM was not considered in this study, as implementing it would require complex run-time coupling of the numerical solution of the cross-section warping problem and the beam, which is avoided by using the PANN material model.}

\revi{For this study}, the \revi{PANN model is} implemented into a mixed isogeometric collocation method~\cite{weegerMixedIsogeometricCollocation2022}. \revi{For this purpose, the analytical expressions of the derivatives of the PANN, which are required to evaluate the stress measures $\vMatStress=\partial  \potRodPANN / \partial \vMatStrain$ and the elasticity tensor $\mathbb{C}=\partial^2 \potRodPANN / \partial \vMatStrain^{2}$ of the beam model, are implemented.}  Since the PANN model requires just one hidden layer, the analytical first and second derivatives are particularly easy to compute and implement~\cite{frankeAdvancedDiscretizationTechniques2023}.
For the results presented, the weights of the point symmetric model $\modelsym$ for beam radius $R=1.0$ corresponding to the training run with the lowest evaluation loss presented in~\cref{subsubsec:study-ii} are used.
\revi{For the isogeometric discretization of the beams, 22 B-spline basis functions with degree 5 and 16 knot spans (elements) are used. As a collocation approach is employed, the assembly of the force vector and tangent stiffness matrix requires 22 evaluations of the constitutive model.}

\paragraph{Beam bending.}

\begin{figure*}[t!]
    \centering
    \begin{subfigure}[t]{.35\linewidth}
        \includegraphics[width=\linewidth]{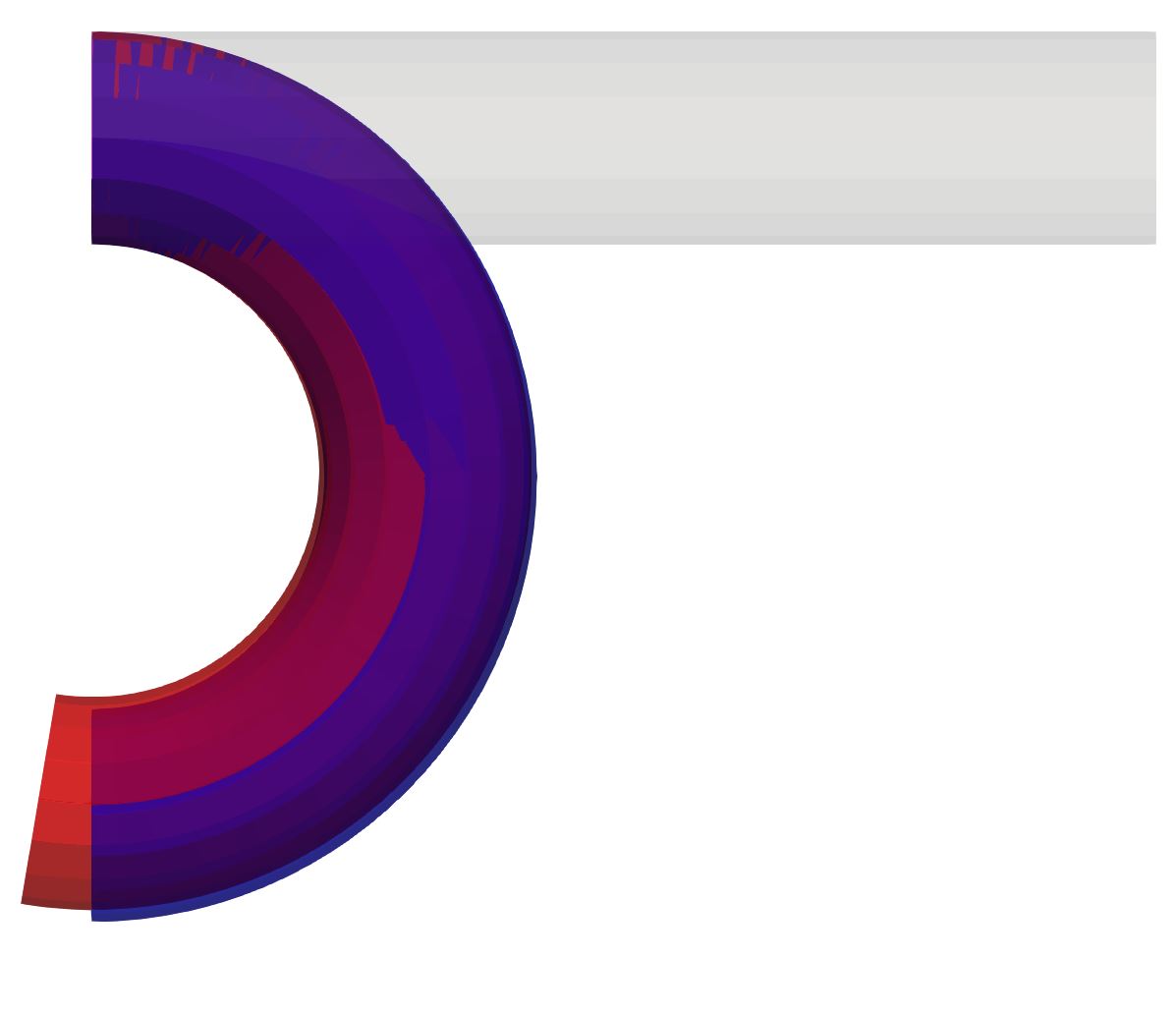}
        \caption{Deformed geometries obtained with the LEM (blue) and PANN (red)}
    \end{subfigure}
    \hspace{0.03\linewidth}
    \begin{subfigure}[t]{.48\linewidth}
        \centering
        \resizebox{0.85\linewidth}{!}{
            \tikzsetnextfilename{fig_simulation_bending}
\begin{tikzpicture}
\begin{axis}[
  name=plot1,
  xlabel={load factor},
  ylabel={strain measures $p_i$},
  transpose legend,
  legend style={at={(0.03,0.97)}, anchor=north west},
  legend columns=2,
  grid=major,
xtick={0,0.125,0.25,0.375,0.5},
xticklabels={0,0.25,0.5,0.75,1},
]
    
    \addplot [color=CPSdarkblue, dashed, very thick, no markers] table [col sep=tab, x=step, y=lin_e3]{fig_simulation_data_bend10M_data.txt};
    \addlegendentry{$\epsilon_3$ (LEM)}
    \addplot [color=CPSgreen, dashed, very thick, no markers] table [col sep=tab, x=step, y=lin_k1]{fig_simulation_data_bend10M_data.txt};
    \addlegendentry{$\kappa_1$ (LEM)}

    \addplot [color=CPSdarkblue, solid, very thick, no markers] table [col sep=tab, x=step, y=nn97_e3]{fig_simulation_data_bend10M_data.txt};
    \addlegendentry{$\epsilon_3$ (PANN)}
    \addplot [color=CPSgreen, solid, very thick, no markers] table [col sep=space, x=step, y=nn97_k1]{fig_simulation_data_bend10M_data.txt};
    \addlegendentry{$\kappa_1$ (PANN)}

\end{axis}
\end{tikzpicture}
        }
        \caption{Strain measures over load factor of applied bending moment}
    \end{subfigure}
    \caption{Application of the PANN model to beam bending and comparison to the LEM.}
    \label{fig:simulation-bending}
\end{figure*}

\revi{For the first comparison of LEM and PANN, a} beam of length $L=10$ and radius $R=1.0$ is clamped on the left and subjected to a bending moment on the right. The resulting deformations and strain states are shown in~\cref{fig:simulation-bending}. While the LEM (blue beam) yields a bend of exactly 180$\text{\textdegree}$ without axial strain, the PANN model (red beam) leads to \revii{$5.28\,\%$} more curvature $\kappa_1$ and a maximum axial strain of $\epsilon_3 = \revii{3.30\,\%}$. These observations align with the constitutive behavior of the DHM discussed in~\cref{subsec:data-generation}, where bending strain leads to a bending moment and a negative normal force $n_3$. Hence, applying a bending moment leads to bending and longitudinal expansion.

\begin{figure*}[t!] 
    \centering
    \begin{subfigure}[t]{.35\linewidth}
        \includegraphics[width=\linewidth]{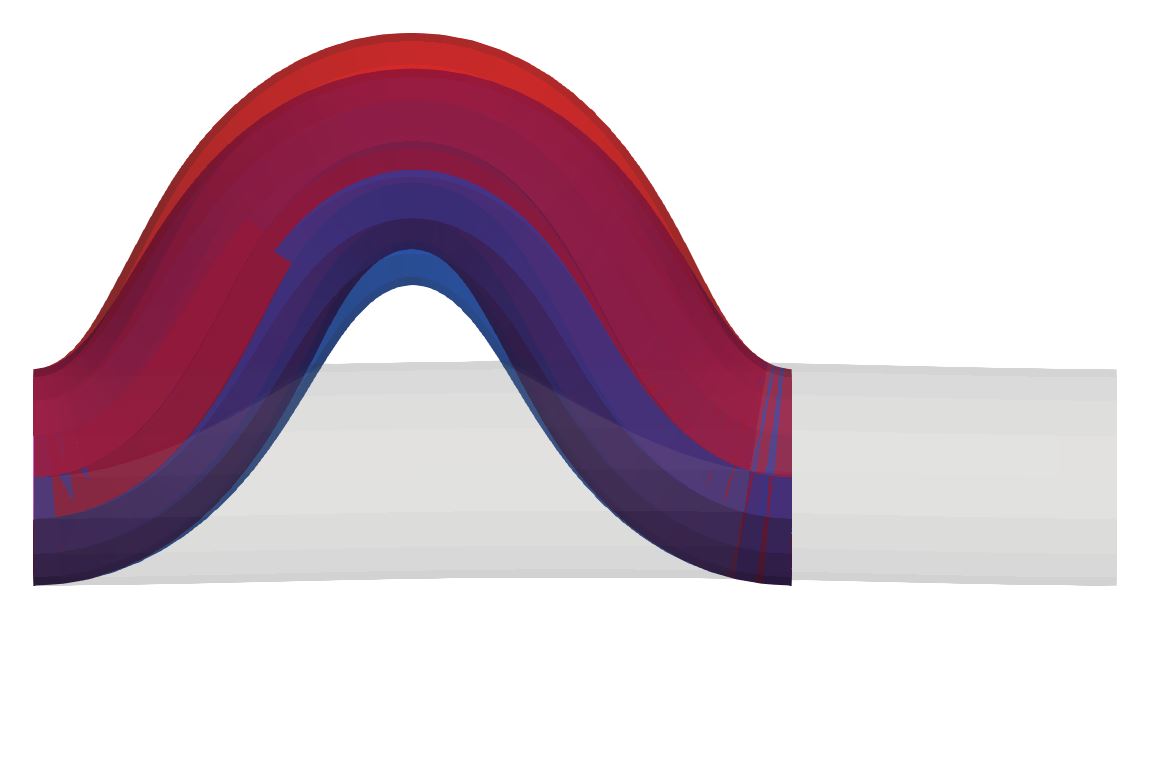}
        \caption{Deformed geometries obtained with the LEM (blue) and PANN (red)}
    \end{subfigure}
    \hspace{0.03\linewidth}
    \begin{subfigure}[t]{.46\linewidth}
        \centering
        \resizebox{0.85\linewidth}{!}{
            \tikzsetnextfilename{fig_simulation_compression}
\begin{tikzpicture}
\begin{axis}[
  name=plot1,
  xlabel={applied displacement  $-u_1$},
  ylabel={stress resultants $q_i$},
  xmin = -0.02,
  xmax = 0.32,
  ymax = 44,
  ymin = -4,
  transpose legend,
  legend style={at={(0.03,0.97)}, anchor=north west},
  legend columns=3,
  grid=major,
  xtick={0,0.05,0.10,0.15,0.2,0.25,0.3},
  xticklabels={0,0.5,1,1.5,2,2.5,3},
]
    
    \addplot [color=CPSdarkblue, dashed, very thick, no markers] table [col sep=tab, x=step, y=0_1.00_nx]{fig_simulation_data_compr10_data_lin.txt};
    \addlegendentry{$n_3$ (LEM)}
    \addplot [color=CPSgreen, dashed, very thick, no markers] table [col sep=tab, x=step, y=0_1.00_my]{fig_simulation_data_compr10_data_lin.txt};
    \addlegendentry{$m_1$ (LEM)}
    \addplot [color=CPSred, dashed, very thick, no markers] table [col sep=tab, x=step, y=0_1.00_mz]{fig_simulation_data_compr10_data_lin.txt};
    \addlegendentry{$m_2$ (LEM)}

    \addplot [color=CPSdarkblue, solid, very thick, no markers] table [col sep=tab, x=step, y=0_1.00_nx]{fig_simulation_data_compr10_data_nn97.txt};
    \addlegendentry{$n_3$ (PANN)}
    \addplot [color=CPSgreen, solid, very thick, no markers] table [col sep=tab, x=step, y=0_1.00_my]{fig_simulation_data_compr10_data_nn97.txt};
    \addlegendentry{$m_1$ (PANN)}
    \addplot [color=CPSred, solid, very thick, no markers] table [col sep=tab, x=step, y=0_1.00_mz]{fig_simulation_data_compr10_data_nn97.txt};
    \addlegendentry{$m_2$ (PANN)}
    
\end{axis}
\end{tikzpicture}
        }
        \caption{Stress resultants over applied compressive displacement}
    \end{subfigure}
    \caption{Application of the PANN model to beam compression with buckling and comparison to the LEM.}
    \label{fig:simulation-compression}
\end{figure*}

\paragraph{Beam compression with buckling.}

For the second case, the beam of length $L=10$ and radius $R=1.0$ is clamped on the left while a compressive axial displacement up to 30\% (3 units) is applied on the right. Before the simulation, a small pre-curvature was applied to ensure all beams buckle in the same direction. The deformed shapes of the beams with the LEM (blue beam) and the PANN model (red beam) are visualized in~\cref{fig:simulation-compression} alongside the stress resultants on the right end of the beam. Like the first test, the PANN model yields slightly larger deformations corresponding to a higher normal force $n_3$ and a higher bending moment $m_1$. Furthermore, the simulation with the PANN model shows some out-of-plane bending, which can be seen from the non-zero $m_2$ that arises around $-u_1=1$.

\begin{figure*}[t!]
    \centering
    \begin{subfigure}[b]{0.4\linewidth}
        \centering
        \includegraphics[width=0.9\linewidth]{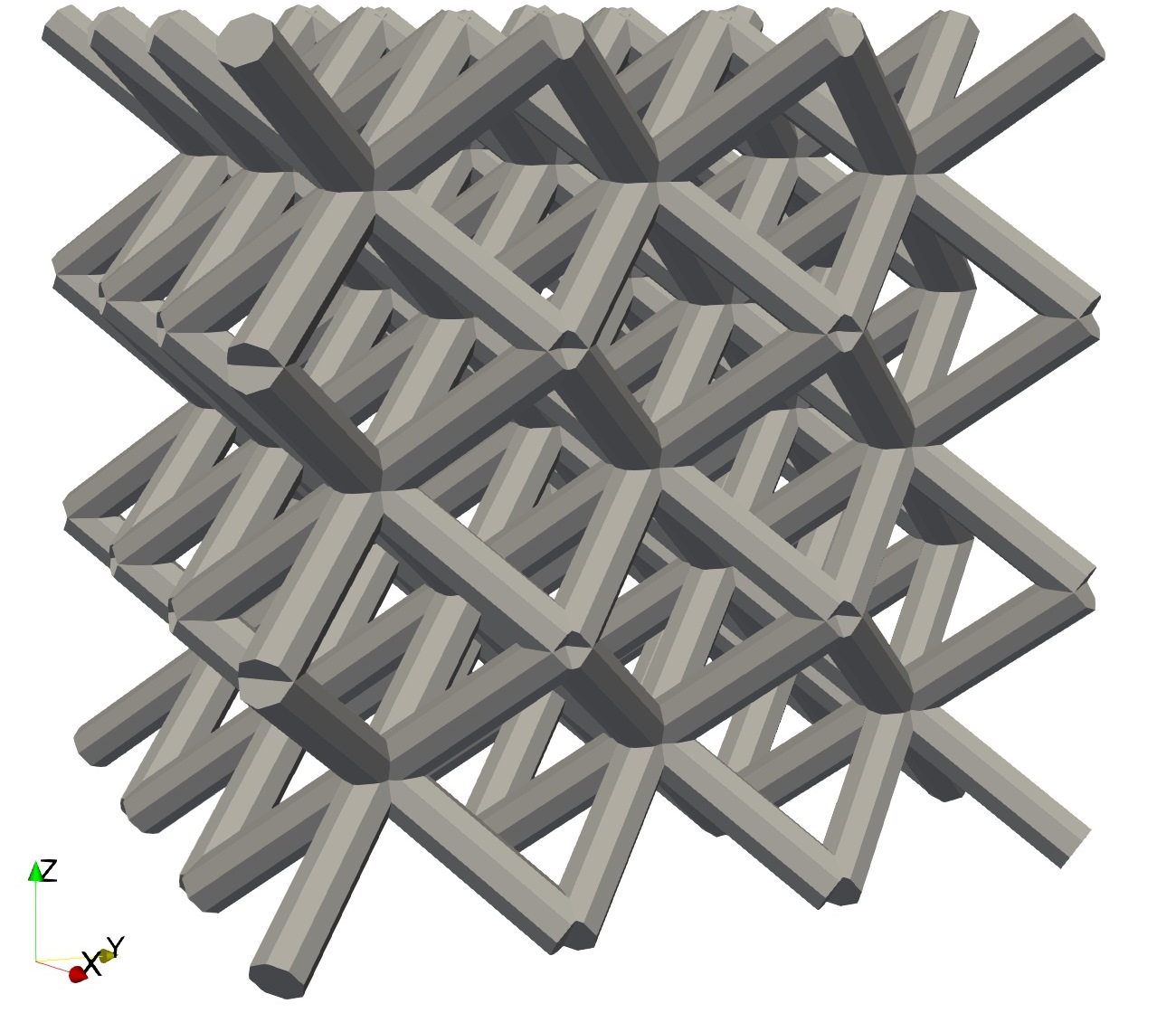}
        \caption{Geometry of BCC lattice}
        \label{fig:bcc-geo}
    \end{subfigure}
 ~
    \begin{subfigure}[b]{0.4\linewidth}
        \centering
        \resizebox{\linewidth}{!}{
            \tikzsetnextfilename{fig_simulation_bcc}
\begin{tikzpicture}
\begin{axis}[
  name=plot1,
  xlabel={applied displacement  $-u_3$},
  ylabel={resultant force $n_3$},
  ymax = 100,
  ymin = -7,
  transpose legend,
  legend style={at={(0.03,0.97)}, anchor=north west},
  legend columns=3,
  grid=major,
  xmin = -0.5,
  xmax = 8,
  xtick={0,1,...,8},
]
    
    \addplot [color=CPSdarkblue, solid, very thick, no markers, x filter/.expression={x<7.55 ? x : nan}] table [col sep=tab, x=du, y=fz]{fig_simulation_data_BCCnn_EL70_data.txt};
    \addlegendentry{LEM}

    \addplot [color=CPSred, solid, very thick, no markers] table [col sep=tab, x=du, y=fz]{fig_simulation_data_BCCnn_NN97_data.txt};
    \addlegendentry{PANN}
    
\end{axis}
\end{tikzpicture}
        }
        \caption{Resultant force vs.\ applied displacement}
        \label{fig:bcc-plot}
    \end{subfigure}
    \vspace*{4mm}

    \begin{subfigure}[b]{0.4\linewidth}
        \centering
        \includegraphics[width=0.9\linewidth]{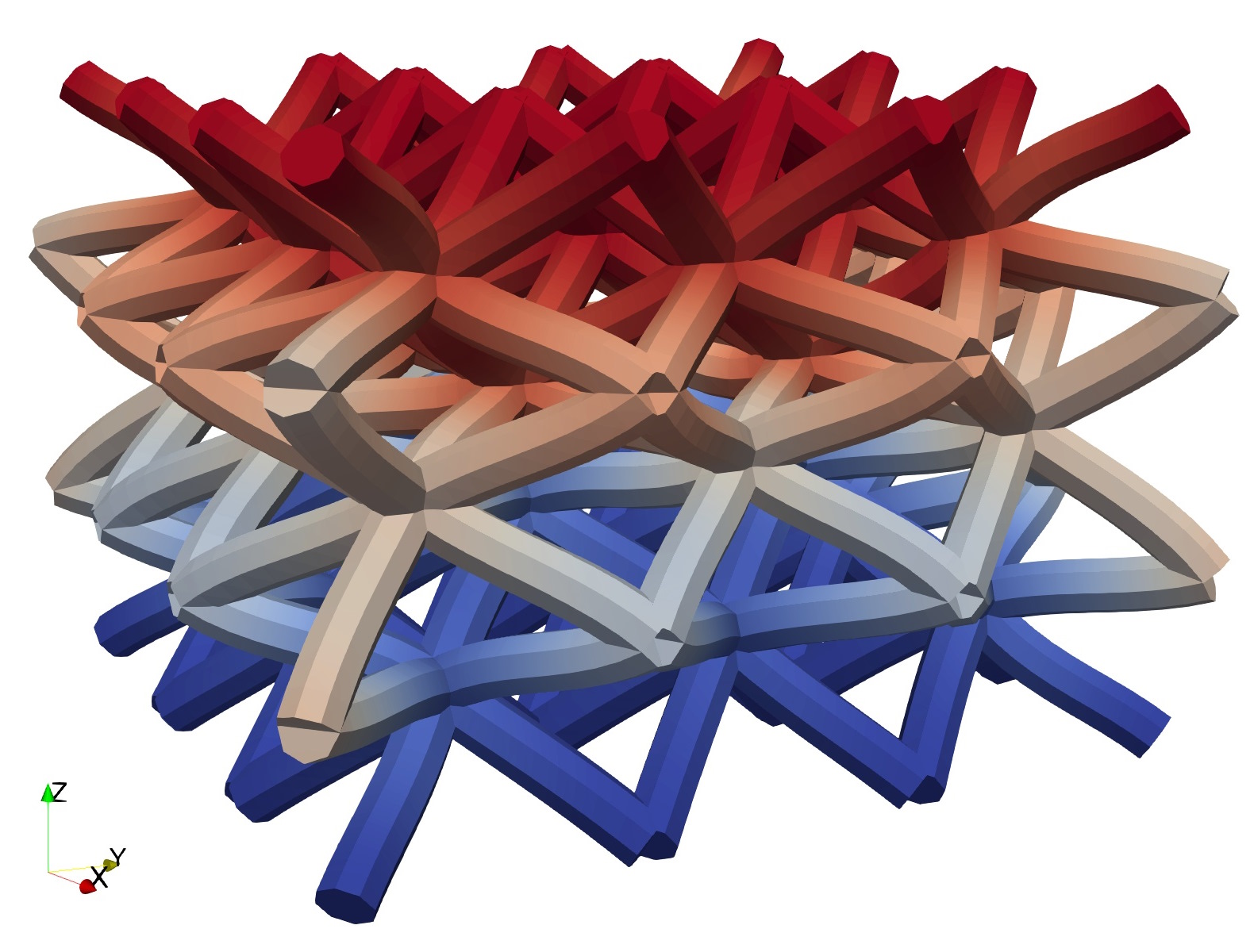}
        \caption{Deformed structure with LEM}
        \label{fig:bcc-def-lem}
    \end{subfigure}
~
    \begin{subfigure}[b]{0.4\linewidth}
        \centering
        \includegraphics[width=0.9\linewidth]{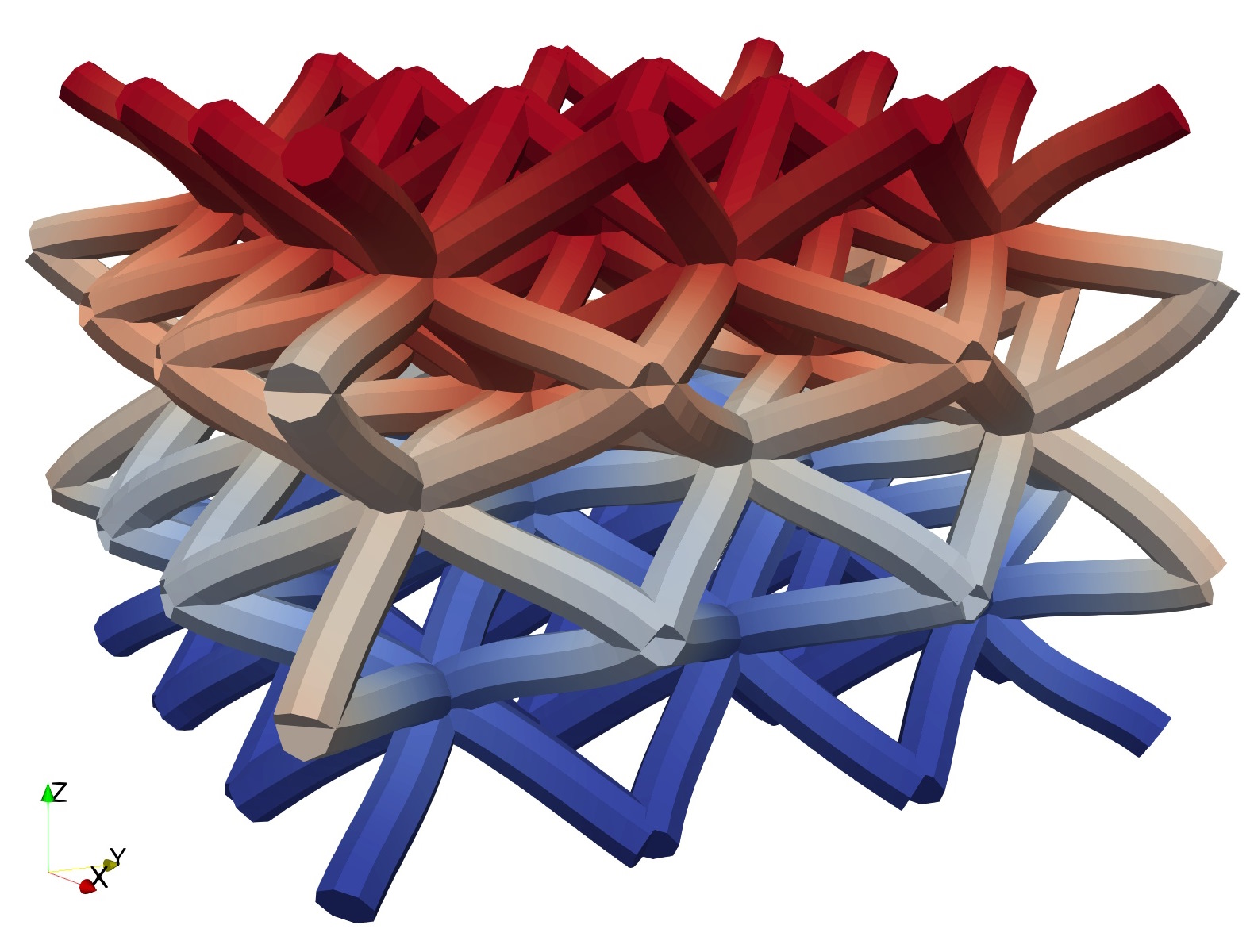}
        \caption{Deformed structure with PANN}
        \label{fig:bcc-def-pann}
    \end{subfigure}

    \caption{\revi{Application of the PANN model to compression of a $3 \times 3 \times 3$ BCC lattice structure with size $30 \times 30 \times 30$ and comparison to the LEM. Non-circular cross-sections result only from the selected visualization.}}
    \label{fig:bcc}
\end{figure*}

\revi{
\paragraph{BCC lattice structure.}

As a final example, the PANN model is applied within the simulation of a lattice structure, which could, for instance, be manufactured from a soft polymer material, c.f.~\cite{weegerInelasticFiniteDeformation2023}. The body-centered cubic (BCC) lattice consists of $3\times 3\times 3$ unit cells of cell size 10, each with 8 beams, i.e., a total of 216 beams. Each beam has length $L=5\sqrt{3}\approx 8.66$ and radius $R=1.0$. Also, the isogeometric discretization is applied with B-splines of degree $5$ and $16$ knot spans, resulting in 22 collocation points per beam.
The structure is compressed by 25\% of its height, i.e., by 7.5 length units, which is realized by clamping all beam ends at the bottom and top of the structure and using displacement boundary conditions.

The initial geometry, deformed structures at 25\% compression, and total resultant force vs. applied displacement curves are shown in~\cref{fig:bcc}. As seen from~\cref{fig:bcc-plot}, the bending-dominated lattice structure buckles at a displacement of around 6.5--7.0, which also leads to beam-to-beam contacts. The PANN also performs reliably in this challenging simulation.
For increasing deformation, the results between LEM and PANN deviate moderately, resulting in a larger displacement and force at the buckling point. It is to be expected that the deviation between the LEM and PANN model would be more pronounced for different load scenarios, cf.~\cref{fig:deformable_vs_rigid_bending}, or for more complex cross-sectional shapes~\cite{leclezioNumericalTwoscaleApproach2023}.
}

\paragraph{\revi{Discussion.}}

In summary, the PANN beam model was demonstrated to be suitable for application in beam simulations. The coupling effects arising from the cross-section warping problem translate to visible changes in the deformation of thick beams. Furthermore, the PANN beam model is stable enough to simulate the challenging scenarios of buckling under compression \revi{as well as compression of a lattice structure with buckling and contacts}. 

\revi{
The PANN can also be considered much more numerically efficient than the DHM for simulating lattice structures or optimizing beams or lattices with many iterations.
For instance, in the lattice simulation above with 216 beams and 22 evaluation points per beam, $216\cdot 22 = 4752$ constitutive model evaluations are required per assembly. For the deformation-controlled simulation up to 25\% compression, 50 load steps were used, which required roughly 6 Newton iterations per load step due to the instabilities and contacts, i.e., a total of $50\cdot 6=300$ assemblies. Overall, this amounts to 1.4 million constitutive model evaluations (of stress resultants and stiffnesses) to carry out one nonlinear lattice simulation. 
Using the PANN instead of the LEM only marginally increases the overall computational effort since constitutive model evaluations contribute only a small part of the numerical operations. On the other hand, in the DHM, 1.4 million cross-sectional warping FEM problems would have to be solved concurrently within the beam simulations.
Even if such simulations could be implemented extremely efficiently by tracing and storing the load paths of all 4752 evaluation points (which would require extensive memory), the numerical effort would still be much higher compared to the only 256 load paths generated for training the PANN $\symmodel$, compare~\cref{subsubsec:study-ii}.
}
\section{Conclusion}
\label{sec:concl}


The present work proposes a neural network-based constitutive model for hyperelastic, shear-deformable geometrically exact beams. The model is physics-augmented, i.e., formulated to fulfill important mechanical conditions relevant to beam constitutive modeling. As such, it aims to bridge the gap between accuracy and efficiency in multiscale beam simulations. This is achieved by applying the PANN beam constitutive model as a surrogate for the warping problem proposed in~\cite{aroraComputationalApproachObtain2019}, enabling efficient modeling of beams with both material and geometrical nonlinearities occurring on the cross-section.


First, in~\cref{sec:nn-model}, physics-augmented neural network (PANN) constitutive models for hyperelastic beams are proposed. For this, the hyperelastic beam potential is represented with an NN, which receives the strains and curvatures of the beam as inputs. Forces and moments are obtained as the gradients of the beam potential, ensuring thermodynamic consistency. To fulfill normalization conditions, the NN potential is complemented by additional projection terms. For beams with \revii{circular or ring-shaped} cross-sections \revi{and no chirality}, \revii{extensions to transverse isotropy and a derived point symmetry constraint are proposed}.
\revii{Additionally, a parameterized model fulfilling all the aforementioned mechanical conditions is formulated for ring-shaped cross-sections with different ratios between the inner and outer radius.}


In~\cref{sec:results}, the proposed models are calibrated to data generated for beams with hyperelastic base materials and deformable cross-sections. Strains and curvatures are sampled using a concentric sampling strategy. Data is generated for beams with circular \revii{and ring-shaped} cross-sections and different \revii{inner and outer} radii.
The proposed PANN models yield excellent predictions of the beam's forces and moments in all studied cases. \revii{In particular, the ring-parameterized PANN model accurately predicts the stress resultants across a large range of ring shapes. Furthermore, the scaling behavior of the proposed models was improved significantly through a data augmentation approach.}
Finally, the straightforward applicability of our approach for beam simulations is demonstrated with several numerical examples for beams and beam structures subject to complex deformation scenarios. 


\medskip

The proposed model can now readily be applied for hyperelastic base materials and subsequent simulation or design optimization of soft, deformable beams, e.g., cables and additively manufactured lattice structures.
Taking inspiration from recent advances in PANN constitutive modeling for continuum bodies~\cite{rosenkranz2024}, future work will be concerned with formulating inelastic PANN constitutive models for beams.


\vspace*{3ex}

\noindent
\textbf{Conflict of interest.} The authors declare that they have no conflict of interest.
\vspace*{1ex}

\noindent
\textbf{Acknowledgment.} 
The authors acknowledge the financial support provided by the Deutsche Forschungsgemeinschaft (DFG, German Research Foundation, project numbers 460684687 and 492770117), the Graduate School of Computational Engineering at TU Darmstadt, and Hessian.AI.
Furthermore, D.K.~Klein acknowledges support by a fellowship of the German Academic Exchange Service (DAAD).
\vspace*{1ex}


\appendix
\numberwithin{equation}{section} 
\revi{\section{Neural network size study}
\label{sec:node-study}

\cref{tab:node-study-pann} provides the calibration and test losses of four $\modelsym$ architectures \revii{with different numbers of nodes in the hidden layer. Every architecture was calibrated ten times according to Study II, cf.~\cref{subsubsec:study-ii}. The losses of the three models with the lowest test loss for every architecture are shown in the table}. While all architectures can accurately fit the training and test data, the calibration and test losses decrease when the node number increases. As the improvement from 32 to 64 nodes is relatively small, the \([32]\) architecture, i.e., one hidden layer and 32 nodes, is chosen for all non-parameterized models in this work.

\begin{table}[ht]
    \centering
    \caption{Architecture study for Study II in~\cref{subsubsec:study-ii}.}
    \label{tab:node-study-pann}
    \begin{tabularx}{\linewidth}{XXXX}
        \toprule
        Model & NN architecture & \multicolumn{2}{c}{Loss} \\
        \cline{3-4}
        \\[-0.35cm] 
        & & Calibration & Test \\
        \midrule
        $\modelsym$ with & \([16]\) & \(\revii{1.34 \cdot 10^{-3}}\) & \(\revii{3.02 \cdot 10^{-3}}\) \\
        & & \(\revii{1.35 \cdot 10^{-3}}\) & \(\revii{3.39 \cdot 10^{-3}}\) \\
        & & \(\revii{1.37 \cdot 10^{-2}}\) & \(\revii{3.49 \cdot 10^{-3}}\) \\
        \cline{3-4}
        \\[-0.35cm] 
        & \([32]\) & \(\revii{2.55 \cdot 10^{-4}}\) & \(\revii{1.02 \cdot 10^{-3}}\) \\
        & & \(\revii{1.93 \cdot 10^{-4}}\) & \(\revii{2.07 \cdot 10^{-3}}\) \\
        & & \(\revii{3.88 \cdot 10^{-4}}\) & \(\revii{2.32 \cdot 10^{-3}}\) \\
        \cline{3-4}
        \\[-0.35cm] 
        & \([64]\) & \(\revii{6.02 \cdot 10^{-5}}\) & \(\revii{6.26 \cdot 10^{-4}}\) \\
        & & \(\revii{7.76 \cdot 10^{-5}}\) & \(\revii{9.09 \cdot 10^{-4}}\) \\
        & & \(\revii{1.26 \cdot 10^{-4}}\) & \(\revii{1.20 \cdot 10^{-3}}\) \\
        \bottomrule
    \end{tabularx}
\end{table}

 \revii{The test and calibration losses of the ring-parameterized PANN model $\modelsym_{\parameter}$ are provided for different NN architectures in~\cref{tab:node-study-pann-param} after being trained and evaluated according to Study IV, cf.~\cref{subsubsec:study-iv}. All architectures fit the test data very well. The best performance is achieved for the \([32, 32]\) architecture, i.e., two hidden layers with 32 nodes per layer.}

\begin{table}[ht]
    \centering
    \caption{Architecture study for Study IV in~\cref{subsubsec:study-iv}.}
    \label{tab:node-study-pann-param}
    \begin{tabularx}{\linewidth}{XXXX}
        \toprule
        Model & NN architecture & \multicolumn{2}{c}{Loss} \\
        \cline{3-4}
        \\[-0.35cm] 
        & & Calibration & Test \\
        \midrule
        $\revii{\modelsym_{\parameter}}$\ with & \(\revii{[16, 16]}\) & \(\revii{7.28 \cdot 10^{-4}}\) & \(\revii{4.60 \cdot 10^{-3}}\) \\
        & & \(\revii{8.39 \cdot 10^{-4}}\) & \(\revii{4.62 \cdot 10^{-3}}\) \\
        & & \(\revii{8.85 \cdot 10^{-4}}\) & \(\revii{4.82 \cdot 10^{-3}}\) \\
        \cline{3-4}
        \\[-0.35cm] 
        & \(\revii{[32, 32]}\) & \(\revii{1.11 \cdot 10^{-4}}\) & \(\revii{1.41 \cdot 10^{-3}}\) \\
        & & \(\revii{3.29 \cdot 10^{-4}}\) & \(\revii{1.60 \cdot 10^{-3}}\) \\
        & & \(\revii{2.54 \cdot 10^{-4}}\) & \(\revii{1.83 \cdot 10^{-3}}\) \\
        \cline{3-4}
        \\[-0.35cm] 
        & \(\revii{[64, 64]}\) & \(\revii{4.14 \cdot 10^{-5}}\) & \(\revii{1.06 \cdot 10^{-3}}\) \\
        & & \(\revii{6.24 \cdot 10^{-5}}\) & \(\revii{1.25 \cdot 10^{-3}}\) \\
        & & \(\revii{8.05 \cdot 10^{-5}}\) & \(\revii{1.36 \cdot 10^{-3}}\) \\
        \bottomrule
    \end{tabularx}
\end{table}

\section{Data study}
\label{sec:data study}

To assess the required amount of calibration data per radius or parameter in the training of the PANN models in \revii{Studies II, III, IV, cf. ~\cref{subsubsec:study-ii,subsubsec:study-iii,subsubsec:study-iv}, respectively, the calibration and test losses for models trained on three different numbers of concentrically sampled load paths with $\epsilon = 0.1$ perturbation, cf.~\cref{subsubsec:sampling} are listed in~\cref{tab:data-study-pann}. All models were calibrated with the same validation and test datasets according to the strategy presented in~\cref{subsubsec:study-ii}. The values for the three best out of ten trained models are listed for every number of paths.} The test losses decrease when increasing the number of paths from 64 to 128. Increasing the number of load paths beyond 128 does not improve accuracy further. However, since the cost of generating the data and calibrating the models is relatively low, it was decided always to use 256 load paths throughout this work, knowing that 128 load paths are sufficient in practice.

\begin{table}[h]
    \centering
    \caption{Data study for Study II, cf.~\cref{subsubsec:study-ii}.}
    \label{tab:data-study-pann}
    \begin{tabularx}{\linewidth}{XXXXX}
        \toprule
        Model & NN architecture & Concentric training paths & \multicolumn{2}{c}{Loss} \\
        \cline{4-5}
        \\[-0.35cm] 
        & & & Calibration & Test \\
        \midrule
        $\modelsym$ with & \([32]\) & 64 ($\epsilon=0.1$) & \(\revii{1.39 \cdot 10^{-4}}\) & \(\revii{2.36 \cdot 10^{-3}}\) \\
        & & & \(\revii{1.78 \cdot 10^{-4}}\) & \(\revii{3.30 \cdot 10^{-3}}\) \\
        & & & \(\revii{2.38 \cdot 10^{-4}}\) & \(\revii{4.00 \cdot 10^{-3}}\) \\
        \cline{4-5}
        \\[-0.35cm] 
        & & 128 ($\epsilon=0.1$) & \(\revii{1.26 \cdot 10^{-4}}\) & \(\revii{5.86 \cdot 10^{-4}}\) \\
        & & & \(\revii{1.53 \cdot 10^{-4}}\) & \(\revii{8.26 \cdot 10^{-4}}\) \\
        & & & \(\revii{1.84 \cdot 10^{-4}}\) & \(\revii{9.30 \cdot 10^{-4}}\) \\
        \cline{4-5}
        \\[-0.35cm] 
        & & 256 ($\epsilon=0.1$) & \(\revii{2.55 \cdot 10^{-4}}\) & \(\revii{1.02 \cdot 10^{-3}}\) \\
        & & & \(\revii{1.93 \cdot 10^{-4}}\) & \(\revii{2.07 \cdot 10^{-3}}\) \\
        & & & \(\revii{3.88 \cdot 10^{-4}}\) & \(\revii{2.32 \cdot 10^{-3}}\) \\
        \bottomrule
    \end{tabularx}
\end{table}
}

\renewcommand*{\bibfont}{\footnotesize}
\printbibliography

\end{document}